\documentclass[twocolumn]{aastex6}
\usepackage{natbib}
\usepackage{amsmath}
\usepackage{longtable}
\usepackage{threeparttablex}
\usepackage{threeparttable}
\usepackage{afterpage}
\usepackage{multirow}
\newcommand\chandra{{\sl Chandra} }
\newcommand\chandranosp{{\sl Chandra}}
\newcommand\xmm{{\sl XMM-Newton}}
\newcommand\nustar{{\sl NuSTAR}}

\newcommand\cosmos{{COSMOS}}
\newcommand\ecdfs{{ECDF-S}}
\newcommand\cdfs{{CDF-S}}
\newcommand\egs{{EGS}}
\newcommand\ser{{Serendip}}

\def\ls{{_<\atop^{\sim}}}
\def\gs{{_>\atop^{\sim}}}
\newcommand\cgs{\rm{erg\,s^{-1}\,cm^{-2}}}

\newcommand{\fluxcgs}{erg~s$^{-1}$~cm$^{-2}$}
\newcommand{\lumcgs}{erg~s$^{-1}$}
\newcommand{\lumcgsin}{\rm erg~s^{-1}}

\newcommand{\lognh}{{\rm log} (N_{\rm H}/{\rm cm^{-2}})}
\newcommand{\lognhsq}{{\rm log} [N_{\rm H}/{\rm cm^{-2}}]}
\newcommand{\lognhonly}{{\rm log} N_{\rm H}}
\newcommand{\nh}{cm$^{-2}$}

\newcommand{\nhsym}{$N_{\rm H}$}
\begin{document}

\title{The \nustar\ Extragalactic Surveys: X-ray spectroscopic analysis of the bright hard-band selected sample}
\shorttitle{X-ray properties of hard-band selected \nustar\ Extragalactic Survey Field sources}
\shortauthors{Zappacosta et al.}
\author{L. Zappacosta\altaffilmark{1},
A.~Comastri\altaffilmark{2}, 
F.~Civano\altaffilmark{3},
S.~Puccetti\altaffilmark{4},
F.~Fiore\altaffilmark{1},
J.~Aird\altaffilmark{5,6}, 
A.~Del~Moro\altaffilmark{7,6},  
G.~B.~Lansbury\altaffilmark{6,5}, 
G.~Lanzuisi\altaffilmark{8,2}, 
A.~Goulding\altaffilmark{9}
J.~R.~Mullaney\altaffilmark{10}, 
D.~Stern\altaffilmark{11}, 
M.~Ajello\altaffilmark{12}, 
D.M.~Alexander\altaffilmark{6}, 
D.~R.~Ballantyne\altaffilmark{13}, 
F.~E.~Bauer\altaffilmark{14,15,16},
W.~N.~Brandt\altaffilmark{17,18,19},
C.-T.~J.~Chen\altaffilmark{17,18},
D.~Farrah\altaffilmark{20}
F.~A.~Harrison\altaffilmark{21}, 
P.~Gandhi\altaffilmark{6,22},
L.~Lanz\altaffilmark{23},
A.~Masini\altaffilmark{2,8},
S.~Marchesi\altaffilmark{12},
C.~Ricci\altaffilmark{14,24,25}
E.~Treister\altaffilmark{14}
}

\altaffiltext{1}{INAF - Osservatorio Astronomico di Roma, via di Frascati 33, 00078 Monte Porzio Catone, Italy}
\altaffiltext{2}{INAF Osservatorio Astronomico di Bologna, via Gobetti 93/3, 40129 Bologna, Italy}
\altaffiltext{3}{Harvard-Smithsonian Center for Astrophysics, 60 Garden Street,
  Cambridge, MA 02138, USA}
\altaffiltext{4}{Agenzia Spaziale Italiana-Unit\'a di Ricerca Scientifica, Via del Politecnico, 00133 Roma, Italy}
\altaffiltext{5}{Institute of Astronomy, University of Cambridge,
  Madingley Road, Cambridge, CB3 0HA, UK}
\altaffiltext{6}{Centre for Extragalactic Astronomy, Department of Physics,
  Durham University, South Road,  Durham, DH1 3LE, UK}
\altaffiltext{7}{Max-Planck-Institut f\"ur Extraterrestrische Physik
  (MPE), Postfach 1312, D85741, Garching, Germany}
\altaffiltext{8}{Dipartimento di Fisica e Astronomia (DIFA), Universit\`a di Bologna, via Gobetti 93/2, 40129, Bologna, Italy}
\altaffiltext{9}{Department of Astrophysical Sciences, Princeton University, Princeton, NJ, 08544, USA}
\altaffiltext{10}{Department of Physics and Astronomy, The University of
  Sheffield, Hounsfield Road, Sheffield, S3 7RH, UK}
\altaffiltext{11}{Jet Propulsion Laboratory, California Institute of
  Technology, 4800 Oak Grove Drive, Mail Stop 169-221, Pasadena, CA
  91109, USA}
\altaffiltext{12}{Department of Physics and Astronomy, Clemson University, Clemson, SC 29634-0978, USA}
\altaffiltext{13}{Center for Relativistic Astrophysics, School of Physics,
  Georgia Institute of Technology, Atlanta, GA 30332, USA}
\altaffiltext{14}{Instituto de Astrof\'{\i}sica, Facultad de F\'{i}sica,
  Pontificia Universidad Cat\'{o}lica de Chile, 306, Santiago 22,
  Chile}
\altaffiltext{15}{Millennium Institute of Astrophysics, Vicu\~{n}a Mackenna 4860, 782
0436 Macul, Santiago, Chile}
\altaffiltext{16}{Space Science Institute, 4750 Walnut Street, Suite 205,
  Boulder, CO 80301, USA}
\altaffiltext{17}{Department of Astronomy and Astrophysics, 
  The Pennsylvania State University, University Park, PA 16802, USA}
\altaffiltext{18}{Institute for Gravitation and the Cosmos, Pennsylvania State University, University Park, PA 16802, USA}
\altaffiltext{19}{Department of Physics, Pennsylvania State University, University Park, PA 16802, USA}
\altaffiltext{20}{Department of Physics, Virginia Tech, Blacksburg, VA 24061, USA}
\altaffiltext{21}{Cahill Center for Astrophysics, 1216 East California
  Boulevard, California Institute of Technology, Pasadena, CA 91125,
  USA}
\altaffiltext{22}{School of Physics and Astronomy, University of Southampton, Highfield, Southampton SO17 1BJ, UK}
\altaffiltext{23}{Department of Physics and Astronomy, Dartmouth College, 6127 Wilder Laboratory, Hanover, NH 03755, USA}
\altaffiltext{24}{Chinese Academy of Sciences South America Center for Astronomy and China-Chile Joint Center for Astronomy, Camino El Observatorio 1515, Las Condes, Santiago, Chile}
\altaffiltext{25}{Kavli Institute for Astronomy and Astrophysics, Peking University, Beijing 100871, China}

\begin{abstract}
  We discuss the spectral analysis of a sample of 63 Active Galactic Nuclei (AGN) detected above a limiting flux of $S(8-24 \rm keV)=7\times10^{-14}$~\fluxcgs\ in the multi-tiered \nustar\ Extragalactic Survey program. 
  The sources span a redshift range $z=0-2.1$ (median $\langle z \rangle=$0.58).
  The spectral analysis is performed over the broad 0.5-24~keV energy range, combining \nustar\ with \chandra\ and/or \xmm\ data and employing empirical and physically motivated models.
This constitutes the largest sample of AGN selected at $>10$~keV to be homogeneously spectrally analyzed at these flux levels. 
We  study the distribution of spectral parameters such as photon index, column density (\nhsym), reflection parameter ($R$) and 10-40~keV luminosity ($L_{\rm X}$).  
Heavily obscured ($ \lognhsq\ge23$) and Compton Thick (CT; $ \lognhsq\ge24$) AGN constitute $\sim$25\% (15-17 sources) and $\sim$2-3\% ( 1-2 sources) of the sample, respectively.
The observed \nhsym\ distribution fairly agrees with  predictions of Cosmic X-ray Background population synthesis models  (CXBPSM).   We estimate the intrinsic fraction of AGN as a function of \nhsym, accounting for the bias against obscured AGN in a flux-selected sample. The fraction 
  of CT AGN relative to $\lognhsq=20-24$ AGN is poorly constrainted, formally in the range 2-56\% (90\% upper limit of 66\%). 
We derived a fraction ($f_{abs}$) of obscured AGN ($\lognhsq=22-24$)  as a function of $L_{\rm X}$  in agreement  with CXBPSM and previous $z<1$ X-ray determinations. Furthermore $f_{abs}$ at $z=0.1-0.5$ and $\log (L_{\rm X}/\lumcgsin)\approx43.6-44.3$  agrees with observational measurements/trends obtained over larger redshift intervals. We report a significant anti-correlation of $R$ with $L_{\rm X}$ (confirmed by our companion paper on stacked spectra) with considerable scatter around the median $R$ values.
\end{abstract}


\section{Introduction}
Over the last decade the advent of \chandra\ and \xmm\ allowed extragalactic
blank-field X-ray surveys to reach sufficient sensitivities (down to
$10^{-17}$~\fluxcgs) and sky coverage (from tenths to several square degrees)
to allow the study of distant populations of Active Galactic Nuclei \citep[AGN,][]{HM2006,L2012,X2012,BA2015,LBX2017}. They resolved most (up
to 80-90\%) of the Cosmic X-ray Background (CXB) at energies below 10~keV \citep[e.g.][]{HM2006,C2017} as a mixture of obscured and unobscured AGN,
in agreement with early population-synthesis model predictions \citep[][]{SW1989,C1995}. The fraction of resolved CXB gradually decreases with energy
being of the order of $\sim50$\% above $\sim$10~keV and only few percents at $>10$~keV with {\it Swift/BAT} \& {\it INTEGRAL} studies \citep[][]{K2007,A2008}. 

The missing unresolved AGN population which is needed to account for
the remaining high energy CXB flux may be made up of a numerically non-negligible
population  of heavily obscured ($\lognhsq\gs23$) non-local AGNs
\citep[][]{W2005,X2012}.
It is therefore crucial to directly investigate the
distribution of the obscured AGN population at the high column densities contributing to the CXB at high energies.  
A population of AGN with  column densities in excess
of $10^{24}$~\nh, called Compton Thick (CT) AGNs, and numerically comparable to the absorbed Compton Thin AGN, has long been posited to be
responsible for the unaccounted 10-25\% of the CXB flux required by population synthesis models in order to reproduce its peak at 20-30~keV
\citep[][hereafter G07]{C1995,G2007}. Recent works though suggest that also less obscured sources may contribute significantly to the missing flux at $>10$~keV once  other relevant high-energy spectral complexities of the AGN spectrum are taken into proper consideration \citep{TUV2009,Ak2012,U2014}. The latter would therefore lessen the need for a contributing sizable population of CT sources.

Given their very large column densities, the most obscured sources can effectively be
detected in the X-rays at rest-frame energies $>5-10$~keV since their primary
continuum is strongly suppressed at softer energies. This can be currently
done (i) locally ($z<0.1$) by targeting bright sources \citep[e.g., $\gs5\times10^{-12}$~\fluxcgs;][]{B2013} Seyfert-type ($L_X\approx5\times10^{43}$~\lumcgs) with hard X-ray ($>10$~keV) surveys such as those performed by {\it Swift/BAT} \& {\it INTEGRAL} \citep[][]{K2007,A2008} and (ii) at high redshifts ($z>1$) with the most sensitive \chandranosp/\xmm\ observations of the deep/medium survey fields \citep[e.g.][]{C2016,LBX2017}. Through either spectral or hardness ratio analysis they allowed to quantify and characterize the obscured Compton Thin ($\lognhsq=22-24$) AGN population and further shed light on the known decreasing trend between the numerical relevance of this population compared to all AGN (absorbed fraction) and the source luminosity \citep[][]{LE1982,G2007,Bu2011,B2015} and its redshift evolution \citep{LF2005,TU2006,B2006,A2015b,B2015,Liu2017}. They also allowed to explore the importance of the CT population although with different constraining power and different non-negligible degrees of bias especially at the highest column densities and lowest luminosities \citep[e.g.][]{Bu2011,B2014,L2015,B2015,R2015}. Indeed the large diversity in the spectral shapes as well as poorly explored observational parameters in low counting regimes\footnote{i.e. at the highest column densities or at the high/low energy spectral boundaries where the instruments are less sensitive.} such as the high energy cut-off and the  reflection strength at high energies \citep[][hereafter BA11]{TUV2009,Ba2011}, the scattered fractions at low energies \citep{BU2012,L2015} or physical parameters such as the Eddington ratio \citep{DB2010}, may further introduce uncertainty or biases, enlarging the possible range of the fraction of CT sources to one order of magnitude \citep[][]{Ak2012} or even significantly reduce their importance \citep{GP2007}. Indeed, given the paucity of CT sources  effectively contributing to the CXB missing flux,  the most recent population-synthesis models try to explain the CXB missing component as  mainly a pronounced reflection contribution from less obscured sources with a reduced contribution by CT AGN \citep[][]{TUV2009,Ba2011,Ak2012}.
Going deeper at high energies while retaining the capability of being greatly less affected by obscuration bias, will enable us to efficiently sample a more distant ($z=0.1-1$) and luminous population (i.e. at the knee of the luminosity function, $L_X\approx10^{44}$~\lumcgs) of obscured sources and better characterize their high energy spectrum, substantially improving constraints on the majority of the obscured AGN contributing to the CXB \citep[e.g., ][]{Gi2013}. 
The {\it Nuclear Spectroscopic Telescope Array}
\citep[\nustar; ][]{H2013} is perfectly tailored for this task.
Indeed, as the first hard X-ray focusing
telescope in orbit, it provides a two order of magnitude increase in sensitivity compared to 
any previous hard X-ray detector. With its higher
sensitivity, \nustar\ has  resolved $\sim35\%$ of the CXB near
its peak \citep[][hereafter H16]{H2016} and is able to probe the
hard X-ray ($>10$~keV) sky beyond the local Universe ($z>0.1$).

The \nustar\ wedding-cake extragalactic survey strategy
focuses on several well-known medium-deep fields with extensive
multi-wavelength coverage. The core of it includes the EGS (Del~Moro et al. in prep.), E-CDFS \citep{M2015}, COSMOS \citep{C2015} fields, and a
wider and typically shallower Serendipitous survey \citep[][L17]{L2017}. A further extension of it with the observation of two deep fields (CDF-N, Del Moro et al. in prep; UDS, Masini et al. submitted) has also recently been completed. 
This multi-tiered program has already detected 676 AGN out to
$z\approx3.4$ \citep[][]{A2013,C2015,M2015,A2015,L2017}, of which 228 are
significantly detected in the hard 8-24~keV \nustar\ band. In
particular, at low redshift, \citet{C2015} presented the spectroscopic identification
of a local ($z\sim0.04$) low-luminosity ($\sim5\times10^{42}$~\lumcgs) CT AGN not
previously recognized by either \chandra\ or \xmm\  with a column density 
\nhsym$\ge10^{24}$~\nh. \citet{L2017b} identified
three similar sources at $z<0.1$ with even higher obscuration in the
\nustar\ Serendipitous Survey. At high-redshift, \citet{DM2014} 
presented the detection of a heavily absorbed (\nhsym$=6\times10^{23}$~\nh) quasar at $z=2$.

The redshift range and the
luminosities probed by the \nustar\ extragalactic survey program are well 
matched to CXB population-synthesis models in terms of characterization of the AGN high energy spectral shape and of the dominant obscured populations contributing to the CXB. In the latter case population synthesis models predict the largest CT AGN contributions from
sources at $z=$0.4-1.2 with luminosities $L_{2-10}\ls10^{44}$~\lumcgs\ \citep[e.g., ][]{Gi2013} and that their contribution to  the residual CXB flux 
may amount to 90\% by $z\sim2$. 
\citep[][]{TUV2009}.
We therefore  expect
\nustar\ to start to evaluate the relative importance of the obscured AGN populations and shed light
on the main aspects contributing to the still unaccounted remaining flux on the peak of the CXB (i.e. heavy absorption versus reflection).

In order to elucidate on these aspects 
in this paper we carry out a systematic broad-band (0.5-24~keV)
spectral analysis of 63  sources detected in the core \nustar\ Extragalactic Survey program and selected  
to have fluxes in the 8-24~keV energy band brighter than $S_{\rm 8-24}=7\times10^{-14}$~\fluxcgs.  We 
complement the \nustar\ data with archival low-energy data from
\chandra\ and \xmm. We perform broadband (0.5-24~keV)
spectral modeling, characterize their spectral properties, obtain a column density distribution,
absorbed/CT fractions and source counts and compare with predictions from population-synthesis 
models and past observational works. A companion paper, \citet[][DM17]{DM2017}, reports on the 
properties of the average X-ray spectra from all  sources detected in the \nustar\ deep and medium survey fields.

This paper is  organized as follow. Section~\ref{sampledes}  presents the sample, with Sections~\ref{datareduction} and \ref{datanalysis} devoted to the  data reduction and spectral characterization of the source properties, respectively. We then discuss the column density distribution (Section~\ref{NHdistribution}),  fraction of CT AGN (Section~\ref{CTfrac}), fraction of absorbed sources as a function of luminosity and redshift  (Section~\ref{fabs}) and source counts  (Section~\ref{counts}). We discuss the results in Section~\ref{discussion} and present the conclusions in Section~\ref{conclusions}. Relevant notes on individual sources are presented in the Appendix.

Throughout the paper we  adopt a flat cosmology with
$\Omega_\Lambda=0.73$ and $H_0=70\,\rm  km\, s^{-1}\, Mpc^{-1}$. Errors are quoted at the  $1\sigma$ level and upper/lower limits at 90\% confidence level (c.l.).
The X-ray luminosities are quoted in the standard (for \nustar\ survey studies) rest-frame 10-40~keV energy band.

\begin{center}
\tabletypesize{\scriptsize}
    \begin{deluxetable*}{lccccccccccr}
 
      \tablecaption{Main properties of the \nustar\ Extragalactic Surveys and relative catalogs
\label{surveys}
      }
      \tablehead{  Survey          &    Total Exp.\tablenotemark{a}   & Pointings Exp.\tablenotemark{b} & Deepest Exp.\tablenotemark{c} & Pointing layout\tablenotemark{d} & Area    &  \multicolumn{2}{c}{Detected Sources}   & \multicolumn{2}{c}{Sensitivity 50\%\tablenotemark{e}}  & Ref.  \\
                                   &      (Ms)         &  (ks)           &  (ks)          &                 & ($\rm deg^2$) & 3-24~keV   &   8-24~keV &  3-24~keV   &   8-24~keV    & 
      }
      \startdata
      \egs                         &  1.6          &     50        &   400        &   $8\times1$    & $0.24$   & 39  &  14  &   $0.39$  &  $0.35$    & DMIP  \\
      \ecdfs                       &  1.5         &  45-50        &   360        &   $4\times4$    & $0.31$   &  49 & 19   &   $ 0.39 $  &  $0.45$    & M15  \\
      \cosmos                      &  3.0            &  20-30        &    90        &   $11\times11$  & $1.73$   &  91 & 32   &   $ 0.77 $  &   $0.93 $   & C15  \\
      Serendipitous\tablenotemark{f}&  2.2          &  20-1000      &   970        &   Random        & $3.91$   &  118  & 38   &   $1.70$  &   $1.35 $ & L17  
      \enddata
\tablenotetext{a}{Total exposure time devoted to the survey;}
\tablenotetext{b}{Average exposure times of the single pointings; Notice that the Serendipitous survey consists of pointings with a large range of exposure times.}
\tablenotetext{c}{Average exposure in the deepest field;}
\tablenotetext{d}{Tiling design of the survey;}
\tablenotetext{e}{Flux reached at 50\% of the survey area coverage. In units of $10^{-13}$~\fluxcgs.}
\tablenotetext{f}{This is a sub-sample of the Serendipitous survey sample presented in L17 (see Section~\ref{serendipitous})}
    \end{deluxetable*}
  \end{center}

\section{Description of the sample}\label{sampledes}
We draw our sample from the high-energy \nustar\ catalogs compiled for the \cosmos\ \citep[][C15]{C2015}, \ecdfs\ \citep[][M15]{M2015}, \egs\ (Del~Moro et al. in prep., DMIP) and Serendipitous Survey fields (L17). In order to have consistent catalogs, the same data-reduction tasks, mosaicing procedures, source detection steps, photometry and deblending algorithm were applied to all survey fields (see C15, M15 and \citealt{A2015} for details). In the following we briefly outline the source identification procedure adopted in each catalog. The identification of the sources was consistently done through a SExtractor-based procedure on false probability maps generated  on the mosaicked images accounting for the corresponding  background maps in three energy bands (3-8, 8-24, 3-24~keV). No positional priors from previous low energy X-ray surveys have been used in the source identification. Through simulated data, a proper threshold to set the significance of each source identification in each band has been adopted and the final balance between completeness and reliability in each catalog has been chosen so that the possible spurious sources down to the limiting flux in each catalog do not exceed the number of 2-3. Further details and description of the procedures regarding deblending, photometry, final catalog building and association to low-energy counterparts are reported in each catalog paper.
For our purposes in order to minimize obscuration bias, we selected objects with relatively bright fluxes in the hard 8-24~keV band. The fluxes adopted for this selection have been estimated from the 8-24~keV counts collected in $30^{\prime\prime}$ apertures\footnote{The fluxes reported in C15 are from  $20^{\prime\prime}$ apertures. They have been estrapolated to $30^{\prime\prime}$ aperture fluxes by assuming a 1.47 constant conversion factor. This factor is obtained as ratio between the fluxes in $30^{\prime\prime}$ and $20^{\prime\prime}$ apertures measured from the on-axis \nustar\ point-spread function.} by the catalog papers by assuming a power-law model with $\Gamma=$1.8. Whenever possible, we complemented \nustar\ data with archival lower energy data from \chandra and \xmm.

\subsection{Deep-medium survey fields}
Given the $12^\prime\times12^\prime$ \nustar\ field of view, the survey fields (\cosmos, \ecdfs\ and \egs) were observed with a mosaicing strategy whereby each neighboring pointing was shifted by half of a field of view. This tile arrangement produces homogenous and continuous coverage in the deep central region with contiguous shallower edges.
The main properties of these surveys are reported in Table~\ref{surveys}. 
\begin{figure}[t!]
   \begin{center}
\includegraphics[width=0.45\textwidth]{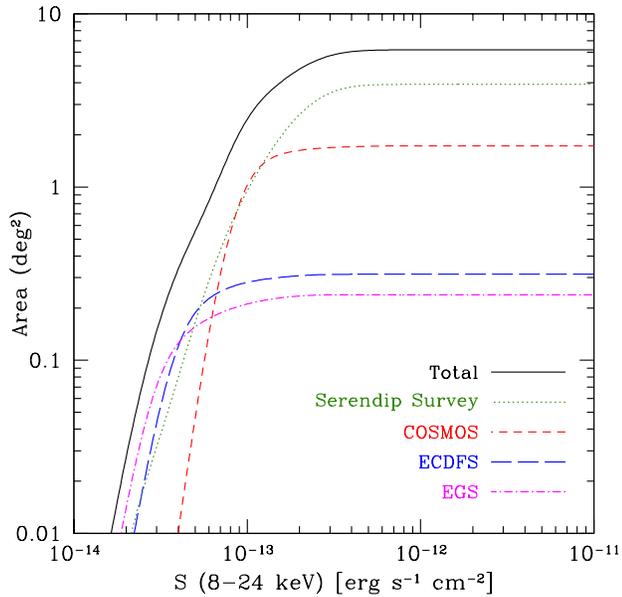}
   \end{center}
\caption{Total and individual sensitivity curves as a function of the hard-band flux for the surveys included in our sample. Black solid curve is for the entire sample, magenta dot-dashed is for \egs, blue long-dashed is for \ecdfs, red short-dashed is for \cosmos and green dotted is for the \ser\ Survey.}
   \label{areas}
\end{figure}

Despite \nustar\ being sensitive up to 79~keV, typical faint sources in the deep surveys are not detected to such high energies.
In the extragalactic survey work to date, we have therefore only considered three energy bands: 3-24~keV (total), 3-8~keV (soft) and 8-24~keV (hard).

Fig.~\ref{areas} reports the 8-24~keV sensitivity curves as a function of hard-band flux for all the fields. The sensitivities at 50\% survey coverage are reported in Table~\ref{surveys}.   
Notice that they are based on the assumption of an unabsorbed $\Gamma=1.8$ power-law spectrum.  This is an approximation which is reasonable for Compton-thin sources given that above 8~keV their spectrum is minimally altered at the highest column densities (i.e. $\lognhsq\gs23$). It may result somewhat inadequate for CT sources whose spectrum substantially deviates from this assumed spectral shape within this hard band. It may therefore give biased results in calculating the intrinsic distribution of physical quantities for the sampled AGN population. We account for this by correcting a-posteriori for this bias (see Sect.~\ref{CTfrac} and Fig.~\ref{absbias}).

\begin{figure*}[t!]
   \begin{center}
     \includegraphics[width=0.47\textwidth]{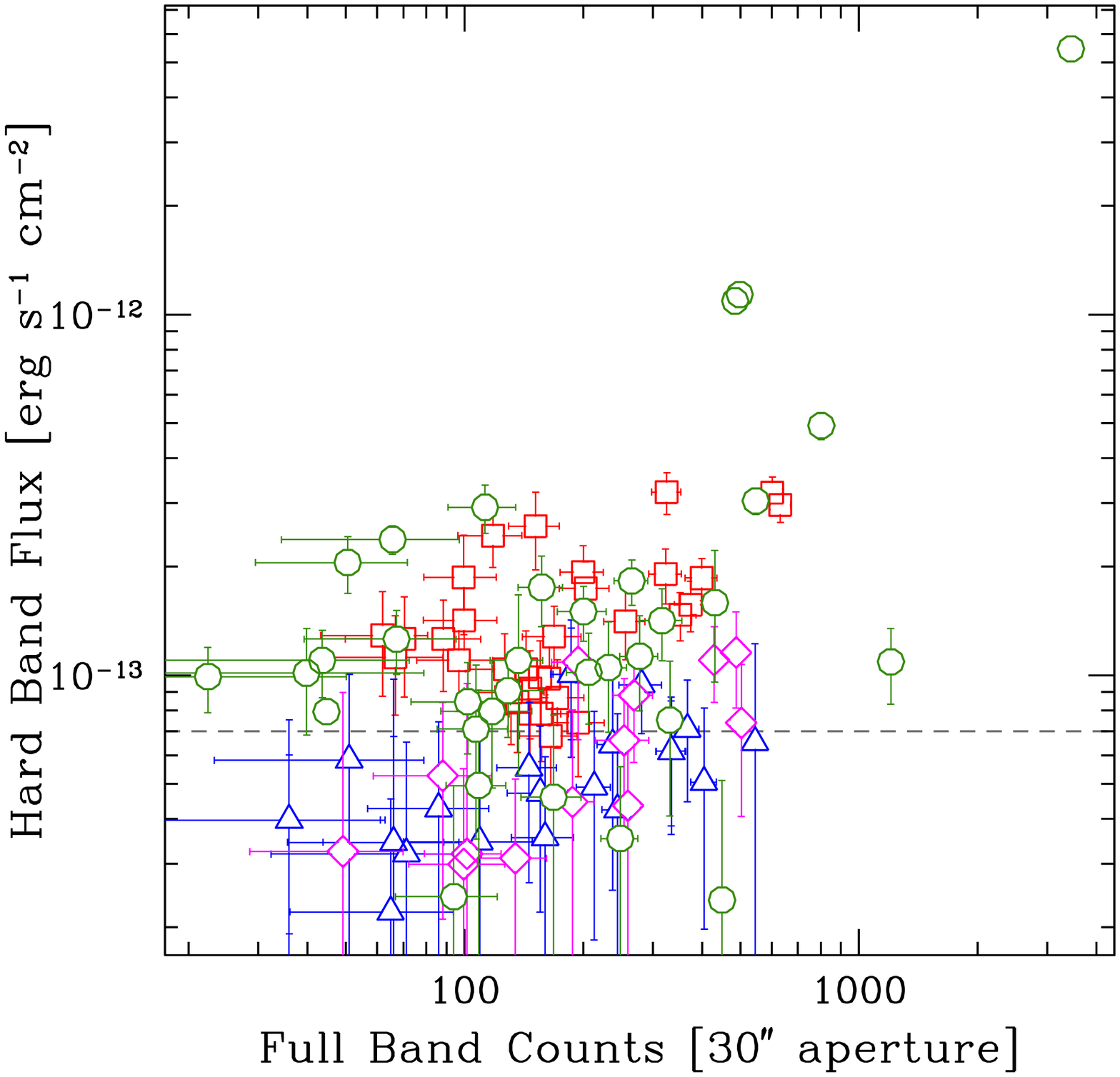}
     \hspace{0.5cm}
\includegraphics[width=0.46\textwidth]{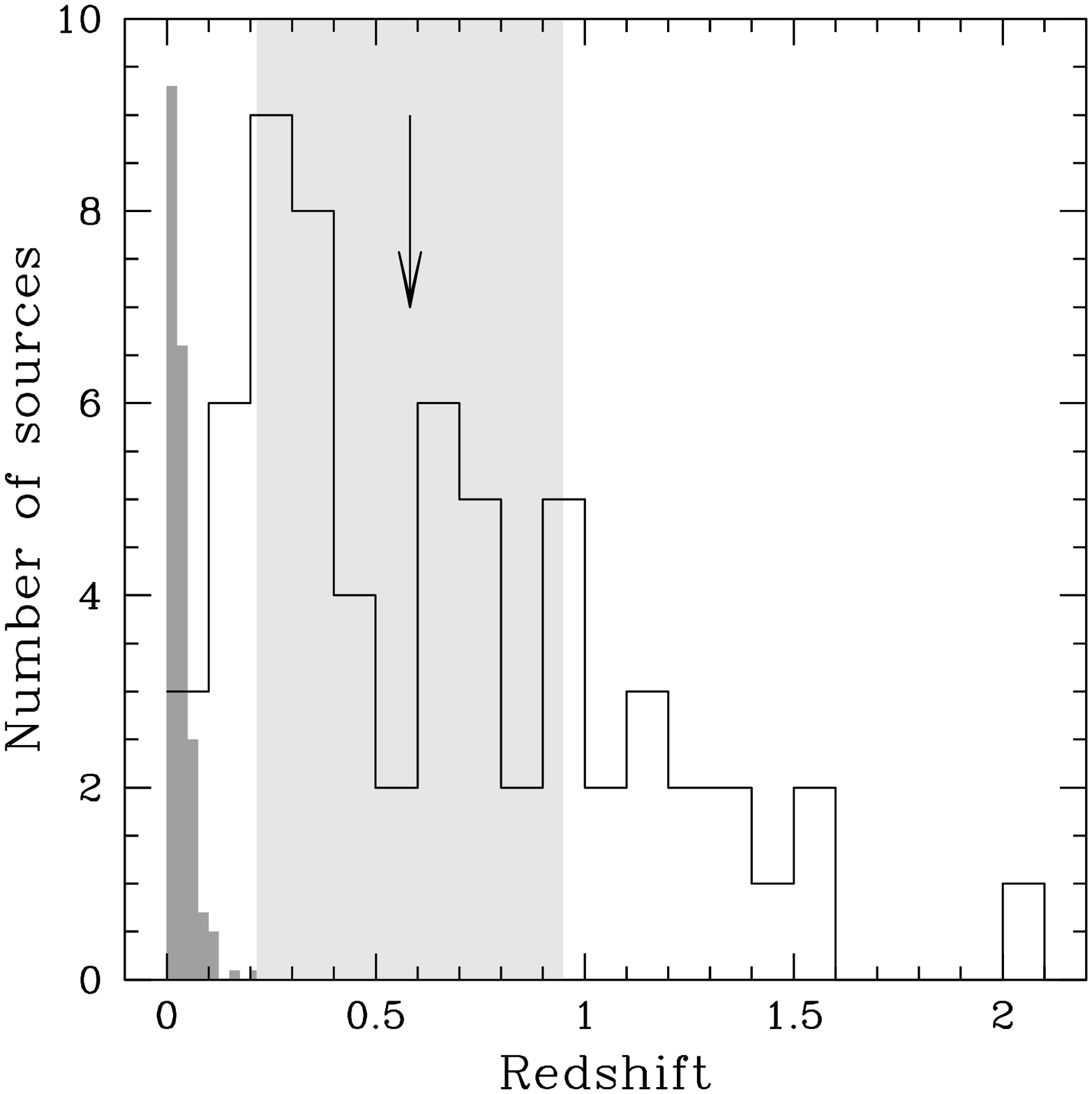} 
   \end{center}
\caption{Left panel: Net counts in the full (3-24~keV) energy band from a $30^{\prime\prime}$ aperture versus the hard (8-24~keV) deblended aperture-corrected flux for the hard-band detected objects in the \nustar-\cosmos\ (red squares), \nustar-\ecdfs\ (blue triangles), \nustar-\egs\ (magenta diamonds) and \nustar-\ser\ (green circles) catalogs. For the COSMOS sources, the flux has been extrapolated from the $20^{\prime\prime}$ apertures reported in C15 assuming a constant conversion factor of 1.47 based on the on-axis \nustar\ point-spread function.
 The horizontal dashed line indicates the  threshold value of $7\times10^{-14}~\cgs$ defining our sample. Right panel:  spectroscopic redshift distribution of our sample (open histogram) compared to the the 199 local sources studied by {\it Swift}-BAT in \citet{Bu2011} (dark grey, with normalization and histogram binning rescaled by a factor of 10). The arrow indicates the median redshift value for the sample ($\langle z \rangle = 0.58$) and the light grey region shows the interquartile range.}
   \label{sample}
\end{figure*}

\subsection{Serendipitous survey fields}\label{serendipitous}
The serendipitous fields considered in this work consist of all fields analyzed as part of the Serendipitous Survey through 2015 January 1. This extends the sample presented by \citet{A2013}, and is a subset of the program presented in  L17. 
The selection criteria adopted are reported in the following and constitute a slight modification to those employed in \citet{A2015}:
\begin{itemize}
\item we minimize Galactic point-source contamination by requiring Galactic latitudes $>20^{\circ}$;
\item to emphasize fields where our serendipitous survey follow-up is currently more complete, we only consider fields accessible from the northern hemisphere by requiring declinations $>-5^{\circ}$;
\item we exclude fields with a large contamination from the primary targets by requiring  $<10^6$~counts within 120$^{\prime\prime}$ of the aimpoint, and that primary targets contribute   $\ls6$\% to the extracted emission of the serendip source within the extraction region.
\end{itemize}
After these cuts, the sky coverage of the serendipitous survey considered here amounts to $\approx4\,\rm{deg^2}$ (see Fig.~\ref{areas}). Further survey details are reported in Table~\ref{surveys}. It is worth noting that despite the Serendipitous Survey having sensitivity  better than \cosmos\ over a wider area and comparable faint source sensitivity to \ecdfs, it also has the disadvantage of having less multi-wavelength coverage. This usually translates to lower redshift completeness (from optical spectroscopy) and a poorer quality X-ray coverage at low energies from \chandra\ and/or \xmm\ (see L17).

\subsection{Selected sample}\label{selsample}
The final catalogs consist of 91, 49, 39 and 118 objects, respectively, from the \nustar\ \cosmos, \ecdfs, \egs\ and the Serendipitous Survey. Of these, 32, 19, 14 and 38, respectively, are significantly detected in the hard-band based on a maximum likelihood estimator (see C15, M15, A15 and L17 for details and the adopted thresholds). These objects are shown in Fig.~\ref{sample} (left panel) which displays the net 3-24~keV counts within a $30^{\prime\prime}$ aperture versus their aperture-corrected photometry in the 8-24~keV energy band. From this combined sample we select sources with hard-band flux $S_{\rm 8-24 keV}\geq7\times10^{-14}$~\fluxcgs. We are sensitive to fluxes larger than this value in 80\% of the surveyed area (see Fig.~\ref{areas}). 
This sub-sample, corresponding to objects above the dashed line in Fig.~\ref{sample}, includes a total of 31, 3, 5 and 24 objects from the four surveys, respectively selected over a total area of $\sim 6~\rm deg^2$. The resulting sample of 63 sources is the focus of the following analysis.
The redshift distribution is reported in the right panel of Fig.~\ref{sample}, compared to the distribution of the 199 local sources detected by {\it Swift}-BAT in the energy range 15-55~keV \citep{Bu2011}. \nustar, with its two orders of magnitude greater sensitivity,  probes sources well beyond the local Universe. 
Table~\ref{tabsample}  reports the position, spectroscopic redshift, \chandra\ and \xmm\  counterparts,  \nustar\ observation IDs, and \nustar\ survey for all 63 sources. When referring to the single sources we use the catalog IDs listed in column 2 prefixed by {\it ecdfs}, {\it egs}, {\it cosmos} and {\it ser} for sources from respectively  the \nustar- \ecdfs, \egs, \cosmos\ and Serendipitous catalogs.

Objects from the deep fields all have unique counterparts at lower energies from either the \chandra\ \citep[][]{L2005,G2012,X2011,C2012,C2016} or \xmm\ \citep[][]{B2010,R2013} surveys of these same fields, with the exception of one source in the \ecdfs\ field ({\it ecdfs5}; this source has two possible counterparts, one at low and one at high-redshift; see Table~\ref{tabsample} and Appendix).

 A few sources have nearby potential contaminants (i.e. inside the \nustar\ extraction radius) in the deep survey fields. Contamination ultimately is unimportant or partially negligible in most cases, as discussed for the affected sources in the Appendix. For some cases ({\it cosmos154} and {\it cosmos181}) we restrict the \nustar\ low energy bound to 4-5~keV, where the contamination becomes less important.  In a few other cases the contamination is such that within the uncertainties it could potentially lower the true hard-band source flux also below the threshold flux used in our sample selection ({\it cosmos107}, {\it cosmos178} and {\it cosmos229}). 
For the Serendipitous Survey sources, most have counterparts from at least \chandra\ or \xmm, the exception being five sources ({\it ser97}, {\it ser285}, {\it ser235}, {\it ser261}, {\it ser409}; see Table~\ref{tabspec}) which have not yet been observed by these satellites. 

\section{Data reduction}\label{datareduction}
\subsection{\nustar}
In order to perform a proper spectral analysis for these low-count point-like 
sources (i.e. from  tens to hundreds of counts; see
Fig.~\ref{sample}), we need to carefully account for: 1. the relatively uniform 
arcmin-scale \nustar\ point spread function \citep[FWHM=$18^{\prime\prime}$; half power
diameter HPD=$58^{\prime\prime}$;][]{H2013}; and 2. the spectrally variable 
and spatially dependent background \citep[for details, see][]{W2014}. In particular, the latter at $<20$~keV is strongly affected by stray light from unfocussed CXB photons reaching the detectors through the open design
of the observatory (called ``aperture background''). 

\begin{center}
\tablewidth{0pt}
\tabletypesize{\scriptsize}
    \begin{deluxetable*}{crrrrrrlr}
      \tablecaption{The bright \nustar\ hard-band spectroscopic sample
\label{tabsample}
      }
      \tablehead{  Name  & $\rm ID_{N}$\tablenotemark{a}    & \multicolumn{1}{c}{R.A.}  & \multicolumn{1}{c}{DEC.} &  \multicolumn{1}{c}{$z$\tablenotemark{b}}  &   \multicolumn{1}{c}{$\rm ID_{C}$\tablenotemark{c}}   &  \multicolumn{1}{c}{$\rm ID_{X}$\tablenotemark{d}}  & \multicolumn{1}{c}{\nustar\ Observation IDs\tablenotemark{e} } & \multicolumn{1}{c}{Catalog}      }
\startdata
  NuSTAR~J100129$+$013636  &   cosmos97  &  150.372537  &  +1.610073  &  0.104  & cid1678  &   2021    &  098001 099001                            &  COSMOS   \\
  NuSTAR~J100249$+$013851  &   cosmos107 &  150.705859  &  +1.647561  &  0.694  & lid1688  &   5496    &  101002 103001                           &  COSMOS   \\
  NuSTAR~J100101$+$014752  &   cosmos129 &  150.256432  &  +1.797837  &  0.907  & cid284  &  54490    &  037002 038001 060001                    &  COSMOS   \\
  NuSTAR~J095815$+$014932  &   cosmos130 &  149.564437  &  +1.825731  &  1.509  & lid961   &   5323    &  090001 091001                            &  COSMOS   \\
  NuSTAR~J095926$+$015348  &   cosmos145 &  149.860885  &  +1.896815  &  0.445  & cid209  &    293    &  012001 018001 062001 063001            &  COSMOS   \\
  NuSTAR~J100055$+$015633  &   cosmos154 &  150.233212  &  +1.942588  &  0.219  & cid1105  &    131    &  029001 030001 035002 036002            &  COSMOS   \\
  NuSTAR~J100024$+$015858  &   cosmos155 &  150.104087  &  +1.982873  &  0.373  & cid358  &      1    &  023001 024001 029001 030001            &  COSMOS   \\
  NuSTAR~J095840$+$020437  &   cosmos178 &  149.668862  &  +2.077021  &  0.340  & cid168  &    417    &  004001 005001 087001 088002            &  COSMOS   \\
  NuSTAR~J100141$+$020348  &   cosmos181 &  150.423842  &  +2.063424  &  0.125  & cid482  &   2608    &  040001 041001 050001 051001            &  COSMOS   \\
  NuSTAR~J095918$+$020956  &   cosmos194 &  149.826665  &  +2.165826  &  1.157  & cid320  &      5    &  003001 004001 009002-B 010001          &  COSMOS   \\
  NuSTAR~J100308$+$020917  &   cosmos195 &  150.785979  &  +2.154816  &  1.470  & lid1646   &   5321    &  109001 111001                            &  COSMOS   \\
  NuSTAR~J095857$+$021320  &   cosmos206 &  149.738507  &  +2.222475  &  1.024  & cid329  &      2    &  003001 004001                            &  COSMOS   \\
  NuSTAR~J100307$+$021149  &   cosmos207 &  150.782581  &  +2.197149  &  0.582  & lid1645   &   5370    &  111001 113001                           &  COSMOS   \\
  NuSTAR~J095817$+$021548  &   cosmos216 &  149.573824  &  +2.263384  &  0.707  & lid633   &  54514    &  086001 087001                           &  COSMOS   \\
  NuSTAR~J100032$+$021821  &   cosmos217 &  150.133457  &  +2.305840  &  1.598  &  cid87  &     18    &  020002 021001 026001 027002            &  COSMOS   \\
  NuSTAR~J095902$+$021912  &   cosmos218 &  149.761712  &  +2.320121  &  0.345  & cid440  &      3    &  002001 003001 008001 009002            &  COSMOS   \\
  NuSTAR~J095909$+$021929  &   cosmos229 &  149.789727  &  +2.324908  &  0.378  & cid420  &     23    &  002001 003001 008001 009002            &  COSMOS   \\
  NuSTAR~J095957$+$022244  &   cosmos232 &  149.990626  &  +2.378889  &  0.931  & cid530  &    212    &  013001 014001 019001 020002            &  COSMOS   \\
  NuSTAR~J100228$+$024901  &   cosmos249 &  150.620155  &  +2.817155  &  0.213  & lid3218   &   5014    &  067001 120001                           &  COSMOS   \\
  NuSTAR~J095945$+$024750  &   cosmos251 &  149.941358  &  +2.797420  &  1.067  & lid545   &   5620    &  076001 077001-A 078001                  &  COSMOS   \\
  NuSTAR~J100238$+$024651  &   cosmos253 &  150.658422  &  +2.780956  &  0.212  & lid484   &   5114    &  120001 121001                           &  COSMOS   \\
  NuSTAR~J095908$+$024310  &   cosmos263 &  149.785525  &  +2.719548  &  1.317  & lid549   &   5230    &  077001 079001 080001                    &  COSMOS   \\
  NuSTAR~J100243$+$024025  &   cosmos272 &  150.682309  &  +2.673758  &  0.669  & lid492   &   5400    &  118001 119001 120001 121001            &  COSMOS   \\
  NuSTAR~J100204$+$023726  &   cosmos282 &  150.520310  &  +2.623974  &  0.519  & lid294   &      7    &  046002 046004 065002 066001            &  COSMOS   \\
  NuSTAR~J095837$+$023602  &   cosmos284 &  149.656383  &  +2.600703  &  0.735  & lid1856   &   2076    &  082001 083001                           &  COSMOS   \\
  NuSTAR~J100232$+$023538  &   cosmos287 &  150.635807  &  +2.593895  &  0.658  & lid280   &   5133    &  046002 116001 117001-B 118001 119001  &  COSMOS   \\
  NuSTAR~J095849$+$022513  &   cosmos296 &  149.704281  &  +2.420472  &  1.108  & cid513  &    126    &  001002 002001                            &  COSMOS   \\
  NuSTAR~J095848$+$022419  &   cosmos297 &  149.700185  &  +2.405449  &  0.375  & cid417  &    135    &  001002 002001                           &  COSMOS   \\
  NuSTAR~J100229$+$023223  &   cosmos299 &  150.624855  &  +2.539961  &  0.432  & lid278   &   5222    &  046002 046004 065002 116001 118001    &  COSMOS   \\
  NuSTAR~J095839$+$022350  &   cosmos322 &  149.662604  &  +2.397242  &  0.356  & lid622   &   1429    &  001002 002001 084001 085001            &  COSMOS   \\
  NuSTAR~J100259$+$022033  &   cosmos330 &  150.747792  &  +2.342593  &  0.044  & lid1791   &   5371    &  113001 115001                            &  COSMOS   \\
  NuSTAR~J033136$-$280132  &   ecdfs5   &   52.901946  &  -28.025645  &  0.141 &  103  &  -  &   001003 001002 001003             & E-CDFS      \\
                         &           &              &              &  1.957 &  100  &  -  &                                    & E-CDFS      \\
  NuSTAR~J033207$-$273736  &   ecdfs20  &   53.032301  &  -27.626858  &  0.976 &  301  & 358 &   013001 013002 014001 014002     & E-CDFS      \\
  NuSTAR~J033328$-$275642  &   ecdfs51  &   53.370361  &  -27.945068  &  0.841 &  712  &  -  &   004001 004002 008001 008002     & E-CDFS      \\
  NuSTAR~J141736$+$523029  &   egs1   &  214.400911  &  +52.508258  &  0.987  &   37  &  -  & 001002 001004 001006 001008  & EGS   \\
  NuSTAR~J141754$+$524138  &   egs9   &  214.475905  &  +52.694030  &  0.464  &  294  &  -  & 001002 001004 001008 002002  & EGS   \\
                $ $        &       &              &              &         &       &     & 002003 002004B 002005 &       \\
  NuSTAR~J142047$+$525809  &   egs26  &  215.196713  &  +52.969305  &  0.201  &  669  &  -  & 006002 006003 006004 006005 & EGS   \\
                $ $        &       &              &              &         &       &     & 007001 007003 007005 007007 &       \\
  NuSTAR~J142052$+$525630  &   egs27  &  215.220076  &  +52.941858  &  0.676  &  622  &  -  & 006002 006004A 006005 007001 & EGS   \\
                $ $        &       &              &              &         &       &     & 007005 007007 &       \\
  NuSTAR~J142027$+$530454  &   egs32  &  215.113227  &  +53.081728  &  0.997  &  863  &  -  & 007001 007003 007005 007007 & EGS   \\
                $ $        &       &              &              &         &       &     & 008001 008002 008003 008004 &       \\
                $ $        &       &              &             &         &     &     &  60002048002                         &                         \\
  NuSTAR~J023228$+$202349  &   ser37  &   38.119089  & +20.397218  &  0.029  &  -  &  -  &  60002047006 60002047004                &  Serendip         \\
  NuSTAR~J035911$+$103126  &   ser77  &   59.798670  & +10.523951  &  0.167  &  -  &  -  &  60061042002                            &  Serendip         \\
  NuSTAR~J051617$-$001340  &   ser97  &   79.073788  &  -0.227904  &  0.201  &  -  &  -  &  60001044004 60001044002                &  Serendip         \\
  NuSTAR~J061640$+$710811  &   ser107 &   94.167546  & +71.136661  &  0.203  &  -  &  -  &  60002048010 60002048006 60002048004    &  Serendip         \\
  NuSTAR~J075800$+$392027  &   ser148 &  119.503085  & +39.341045  &  0.095  &  -  &  -  &  60001131002                            &  Serendip         \\
  NuSTAR~J081909$+$703930  &   ser153 &  124.789365  & +70.658570  &  1.278  &  -  &  -  &  30001031005 30001031003 30001031002    &  Serendip         \\
  NuSTAR~J095512$+$694739  &   ser184 &  148.800066  & +69.794361  &  0.675  &  -  &  -  &  80002092011 80002092009 80002092008    &  Serendip         \\
                           &       &              &             &         &     &     &  80002092007 80002092006 80002092004       &                         \\
                           &       &              &             &         &     &     &  80002092002                               &                         \\
  NuSTAR~J102345$+$004407  &   ser213   &  155.938116  &  +0.735278  &  0.300  &  -  &  -  &  30001027006                          &  Serendip         \\
  NuSTAR~J102628$+$254417  &   ser215  &  156.619145  & +25.738177  &  0.827  &  -  &  -  &  60001107002                           &  Serendip         \\
  NuSTAR~J110740$+$723234  &   ser235  &  166.919819  & +72.542882  &  2.100  &  -  &  -  &  60002042004 60002042002               &  Serendip         \\
  NuSTAR~J112829$+$583151  &   ser243  &  172.122122  & +58.530861  &  0.410  &  -  &  -  &  50002041003 50002041002               &  Serendip         \\
  NuSTAR~J115912$+$423242  &   ser254  &  179.802748  & +42.545158  &  0.177  &  -  &  -  &  60001148004 60001148002 60061217006   &  Serendip         \\
                $ $        &       &              &             &         &     &     &  60061217004B 60061217002                  &                         \\
  NuSTAR~J120613$+$495712  &   ser261 &  181.555033  & +49.953531  &  0.784  &  -  &  -  &  60061357002                            &  Serendip        \\
  NuSTAR~J121358$+$293608  &   ser267  &  183.494820  & +29.602344  &  0.131  &  -  &  -  &  60061335002                           &  Serendip        \\
  NuSTAR~J121425$+$293610  &   ser273  &  183.607905  & +29.603048  &  0.308  &  -  &  -  &  60061335002                           &  Serendip        \\
  NuSTAR~J122751$+$321222  &   ser285  &  186.964887  & +32.206371  &  0.733  &  -  &  -  &  60001108002                           &  Serendip        \\
  NuSTAR~J134513$+$554751  &   ser318  &  206.304766  & +55.797766  &  1.167  &  -  &  -  &  60002028002                           &  Serendip        \\
  NuSTAR~J143026$+$415959  &   ser335  &  217.610385  & +41.999984  &  0.352  &  -  &  -  &  60001103002                           &  Serendip        \\
  NuSTAR~J151508$+$420837  &   ser359  &  228.786883  & +42.143734  &  0.289  &  -  &  -  &  60061348002                           &  Serendip        \\
  NuSTAR~J151654$+$561744  &   ser363  &  229.225216  & +56.295566  &  1.310  &  -  &  -  &  30002039005A 30002039003 30002039002  &  Serendip        \\
  NuSTAR~J171309$+$573421  &   ser382  &  258.288435  & +57.572549  &  0.243  &  -  &  -  &  60001137002                           &  Serendip        \\
  NuSTAR~J181429$+$341055  &   ser401  &  273.621211  & +34.181958  &  0.763  &  -  &  -  &  60001114002                           &  Serendip        \\
  NuSTAR~J182615$+$720942  &   ser409 &  276.563078  & +72.161734  &  1.225  &  -  &  -  &  60161687002                            &  Serendip        \\
  NuSTAR~J204020$-$005609  &   ser451   &  310.087027  &  -0.936058  &  0.601  &  -  &  -  &  30001120005 30001120004 30001120003  &  Serendip        \\
                           &           &              &             &         &     &     &  30001120002                           &                         
  \enddata
    \tablenotetext{a}{Identification name for each source. This is made from a prefix indicating the source parent catalog plus the ID from \nustar\ parent catalogs (Section~\ref{sampledes}). The prefixes of each parent catalog are {\it cosmos} for \cosmos, {\it ecdfs} for \ecdfs\, {\it egs} for \egs\ and {\it ser} for the Serentipitous Survey.}
\tablenotetext{b}{All the redshifts are spectroscopic. They are taken from: \citealt{B2010} (\cosmos), \citealt{L2005}, \citealt{X2011} and \citealt{R2013} (\ecdfs), \citealt{N2015} (\egs) and L17 (\nustar\ Serendipitous Survey).}
\tablenotetext{c}{\chandra\ IDs are from \citealt{E2009} and \citealt{C2016} (\cosmos, with prefix {\it cid} and {\it lid} respectively), \citealt{L2005} (\ecdfs), \citealt{N2015} (\egs).}
\tablenotetext{d}{\xmm\ IDs are from \citealt{B2010} (\cosmos) and \citealt{R2013} (\ecdfs).}
\tablenotetext{e}{To obtain the full \nustar\ observation IDs for the \cosmos, \ecdfs\ and \egs\ fields, the six digit survey identification numbers  60021, 60022 and 60023 must be prefixed, respectively.}
    \end{deluxetable*}
\end{center}

Given the flux levels of the sources in our sample, it is necessary to maximize and carefully account for their contribution relative to the backgrounds (especially with respect to the spatially dependent ``aperture background''). 
We therefore optimize the spectral extraction radius to maximize the signal-to-noise ratio (SNR) and, within the Poissonian uncertainties, the number of collected net counts.
To do this we started with the level 2 data products and simulated background maps where the latter were created using the software {\sc nuskybgd} \citep[][]{W2014} as described in C15 and M15. The simulated background maps reproduce the ``aperture background''  across the FoV and the normalization of the total background in each observation.
In detail  we determined from all the observations pertaining to a given source, the total counts in increasing circular apertures centered on the source position, calculating both source+background counts ($S$) from the event files and  background counts alone ($B$) from the simulated maps. Then we calculated the radial profile for the net source counts  
$N(<r)=S(<r)-B(<r)$
and 
$SNR(<r)=\frac{N(<r)}{\sqrt{N(<r)+2B(<r)}}$. 
The radius for spectral extraction $r_{ex}$ is chosen as the radius which maximizes the SNR profile and, within its $\pm1\sigma$ range, maximizes also $N$. 
In the few (nine for COSMOS and one for ECDFS) cases in which a source is blended with a nearby source (closer than $2$~arcmin), we further reduced $r_{ex}$ so that the source flux  from the contaminating source is reduced, within the aperture, to levels of $5-6\%$. Table~\ref{tabspec} reports $r_{ex}$ values for all the sources in our sample.

We used the task ``nuproducts'' in NUSTARDAS v.1.4.1 with the \nustar\ calibration database (CALDB version 20150123) for the spectral extractions and the creation of relative response files.

The background spectrum for each source spectrum was simulated from the best-fit models of the background across the detectors obtained with {\sc nuskybgd}. This software performs iterative joint fits of the observed backgrounds across the field extracted in $\gs3$~arcmin apertures placed in each chip of each focal plane module. The joint modeling aims to determine the normalization of the different background components and hence characterize them at the position of the source. The fits are performed using spectral models of the instrumental (continuum + line activation due to particle background), cosmic focused (CXB) and cosmic unfocussed background (straylight) components and information on their spatial dependence across the detectors.
We checked each best-fit to ensure that no significant spatial or spectral residuals were present. After this procedure we are in principle able to well reproduce the background spectrum at each position of the detector. We further verify this by creating background-subtracted images and visually inspect them for spatial gradients indicative of poor background modeling.
As a final step, the best fit spectral model is used by {\sc nuskybgd} to simulate the background within the source extraction aperture but using a 100 times higher exposure time to ensure high SNR. 

We then co-added for each source and in each detector spectra, simulated backgrounds and response files. Table~\ref{tabspec} reports \nustar\ net counts and total exposure time collected for each source.

\subsection{\xmm\ and \chandra}
For the \ecdfs\ and \cosmos\ fields we employed all spectra reduced and extracted by previous works. Specifically, for the deep \ecdfs\ field we used \chandra data reduced by \citet{L2005} and \citet{X2016}, and extracted spectra following procedures discused in \citet{DM2014}. For {\it ecdfs20}, which only has an \xmm\ spectrum,  the data reduction and spectral extraction are from \citet{R2013} and \citet[][]{G2013}. For the \cosmos\ field we use spectra reduced and extracted for \xmm\ by \citet{M2007} and for \chandranosp\ by \citet{L2013}, with the only exception being source {\it cosmos330} for which a spectrum from the \cosmos-Legacy field has been used \citep{M2016a}. 

For the Serendipitous Survey fields we reduced and extracted  the \chandra and \xmm\ data.
In the selection of archival observations, we only use data from observations in which CCD detectors are primary instruments (i.e., we exclude \chandra grating observations). In the case of \xmm\ we almost always only use data from PN, the  exception being  {\it ser107}  which was located in a CCD gap of the PN camera. For this source we use the MOS data. For the \chandra\ data we used both ACIS-S and ACIS-I detectors whenever available (see Table~\ref{serlowenergy} for details). 
When multiple archival datasets were available we  chose the data closest in time to the \nustar\ observation, if available, in order to minimize  source variability. Table~\ref{serlowenergy} reports the selected observations for each source.

\begin{center}
    \begin{deluxetable}{r|cc|cc}
      \tablecaption{Low-energy observations used for serendipitous sources
        \label{serlowenergy}
      }
\tablehead{ ID    &    \multicolumn{2}{c|}{\chandra}    & \multicolumn{2}{c}{\xmm}     \\
    &    ObsID    &  Notes    &   ObsID  & Notes  }
\startdata
ser37   &  -         &   -  &   0604210201   &  - \\
        &  -         &   -  &   0604210301 &  - \\
ser77   &  10234     &  1  &   0064600101   & -\\
        &   -        &   -  &    0064600301  & -\\
ser107  &  -         &   -  &   0111220201    &  2  \\
ser148  &  -         &   -  &   0406740101  &  - \\
ser153  &  -         &   -  &   0724810301  & 3\\
ser184  &  10542     &  4   &   0657801901   &  3 \\
        &   10543    &  4   &    0657802101   &  3 \\
        &   10544    &  4   &    0657802301  &  3 \\
        &   10925    &  4   &   -  &  - \\
        &   11800    &  4   &   -  &  - \\
        &   11104    &  4   &   -  &  - \\
ser213  &  -  &   -  &   0203050201  &  - \\
ser215  &  12167  &  5  & - & - \\
ser243  &  15077  &  3,5 &   -  &   -  \\
        &   15619 &  3,5 &   -  &   -  \\
ser254  &  -  &   -  &   0744040301  & 6 \\
        &  -  &   -  &    0744040401 &   6 \\
ser267  &  14042  &  1 &   -  &  - \\
ser273  &  14042  &  1 &   -  &  - \\
ser318  &  -  &   -  &   0722610201    & 3 \\
ser335  &  -  &   -  &   0111260101   & - \\
        &  -  &   -  &   0111260701   & - \\
        &  -  &   -  &   0212480701  & - \\
ser359  &  -  &   -  &   0651850501  &  - \\
ser363  &  -  &   -  &   0724810201   &  3 \\
        &  -  &   -  &    0724810401  &  3  \\
ser382  &  -  &   -  &  0764910201  &  - \\
ser401  &  -  &   -  &   0693750101  &  - \\
ser451  &  -  &   -  &   0111180201    & -  
\enddata
\tablecomments{Notes: (1) \chandranosp/ACIS-I detector; (2) \xmm/MOS data; (3) Observations chosen to be closest in time to the \nustar\ data; (4) the source is on the \chandra/ACIS-S2 chip; (5) the source is on the \chandra/ACIS-S3 chip; (6) see \citet{R2016} for details on data reduction and spectral extraction;   }
    \end{deluxetable}
\end{center}

We reduced the \chandra\ data using CIAO v.~4.7\footnote{http://cxc.harvard.edu/ciao4.7/} with CALDB v.~4.6.7. We re-processed the data using the {\sc chandra\_repro} script to produce new re-calibrated level=2 event files. The spectral extraction was done with the script {\sc specextract}, which automates the creation of source and background spectral files and the relative ARF and RMF. The source and background spectral extractions were performed on user-selected  circular and annular concentric regions, respectively, in order to maximize the source flux and avoid point source contamination to background measurements. We finally combined the resulting spectra using the FTOOLs script {\sc addascaspec},  available in HEASOFT~v.~6.16\footnote{http://heasarc.gsfc.nasa.gov/docs/software/lheasoft/}, and produced combined RMFs and ARFs using the tasks {\sc addrmf} and {\sc addarf}. The resulting exposure times and collected net-source counts are reported in Table~\ref{tabspec}.
  
For the \xmm\ data we used SAS v14.0.0\footnote{https://www.cosmos.esa.int/web/xmm-newton/sas}. 
For each observation we screened the event files for time intervals impacted by soft proton flares by adopting an observation dependent 10-12~keV count-rate threshold ($0.4\pm0.1 ~\rm counts~s^{-1}$ being the average and $1\sigma$ standard deviation of the applied threshold), above which data were removed. For the spectral extraction and creation of response files we followed the standard procedures outlined in the \xmm\ science threads\footnote{http://www.cosmos.esa.int/web/xmm-newton/sas-threads}. We extracted events with pattern $\le 4$ for the PN camera and $\le 12$ for the MOS detectors.  We combined the MOS1 and MOS2 spectra using the SAS task {\sc epicspeccombine}. For sources with more than one data set, we produced combined source spectra, background spectra, ARF and RMF as per the \chandra data.  
Exposure times and net-source counts for each source are reported in Table~\ref{tabspec}.

For the \egs\ field, \chandra data products from \citet{G2012} have been used. The spectral extraction, specifically carried out for this work\footnote{CIAO v.~4.8 with CALDB v.~4.7.0 has been used.}, has been performed using {\sc specextract} for each individual observation. Background regions were taken from annuli with $1.3*r_{90,psf}$—--$3.0*r_{90,psf}$ (with the latter being the radius enclosing 90\% of the point spread function) with other detected sources masked out. Spectra and backgrounds were combined for the different observations using {\sc combine\_spectra} in CIAO.

\section{Data analysis}\label{datanalysis}
We performed the spectral analysis using  XSPEC v.~12.8.2 using the Cash statistic \citep[][]{C1979}  with the direct background subtraction option \citep[][]{WLK1979}. In the limit of a large number of counts per bin, the distribution of this statistic, called the $W$ statistic ($Wstat$), approximates the $\chi^2$ distribution with $N-M$ degrees of freedom ($dof$, where $N$ is the number of independent bins and $M$ is the number of free parameters). We performed all our modeling with spectra binned to 5 net-counts (i.e., background subtracted) per bin with the exception of sources with low number of counts (i.e. $\ls50$ counts from both \nustar\ detectors) for which we resorted to a finer binning of 1 net-count per bin.

The spectral modeling has been performed: (a) for the \nustar-only data in the energy range $3-24$~keV assuming a power-law, with absorption and reflection (Sect.~\ref{nustaronly}), and (b) jointly together with \xmm\ and \chandra\ over the broader $0.5-24$~keV energy band using more complex models (Sect.~\ref{broadband}).

Notice that despite the spectral analysis being performed up to 24~keV, on average, our spectra are sensitive to slightly lower energies. The median and semi-interquartile range for the highest energy bin in the FPMA and FPMB spectra are $19.6\pm3.0$~keV and $17.9\pm2.4$~keV, respectively.

\subsection{\nustar\ spectral analysis }\label{nustaronly}
\begin{figure}[!t]
   \begin{center}
\includegraphics[width=0.45\textwidth]{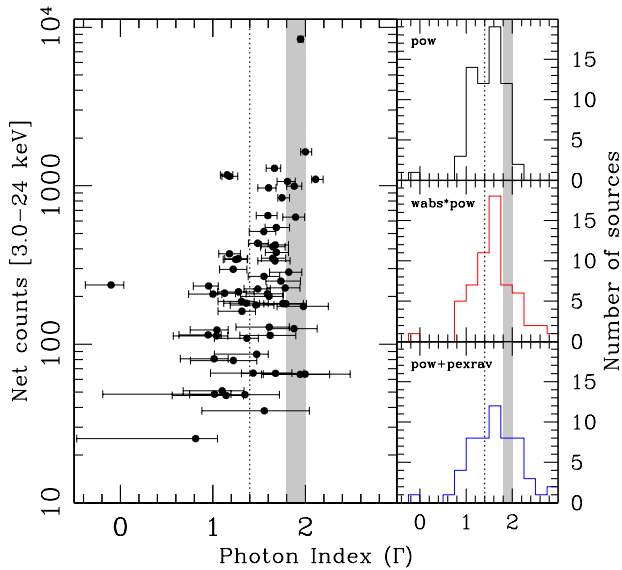}
   \end{center}
\caption{Left panel: 3-24~keV net counts vs. photon index for the \nustar-only joint fit derived using a power-law model (black dots). Right panels (from top to bottom): black, red and blue histograms report the distribution for the power-law,  absorbed power-law and power-law plus reflection component models, respectively.  Grey regions represent the canonical range of $\Gamma$ values measured in the literature for the power-law component. The dotted line represents the 3-15~keV slope of the CXB as measured by {\it HEAO-1} \citep[][]{M1980}. } 
   \label{gammacounts}
\end{figure}

For the \nustar-only analysis ($3-24$~keV) we first used a power-law model. We freeze the cross-calibration between FPMA and FPMB since, given the few percent level of accuracy measured by \citet{MK2015} and the limited counts of the majority of our spectra (up to a few hundreds) we do not expect to distinguish these small calibration levels (i.e., the statistical uncertainties exceed the systematic ones).
The left panel of Fig.~\ref{gammacounts} presents the power-law photon index $\Gamma$ values plotted against the net counts in the $3-24$~keV band. 
The $\Gamma$ values are, on average, flatter than the canonical 1.8-2 values \citep[e.g.][]{P2005,D2008} with a mean (median) of 1.5 (1.6). The distribution of $\Gamma$ is reported by the black histogram in the right-upper panel. 
Spectra with fewer counts than the median  
have slightly flatter photon indices than high-count sources, $\langle\Gamma_{\rm low}\rangle=1.4\pm0.2$ compared to $\langle\Gamma_{\rm high}\rangle=1.7\pm0.3$. The average hardening of faint sources agrees with the shape of the CXB in approximately the same energy range \citep[][]{M1980} as already found at lower energies \citep[e.g., ][]{M2000,G2001,B2001}.  
There is one notable outlier with a negative $\Gamma$ value, the CT thick source in the COSMOS field reported by C15 ({\it cosmos330} in Table~\ref{tabsample}). The average flat values of $\Gamma$ point to a more complex spectral shape  in the \nustar\ energy band.
  \begin{center}
    \begin{deluxetable*}{r|rrr|rrr|rrr|rrr}
      \tablecaption{Spectral extraction parameters
\label{tabspec}
      }
      \tablehead{\nustar\ ID &            \multicolumn{3}{c}{\nustar\ FPMA}    & \multicolumn{3}{c}{\nustar\ FPMB}    & \multicolumn{3}{c}{\chandra} & \multicolumn{3}{c}{\xmm}  \\
        & net cts  &  exp\,\tablenotemark{a}   &$r_{ex}$\,\tablenotemark{b} & net cts & exp\,\tablenotemark{a}    &$r_{ex}$\,\tablenotemark{b} & net cts     &  exp\,\tablenotemark{a}  & Ref. & net cts &  exp\,\tablenotemark{a} & Ref.  }
\startdata
cosmos97   &  246  &   57.1  &  60   &   268  &   57.0  &  60   &       392  &    47.1  &  3 &   675  &   40.1   & 2   \\
cosmos107  &   26  &   52.8  &  25   &    21  &   52.7  &  25   &        -   &      -   &  - &    73  &   20.5   & 2   \\
cosmos129  &   21  &   71.5  &  30   &    27  &   71.5  &  25   &        46  &   186.7  &  3 &    54  &   57.1   & 2   \\
cosmos130  &  111  &   49.6  &  40   &   103  &   49.5  &  40   &        -   &      -   &  - &  1498  &   37.5   & 2   \\
cosmos145  &  109  &  103.2  &  50   &    77  &  103.1  &  50   &       603  &   189.5  &  3 &   329  &   45.8   & 2   \\
cosmos154  &  114  &  100.1  &  45   &   119  &   99.5  &  45   &       185  &   189.4  &  3 &    -   &     -    & -   \\
cosmos155  &  531  &  104.8  &  60   &   534  &  104.1  &  55   &      5127  &   184.8  &  3 &  3520  &   46.1   & 2   \\
cosmos178  &  110  &   97.8  &  60   &    67  &   97.7  &  40   &       402  &    93.2  &  3 &   218  &   32.0   & 2   \\
cosmos181  &   75\,\tablenotemark{c}  &   95.0  &  40   &    55\,\tablenotemark{c}  &   94.9  &  40   &        62  &    92.4  &  3 &    99  &   53.5   & 2   \\
cosmos194  &   86  &   73.1  &  35   &   182  &   99.1  &  40   &      3174  &   185.9  &  3 &  2383  &   53.3   & 2   \\
cosmos195  &   82  &   48.7  &  40   &   100  &   48.7  &  40   &        -   &      -   &  - &   682  &   17.5   & 2   \\
cosmos206  &  149  &   46.6  &  45   &   101  &   46.6  &  30   &      1920  &    91.5  &  3 &  2957  &   43.3   & 2   \\
cosmos207  &   58  &   49.7  &  35   &    65  &   49.6  &  45   &        -   &      -   &  - &   175  &   18.7   & 2   \\
cosmos216  &   45  &   51.5  &  45   &    34  &   51.4  &  40   &        88  &   140.5  &  3 &   104  &   56.2   & 2   \\
cosmos217  &   44  &  114.4  &  20   &    42  &  114.3  &  20   &       945  &   188.5  &  3 &   821  &   63.5   & 2   \\
cosmos218  &  500  &   98.6  &  55   &   470  &   98.5  &  55   &      2170  &    88.9  &  3 &  3417  &   39.6   & 2   \\
cosmos229  &   56  &   98.6  &  25   &    53  &   98.5  &  25   &       316  &   133.9  &  3 &  1096  &   60.8   & 2   \\
cosmos232  &   75  &  111.8  &  40   &    38  &  111.7  &  20   &       240  &   185.6  &  3 &   115  &   63.7   & 2   \\
cosmos249  &   23  &   52.0  &  25   &    28  &   51.9  &  20   &        -   &      -   &  - &   138  &   32.9   & 2   \\
cosmos251  &   74  &   78.0  &  30   &    51  &   52.1  &  30   &        -   &      -   &  - &   721  &   29.1   & 2   \\
cosmos253  &   71  &   52.9  &  40   &    43  &   52.8  &  40   &        -   &      -   &  - &   191  &   32.5   & 2   \\
cosmos263  &  239  &   76.7  &  55   &   184  &   76.5  &  55   &        -   &      -   &  - &   514  &   12.9   & 2   \\
cosmos272  &  110  &  106.4  &  40   &   113  &  106.2  &  40   &        -   &      -   &  - &   189  &   36.9   & 2   \\
cosmos282  &  158  &   83.1  &  40   &   177  &   82.9  &  45   &        -   &      -   &  - &  2790  &   46.0   & 2   \\
cosmos284  &   38  &   52.6  &  40   &    43  &   52.5  &  30   &        -   &      -   &  - &  1159  &   73.3   & 2   \\
cosmos287  &   88  &  111.7  &  30   &    93  &  107.1  &  40   &        -   &      -   &  - &   946  &   31.3   & 2   \\
cosmos296  &   27  &   45.6  &  20   &    38  &   45.5  &  25   &       658  &    91.6  &  3 &   893  &   45.9   & 2   \\
cosmos297  &   78  &   45.6  &  30   &    50  &   45.5  &  35   &       402  &    93.2  &  3 &  1056  &   64.3   & 2   \\
cosmos299  &  167  &  110.7  &  40   &   176  &  110.5  &  40   &        -   &      -   &  - &   111  &   38.2   & 2   \\
cosmos322  &  114  &   99.1  &  40   &    93  &   99.0  &  40   &        -   &      -   &  - &   613  &   51.8   & 2   \\
cosmos330  &  131  &   52.8  &  55   &   105  &   52.7  &  55   &       183  &   147.9  &  1 &    80  &   39.9   & 2   \\
ecdfs5  &     82  &   89.5  &  30   &   127  &   89.1  &  40   &       532  &  240.3  &  4,5 &    -   &     -    & -    \\
ecdfs20 &    149  &  192.6  &  20   &   136  &  192.6  &  20   &      1204  &  357.9  &  4,5 &  3743  &  1862.2   & 6   \\
ecdfs51 &    170  &  185.8  &  45   &   179  &  185.6  &  45   &      2079  &  245.6  &  4,5 &    -   &     -    & -   \\
egs1    &    180  &  149.4  &  55   &   166  &  205.5  &  35   &      2894  &  197.5  &  7 &    -   &     -    & -   \\
egs9    &    453  &  302.8  &  50   &   387  &  354.2  &  40   &      4195  &  719.0  &  7 &    -   &     -    & -   \\
egs26   &    593  &  400.5  &  55   &   578  &  399.6  &  55   &      2531  &  683.9  &  7 &    -   &     -    & -   \\
egs27   &    294  &  300.2  &  55   &   250  &  251.3  &  50   &     10696  &  683.9  &  7 &    -   &     -    & -   \\
egs32   &    614  &  390.4  &  60   &   484  &  389.6  &  40   &      5448  &  683.9  &  7 &    -   &     -    & -   \\
ser37   &    267  &   49.1  &  20   &   106  &   30.1  &  15   &        -   &     -   &  - &   956  &   28.7   & 8   \\
ser77  &    104  &   27.3  &  55   &    97  &   27.3  &  45   &       160  &   31.7  &  8 &    42  &   12.5   & -   \\
ser97   &    290  &  120.8  &  35   &   344  &  120.6  &  35   &        -   &     -   &  - &    -   &     -    & -   \\
ser107  &     80  &  127.6  &  30   &   130  &  127.4  &  50   &        -   &     -   &  - &    50\,\tablenotemark{d}  &   49.8\,\tablenotemark{d}   & 8   \\
ser148  &   4105  &   38.7  &  90   &  4299  &   42.4  &  85   &        -   &     -   &  - &  6316  &   10.9   & 8   \\
ser153    &    321  &  142.5  &  50   &   327  &  221.7  &  50   &        -   &     -   &  - &     5  &    5.8   & 8   \\
ser184  &    635  &  943.6  &  40   &  1002  &  974.7  &  45   &      5717  &  381.9  &  8 &   251  &   21.2   & 8   \\
ser213    &    637  &   94.3  &  55   &   510  &   94.1  &  45   &        -   &     -   &  - &    52  &   11.3   & 8   \\
ser215   &     99  &   59.4  &  45   &    74  &   59.3  &  35   &        55  &    5.0  &  8 &    -   &     -    & -   \\
ser235   &    211  &  123.1  &  50   &   169  &  122.8  &  50   &        -   &     -   &  - &    -   &     -    & -   \\
ser243   &    172  &   69.2  &  20   &   125  &   69.1  &  20   &       674  &   90.4  &  8 &    -   &     -    & -   \\
ser254   &    495  &   90.2  &  45   &   795  &  142.1  &  50   &        -   &     -   &  - &   439  &   36.5   & 9   \\
ser261  &     11  &   22.4  &  25   &    14  &   22.4  &  35   &        -   &     -   &  - &    -   &     -    & -   \\
ser267   &     27  &   20.4  &  40   &    38  &   20.3  &  35   &        45  &    5.0  &  8 &    -   &     -    & -   \\
ser273   &    521  &   20.4  &  70   &   473  &   20.3  &  60   &      1028  &    5.0  &  8 &    -   &     -    & -   \\
ser285   &     33  &   23.2  &  35   &    33  &   23.1  &  35   &        -   &     -   &  - &    -   &     -    & -   \\
ser318   &    123  &   67.7  &  50   &   104  &   68.0  &  40   &        -   &     -   &  - &   129  &    3.0   & 8   \\
ser335   &     93  &   49.2  &  40   &    87  &   49.1  &  35   &        -   &     -   &  - &   996  &   23.9   & 8   \\
ser359   &     56  &   23.9  &  50   &    58  &   23.8  &  60   &        -   &     -   &  - &   151  &   18.1   & 8   \\
ser363   &    293  &  224.7  &  50   &   119  &  110.1  &  50   &        -   &     -   &  - &   898  &   32.8   & 8   \\
ser382   &     33  &   49.4  &  30   &    32  &   50.0  &  25   &        -   &     -   &  - &    190   &     19.4    & 8   \\
ser401   &     17  &   21.3  &  20   &    21  &   21.3  &  20   &        -   &     -   &  - &   158  &   22.8   & 8   \\
ser409  &     23  &   32.3  &  25   &    26  &   32.2  &  20   &        -   &     -   &  - &    -   &     -    & -   \\
ser451    &    261  &   95.1  &  60   &   173  &   98.7  &  60   &        -   &     -   &  - &     5  &    7.6   & 8   
\enddata
  \tablecomments{References: (1) \citealt{M2016a}; (2) \citealt{M2007}; (3) \citealt{L2013}; (4) \citealt{L2005}; (5) \citealt{X2016}; (6) \citealt{R2013}; (7) \citealt{G2012}; (8) this work and (9) data reduction as in \citealt{R2016}}
  \tablenotetext{a}{Exposure time in ksec.}
  \tablenotetext{b}{Extraction radius in arcsec.}
  \tablenotetext{c}{Counts in the energy range 4.5-24~keV. See Appendix for details on this source.}
  \tablenotetext{d}{MOS spectrum; the source falls in a chip gap in PN.}

    \end{deluxetable*}
  \end{center}

In order to identify  a more suitable model    which would bring the power-law photon indices to the canonical 1.8-2 values, we explored two modifications to the simple power-law model. We first allowed for low-energy photoelectric absorption by the circumnuclear interstellar matter using the {\sc zwabs} model in XSPEC. Given the 3~keV lower bound of the \nustar\ energy range, this model modification did not change appreciably the distribution of $\Gamma$, producing a median of 1.6 and only a few outliers ($\sim10$\% of the sample) at values much larger than 3 (see red dashed histogram in Fig.~\ref{gammacounts}). An alternative modification is the inclusion, beside the simple power-law component, of an additional cold Compton-reflection component to account for the  disk/torus reflectors. This component is particularly important in the \nustar\ hard-energy band. We used the {\sc pexrav} model \citep[][]{MZ1995} which assumes that the reflector is an infinite slab with infinite optical depth illuminated by the primary power-law continuum, subtending an angle $\Omega=2\pi R$, where $R$ is the reflection parameter. For a source of isotropic emission $\Omega=2\pi$, hence $R=1$. 
We tied both the photon index and the normalization of the reflection model to those of the primary power-law and let $R$ vary. In our modeling throughout the paper we set this parameter in XSPEC to be negative, as for {\sc pexrav} this will switch on the reflection-only solution as opposed to the reflection+power-law solution activated by positive values. Throughout the text we quote the absolute value of the parameter. We left the  abundance at its default solar value, $\cos\theta=0.45$ (i.e., inclination angle $\theta\sim63$~deg, the default value in the model), and set the  exponential cut-off ($E_{\rm c}$) for the incident power-law primary continuum at 200~keV (as assumed by G07 and consistent with recent determinations by \nustar; see \citealt[][]{F2015} for a compilation). This additional component shifts the mean and median photon index to higher values  ($\Gamma=1.8$ and $\Gamma=1.7$, respectively), but at the cost of increasing the dispersion of the distribution  (see blue histogram in Fig.~\ref{gammacounts}). There is no trend in the median $\Gamma$ with the number of counts except for the dispersion with low-count sources having an interquartile range of 1.2 as opposed to high counts sources which have an interquartile range of 0.6. Histograms of the $\Gamma$ distribution for the three models are reported in the right panels of Fig.~\ref{gammacounts}.

\begin{figure*}[t!]
   \begin{center}
\includegraphics[height=0.8\textheight,angle=90]{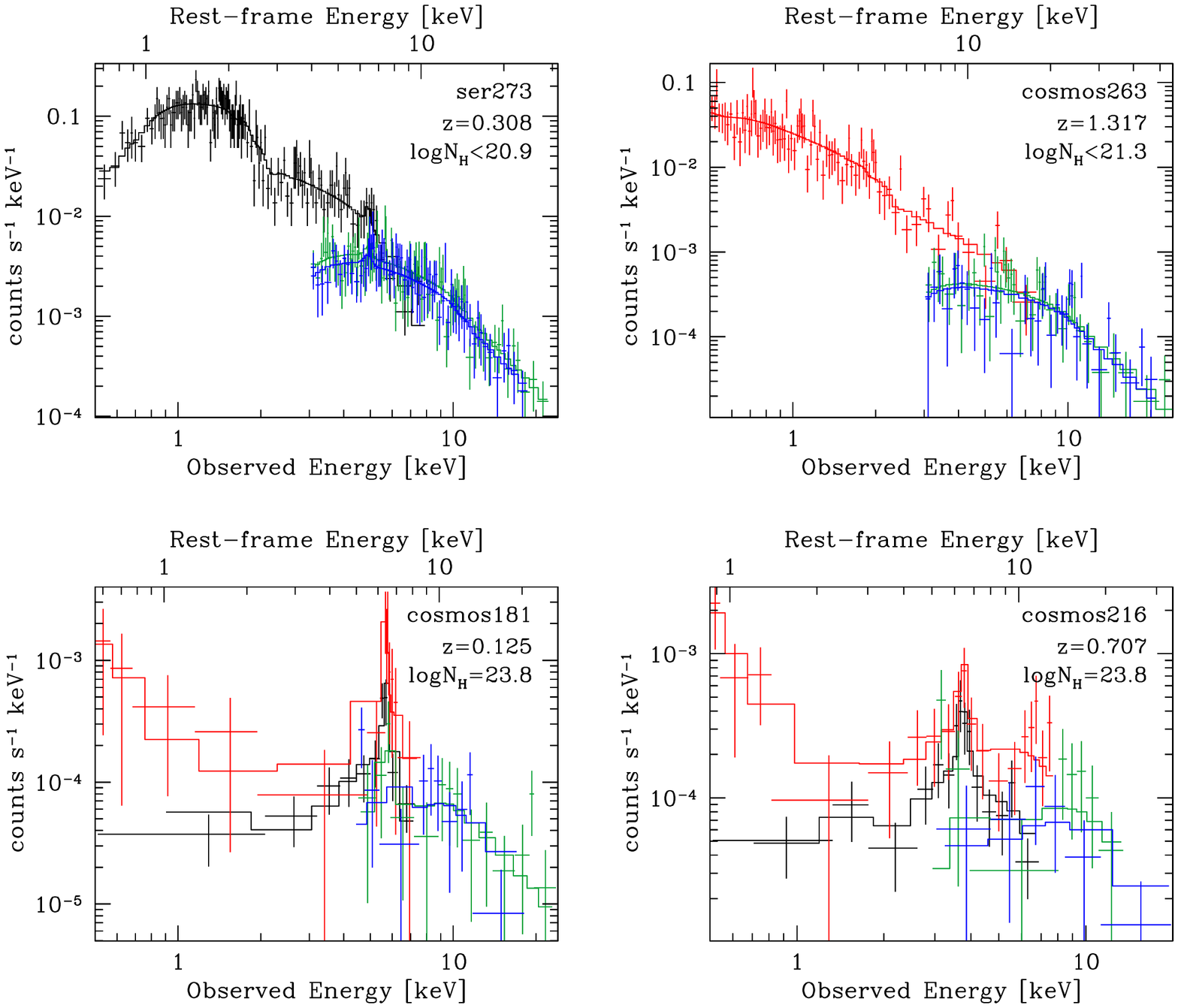}
   \end{center}
\caption{Examples of broad-band spectra for four sources with best-fit baseline models (solid lines). Black, red, green and blue spectra refer to \chandranosp, \xmm, \nustar-FPMA and \nustar-FPMB, respectively. Upper/lower spectra are for unabsorbed/absorbed sources. Spectra on the left/right are for sources with redshifts lower/higher than $\langle z \rangle$. }
   \label{spectrasample}
\end{figure*}

\subsection{Joint broad band analysis}\label{broadband}
In order to improve the modeling and obtain tighter constraints on the spectral parameters, we added lower energy data from \xmm\ and \chandra, thereby extending the spectral range down to E$=0.5$~keV (observed frame).  Table~\ref{tabspec} reports details on  the low energy data used for each source.

We first consider an empirical model (hereafter called {\it baseline model}) expressed in XSPEC as:
\smallskip
\smallskip
\smallskip

  \textsc{constant$\times$wabs(powerlaw$_{\rm sc}$+zwabs$\times$powerlaw + zgauss + pexrav)}\\

  \smallskip
\noindent where  {\sc powerlaw} represents the primary coronal component modified  at low energies with photoelectric absorption (model {\sc zwabs}) and complemented at high energies with the addition of a cold Compton-reflection component (model {\sc pexrav}). We further add at low-energy a power-law (\textsc{powerlaw$_{\rm sc}$}) to account, when needed, for residual low energy flux for absorbed sources (hereafter called scattered component) consisting either of primary component flux scattered outside the nuclear absorbing region or of circumnuclear photoionized gas.  At high energy we add a  line ({\sc zgauss}) to account for neutral Fe~K$\alpha$ emission at 6.4~keV produced by the surrounding reflecting cold medium and let its normalization free to vary. The entire model is modified by photoelectric absorption ({\sc wabs}) from  Galactic interstellar gas using values reported by \citet{K2005} at the position of each source.
\begin{figure}[t!]
   \begin{center}
\includegraphics[width=0.45\textwidth]{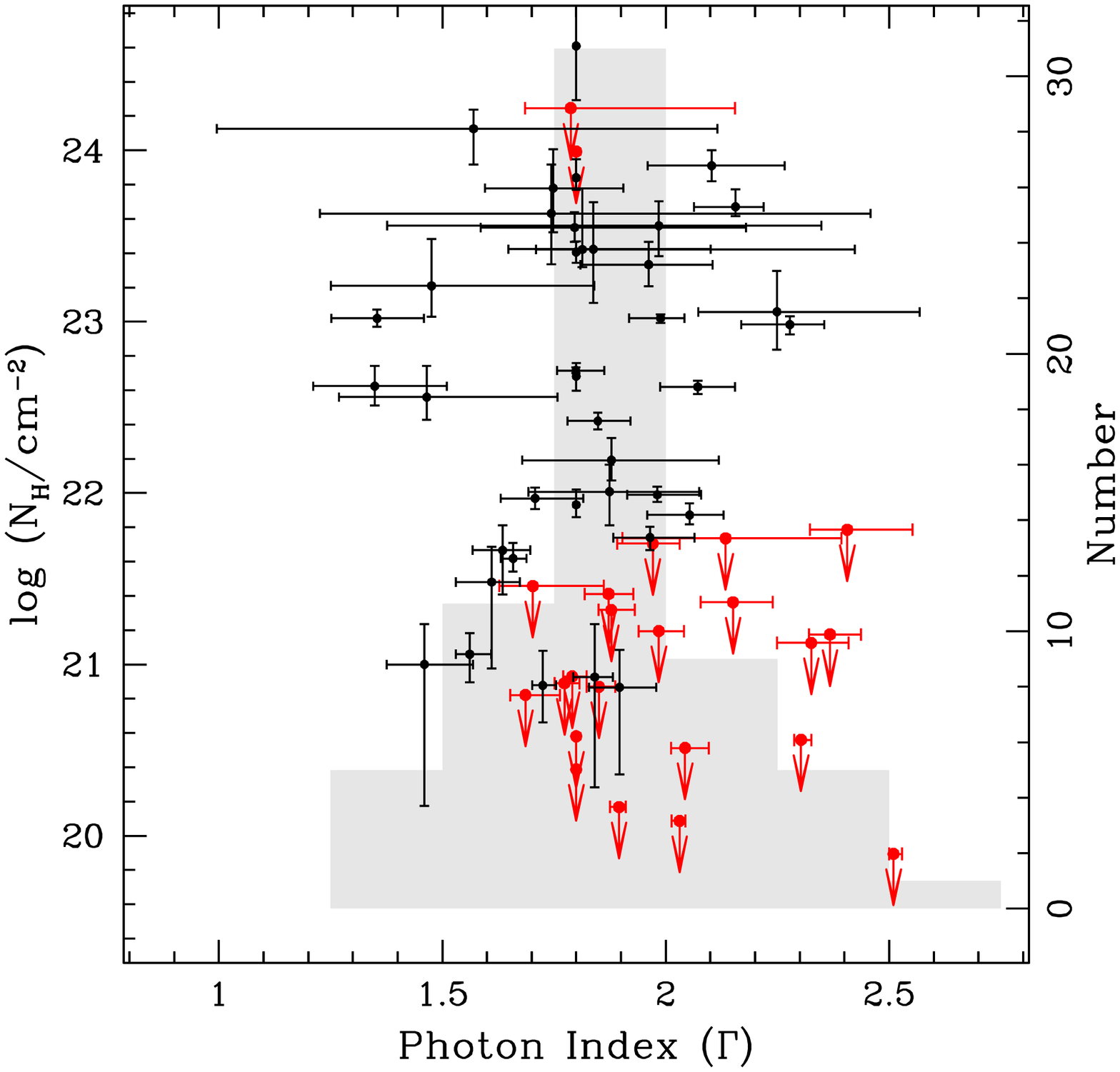}
   \end{center}
\caption{Intrinsic column density (left-hand y-axis) versus photon index from the baseline modeling. Red arrows are 90\% c.l. upper limits on \nhsym. The shaded grey histogram reports the distribution of $\Gamma$ (right-hand y-axis). Data points without error-bars in $\Gamma$ are sources for which the value of $\Gamma$ was fixed to a value of 1.8 during the modeling because of poor constraints due to a combination of limited statistics and large column densities.} 
   \label{g_vs_nh}
\end{figure}
The \textsc{constant} accounts for instrument intercalibration and possible source flux variability, as well allowing for a crude accounting of possible contamination from blended sources inside  differing  extraction radii. We left the constant free to vary between satellites, but always tied between the two \nustar\ FPMs\footnote{For the 58 sources with low-energy spectral data, the difference between the estimated best-fit constants is reasonably low being smaller than a factor of $\sim2$ for the majority of the sources (50). Four sources ({\it cosmos129}, {\it cosmos229}, {\it cosmos297} and  {\it ser77}) show variations larger than a factor of 2-3 in both \xmm\ and \nustar\ clearly pointing to source variability as main cause of discrepancy. The remaining four sources {\it cosmos249}, {\it cosmos263}, {\it ser148} and {\it ecdfs5} have variations by factors 2-4 with the latter showing the largest variation which is possibly due to contamination from a nearby source (see Appendix for details).} as done in Section~\ref{nustaronly}. We left the slope and the normalization of the scattered component free to vary. As in Sect.~\ref{nustaronly} we used the reflection-only component from {\sc PEXRAV} and tied both $\Gamma$ and normalization to the corresponding parameters of the primary component. Other {\sc PEXRAV} parameters are set to the default values as reported in Sect.~\ref{nustaronly}. In this way our fits with the baseline model are performed with 5 free parameters. In case of joint fit performed with one or two additional low-energy datasets, one or two intercalibration constants need to be accounted as additional free parameters, respectively. Furthermore in case of sources with soft-excess component two additional free parameters need to be considered for the slope and normalization of the scattered component. 
In order to speed up our modeling which, using {\sc pexrav}, can be quite time consuming, the error estimation on all parameters was obtained with the reflection strength parameter $R$ and calibration constants fixed to their best-fit values. For error estimation in $R$, we left free to vary only \nhsym, $\Gamma$ and normalization of the primary power-law component.
Best-fit spectral parameters are reported in Table~\ref{specparambase} along with fluxes in the 8-24~keV and 3-24~keV bands, and 10-40~keV unabsorbed and intrinsic coronal luminosities inferred from the best-fit baseline model (see Section~\ref{lum1040} for details). Fig.~\ref{spectrasample} shows broad-band spectra for four sources along with their best-fit model. For the few sources exhibiting extreme $\Gamma$ values below 1.3 or above $\sim2.5$ or reflection parameter larger than $\sim10$, we redid the fits with $\Gamma$ fixed to 1.8. These sources are {\it cosmos129},  {\it cosmos232}, {\it cosmos253},  {\it cosmos282}, {\it ser285}, {\it ser77} and {\it ser261}.
In three cases, mainly unabsorbed sources with high-quality spectra, the baseline parametrization in the soft ({\it ser148}) and broad-band ({\it ser37, egs26}) was inadequate. Indeed, in these energy ranges absorbed power-law models return slopes in the range $\Gamma=0.2-1.2$ with very little absorption.   We therefore further modified the absorbed primary power-law by further applying absorption from  a partial covering cold ({\sc zpcfabs} in XSPEC) or partially ionized ({\sc zxipcf}) medium. Details on these sources are reported in the Appendix.

\subsection{Absorption and photon index from the primary power-law}
The distribution of the measured $\Gamma$  peaks at around 1.8-2, with a  mean  value of $1.89\pm0.26$, as reported in Fig.~\ref{g_vs_nh}. 
Best-fit column density values range from $\lesssim 10^{21}$~\nh\ to $\gtrsim10^{24}$~\nh. We have upper limits for 23 sources. 
Twenty are unabsorbed sources with \nhsym\ upper limits  $<10^{22}$~\nh. The two remaining sources have \nhsym\ upper limits reaching into the heavily absorbed regime ($\sim10^{23}-10^{24}$~\nh). These sources, {\it ser285} and {\it ser235}, have low-count \nustar\ data and no lower energy data available. For only one source  with \nustar-only data ({\it ser409}) we cannot constrain its \nhsym\ value even when fixing $\Gamma=1.8$. For 17 sources, $\sim$27\% of the sample, we measure \nhsym$\ge10^{23}$~\nh. Two sources ($\sim3$\% of the sample), {\it cosmos330} and {\it ser261}, exhibit CT column densities. The former is the CT AGN discovered by C15. 
Fig.~\ref{g_vs_nh} shows $\Gamma$ as a function of intrinsic \nhsym. Error bars in $\Gamma$ tend to be larger for obscured sources (i.e., $N_H\gtrsim10^{22}$~\nh).


\subsection{Luminosity in the 10-40~keV energy band}\label{lum1040}
In the last two columns of Table~\ref{specparambase} we report the
10-40~keV luminosities from the baseline model. They are unabsorbed luminosities ($L_{u,X}$, penultimate column) and intrinsic luminosities ($L_{i,X}$, last column). 
The unabsorbed luminosity is estimated by simply removing the Galactic and intrinsic absorption components from the best-fit baseline model.
The intrinsic coronal luminosities are computed from the unabsorbed coronal power-law component by simply removing the reflection contribution to the best-fit baseline model.
The uncertainties in $L_{i,X}$ due to parameter degeneracy in our modeling are estimated by fitting the baseline model with $R$ fixed to its lower and upper error bounds. 

\begin{figure}[t!]
   \begin{center}
\includegraphics[width=0.45\textwidth]{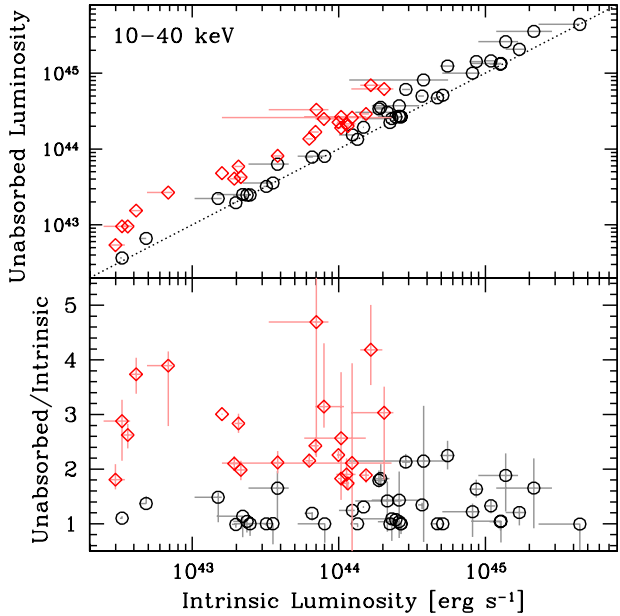}
   \end{center}
\caption{Upper panel:  intrinsic 10-40~keV luminosity measured on the coronal primary component only (i.e., unabsorbed luminosity computed after removing the Compton-reflection term) versus unabsorbed 10-40~keV luminosity measured on the unabsorbed baseline best-fit model (i.e., including the Compton-reflection term). Uncertainties in the baseline spectral modeling are reported in light colors.  The dotted line shows a relation with slope 1. Lower panel: ratio between unabsorbed and intrinsic luminosities. Red diamonds show the sources exhibiting $R>1$. } 
   \label{lumcfr}
\end{figure}

In the context of the baseline parametrization, $L_{i,X}$ is supposed to reflect more closely the true X-ray radiative output of the primary (direct) X-ray emitting nuclear source. Notice though that the planar geometry assumed in \textsc{pexrav} is an approximate description of the cold reflector which, according to unification schemes, has a toroidal geometry.  
In any case, in the 10-40~keV band the additional reflection contribution can become relevant compared to the intrinsic coronal one, especially for sources with low-luminosity  and large reflection strengths. Including the reflection  term in the luminosity calculation may lead to a ``double counting'' of the intrinsic X-ray radiative output. Indeed in this case the estimate of $L_{u,X}$ would include  both the 
primary coronal power-law component and the primary coronal photons reflected from the circum-nuclear material back to the observer. This overestimation of the intrinsic luminosity is negligible (10-30\% for $R$=1-6) in the 2-10~keV band where the reflection component is a few percent of the primary emission. 
The upper panel of Fig.~\ref{lumcfr}  shows, for the 10-40~keV band, the overestimate of the ``unabsorbed luminosities'' including the reflection component compared to the intrinsic luminosities derived from the unabsorbed primary component only.
In the lower panel we report the ratio between these two quantities in order to quantify better the level of overestimation. $L_{u,X}$ can be larger by factors up to $\sim2-4$ and the majority of those sources are those with best-fit $R>1$ (red diamonds)  at low luminosity (i.e., $L_{i,X}\ls2\times10^{44}$~\lumcgs). This is due to an induced dependence between $R$ and luminosity which, if not accounted for, may lead to a biased view of the relationship between luminosity and reflection strength (see Section~\ref{Rvs} and Fig.~\ref{R_vs_NH}, bottom panels). Notice that few sources at higher luminosities ($\gs2\times10^{44}$~\lumcgs) have overestimates of a factor of $\sim2$ even though they have low reflection strenghts (i.e., $R<1$). This is due to the fact that the $L_{u,X}/L_{i,X}$  ratio in the observed 10-40~keV energy range is an increasing function of the redshift\footnote{Indeed the redshift progressively shifts to lower energies (i.e., outside the band) portions of the spectrum where the decreasing primary component still significantly contribute to the total flux.} and our sample, selected in flux, contains, on average, higher luminosity sources at higher redshifts. 

In order to keep the {\it baseline} parametrization simple and suitable for low SNR spectra we did not include a Compton-scattering term which can become important for the most obscured sources. This may lead to an underestimate of the true luminosity for the most obscured sources. We compared our unabsorbed values with the best-fit values obtained by adding a Compton-scattering term parametrized with {\sc cabs} for the \cosmos\ sources  with $\lognh\gs24$, i.e. those for which we have the best-quality broad-band data. We obtained, on average, larger luminosities with values ranging from $<0.1$~dex for the less obscured sources up to $\sim0.4$~dex for the most obscured ones. However, {\sc cabs} approximates the Compton-scattering by only accounting for the scattering of the photons outside of the beam and neglecting photons reflected by surrounding material into the line-of-sight. Hence more appropriate luminosity values may be estimated by accounting for the geometry of the obscurer.
For this reason we compared our values with those obtained with the torus modelings employed in Section~\ref{torus} which self-consistently account for Compton-scattering effects due to the toroidal geometry of the obscurer. We found that in the range $\lognh\approx23-24.5$, the 10-40~keV  luminosity  is underestimated on average by at most $\sim 0.1$~dex with only two exceptions in our sample: {\it cosmos129} and {\it ser261}. These sources are among the most obscured sources in our sample and for them we are finding $L_{u,X}$ underestimated by  0.2~dex and 0.3-0.4~dex, respectively. No significant difference is found for less-obscured sources.


\begin{center}
\tablewidth{0pt}
\tabletypesize{\scriptsize}
  \begin{deluxetable*}{lrrrrrcccc}
    \tablecaption{Best-fit parameters for baseline model
\label{specparambase}
    }
    \tablehead{\nustar\ ID &     stat &   dof &    $\Gamma$          &   $\lognhonly$\,\tablenotemark{a}              &  \multicolumn{1}{c}{$R$}   &  $S_{\rm 8-24}$\,\tablenotemark{b}   &  $S_{\rm 3-24}$\,\tablenotemark{b}  &  $\log L_{u,X}$\,\tablenotemark{c}  &  $\log L_{i,X}$\,\tablenotemark{d}  }
\startdata
cosmos97                  &  257.1& 262& $ 2.07_{-0.08}^{+0.08}$ & $ 22.62_{-0.04}^{+0.04}$ & $  5.89_{-1.41}^{+1.75}$ &  3.8&  5.4& 43.2& $ 42.62_{-0.04}^{+0.03}$  \\
cosmos107                 &   19.0&  16& $ 1.74_{-0.52}^{+0.71}$ & $ 23.63_{-0.30}^{+0.29}$ & $ <0.91$ &  0.9&  1.3& 44.4& $ 44.36_{-0.20}^{+0.00}$ \\ 
cosmos129                 &   16.0&  24& $  1.8              $ & $ 23.84_{-0.07}^{+0.11}$ & $ <0.32$ &  0.5&  0.7& 44.4& $ 44.42_{-0.04}^{+0.00}$  \\
cosmos130                 &  256.1& 247& $ 1.87_{-0.05}^{+0.06}$ & $ <21.41$ & $  0.39_{-0.18}^{+0.36}$ &  1.6&  2.5& 45.4& $ 45.14_{-0.19}^{+0.08}$  \\
cosmos145                 &  150.6& 170& $ 1.71_{-0.08}^{+0.11}$ & $ 21.97_{-0.06}^{+0.06}$ & $  0.83_{-0.73}^{+1.64}$ &  0.7&  1.0& 43.8& $ 43.58_{-0.20}^{+0.08}$  \\
cosmos154                 &   76.3&  70& $ 1.80_{-0.21}^{+0.38}$ & $ 23.55_{-0.08}^{+0.09}$  & $ <0.34$ &  1.6&  2.1& 43.5& $ 43.51_{-0.02}^{+0.00}$ \\ 
cosmos155                 &  788.7& 792& $ 2.03_{-0.02}^{+0.01}$ & $ <20.09$ & $  1.24_{-0.38}^{+0.27}$ &  3.3&  5.3& 44.3& $ 44.06_{-0.03}^{+0.01}$  \\
cosmos178                 &  143.9& 127& $ 1.46_{-0.08}^{+0.11}$ & $ 21.00_{-0.82}^{+0.24}$ & $ <1.96$ &  0.7&  1.0& 43.6& $ 43.55_{-0.21}^{+0.00}$ \\ 
cosmos181                 &   38.0&  45& $ 2.10_{-0.14}^{+0.16}$ & $ 23.91_{-0.09}^{+0.09}$ & $  0.76_{-0.37}^{+0.84}$ &  1.0&  1.2& 42.8& $ 42.69_{-0.00}^{+0.05}$\\  
cosmos194                 &  581.2& 587& $ 1.79_{-0.02}^{+0.03}$ & $ <20.93$ & $  0.38_{-0.15}^{+0.15}$ &  1.7&  2.6& 45.2& $ 44.94_{-0.04}^{+0.04}$  \\
cosmos195                 &  139.3& 138& $ 1.61_{-0.08}^{+0.06}$ & $ 21.48_{-0.50}^{+0.21}$ & $  0.31_{-0.22}^{+0.51}$ &  2.4&  3.7& 45.5& $ 45.33_{-0.26}^{+0.12}$  \\
cosmos206                 &  547.4& 571& $ 1.77_{-0.02}^{+0.03}$ & $ <20.89$ & $  0.23_{-0.13}^{+0.16}$ &  2.3&  3.6& 45.2& $ 45.04_{-0.05}^{+0.03}$  \\
cosmos207                 &   57.6&  50& $ 1.70_{-0.07}^{+0.16}$ & $ <21.46$ & $  3.88_{-1.56}^{+4.84}$ &  2.0&  2.7& 44.5& $ 43.85_{-0.33}^{+0.08}$  \\
cosmos216                 &   23.8&  41& $ 1.75_{-0.15}^{+0.16}$ & $ 23.78_{-0.26}^{+0.23}$ & $  1.97_{-1.39}^{+1.33}$ &  1.0&  1.2& 44.4& $ 43.90_{-0.06}^{+0.14}$  \\
cosmos217                 &  321.3& 267& $ 1.85_{-0.07}^{+0.07}$ & $ 22.42_{-0.05}^{+0.05}$ & $ <0.40$ &  0.5&  0.9& 45.0& $ 44.91_{-0.21}^{+0.05}$  \\
cosmos218                 &  674.8& 695& $ 1.90_{-0.02}^{+0.02}$ & $ <20.17$ & $  1.17_{-0.11}^{+0.30}$ &  3.6&  5.5& 44.3& $ 44.02_{-0.02}^{+0.00}$  \\
cosmos229                 &  231.4& 221& $ 2.04_{-0.03}^{+0.05}$ & $ <20.51$ & $  2.71_{-0.61}^{+0.67}$ &  0.8&  1.1& 43.7& $ 43.20_{-0.00}^{+0.00}$  \\
cosmos232                 &   62.3&  76& $  1.8                $ & $ 23.41_{-0.06}^{+0.06}$ & $  0.58_{-0.30}^{+0.27}$ &  0.7&  0.9& 44.5& $ 44.27_{-0.03}^{+0.04}$  \\
cosmos249                 &   23.4&  27& $ 1.98_{-0.61}^{+0.36}$ & $ 23.56_{-0.18}^{+0.14}$ & $ <0.95$ &  1.0&  1.4& 43.3& $ 43.30_{-0.03}^{+0.00}$  \\
cosmos251                 &  138.0& 127& $ 2.37_{-0.05}^{+0.07}$ & $ <21.17$ & $  1.17_{-0.49}^{+0.88}$ &  0.8&  1.3& 44.8& $ 44.31_{-0.22}^{+0.06}$  \\
cosmos253                 &   49.5&  52& $  1.8                $ & $ 22.68_{-0.08}^{+0.08}$ & $  4.99_{-1.43}^{+5.19}$ &  1.4&  1.8& 43.4& $ 42.84_{-0.15}^{+0.03}$  \\
cosmos263                 &  160.8& 153& $ 1.88_{-0.03}^{+0.05}$ & $ <21.32$ & $  0.11_{-0.10}^{+0.17}$ &  1.7&  2.8& 45.3& $ 45.23_{-0.09}^{+0.04}$  \\
cosmos272                 &   81.9&  68& $ 1.35_{-0.14}^{+0.16}$ & $ 22.62_{-0.11}^{+0.12}$ & $  0.02_{-0.00}^{+0.39}$ &  1.1&  1.6& 44.4& $ 44.39_{-0.06}^{+0.00}$  \\
cosmos282                 &  390.2& 377& $  1.8                $ & $ <20.39$ & $  0.42_{-0.18}^{+0.18}$ &  1.5&  2.3& 44.3& $ 44.17_{-0.00}^{+0.00}$  \\
cosmos284                 &  184.4& 186& $ 2.05_{-0.09}^{+0.08}$ & $ 21.87_{-0.06}^{+0.07}$ & $  1.30_{-0.84}^{+1.66}$ &  0.9&  1.4& 44.4& $ 44.02_{-0.25}^{+0.17}$  \\
cosmos287                 &  162.3& 164& $ 2.33_{-0.08}^{+0.08}$ & $ <21.13$ & $  1.24_{-0.54}^{+1.01}$ &  0.7&  1.2& 44.2& $ 43.84_{-0.03}^{+0.02}$  \\
cosmos296                 &  202.5& 234& $ 2.15_{-0.07}^{+0.09}$ & $ <21.36$ & $  1.57_{-0.80}^{+1.06}$ &  0.8&  1.3& 44.8& $ 44.22_{-0.07}^{+0.08}$  \\
cosmos297                 &  272.9& 247& $ 1.98_{-0.07}^{+0.10}$ & $ 21.99_{-0.04}^{+0.05}$ & $  1.55_{-0.94}^{+2.11}$ &  1.3&  2.0& 43.9& $ 43.58_{-0.04}^{+0.04}$  \\
cosmos299                 &   99.1&  71& $ 2.16_{-0.09}^{+0.06}$ & $ 23.67_{-0.05}^{+0.10}$ & $  0.28_{-0.21}^{+0.20}$ &  1.7&  2.3& 44.2& $ 44.09_{-0.09}^{+0.02}$  \\
cosmos322                 &  151.9& 137& $ 1.96_{-0.08}^{+0.10}$ & $ 21.74_{-0.07}^{+0.06}$ & $  2.58_{-1.08}^{+1.67}$ &  1.1&  1.5& 43.8& $ 43.32_{-0.03}^{+0.03}$  \\
cosmos330                 &   81.2&  81& $ 1.57_{-0.57}^{+0.55}$ & $ 24.13_{-0.21}^{+0.11}$ & $  0.22_{-0.17}^{+0.30}$ &  3.6&  3.8& 42.6& $ 42.52_{-0.00}^{+0.01}$ \\ 
ecdfs5                    &  151.7& 104& $ 1.90_{-0.07}^{+0.08}$ & $ 20.87_{-0.51}^{+0.22}$ & $  3.17_{-2.09}^{+11.26}$ &  1.2&  1.8& 43.0& $ 42.56_{-0.04}^{+0.03}$ \\ 
ecdfs20                   &  695.6& 717& $ 1.99_{-0.07}^{+0.05}$ & $ 23.02_{-0.03}^{+0.02}$ & $  0.74_{-0.15}^{+0.15}$ &  1.1&  1.6& 44.8& $ 44.46_{-0.03}^{+0.03}$  \\
ecdfs51                   &  318.2& 229& $ 1.98_{-0.05}^{+0.06}$ & $ <21.20$ & $  1.05_{-0.60}^{+0.93}$ &  0.7&  1.1& 44.5& $ 44.19_{-0.03}^{+0.02}$  \\
egs1                      &  220.1& 268& $ 1.84_{-0.05}^{+0.04}$ & $ 20.93_{-0.64}^{+0.31}$ & $  0.24_{-0.12}^{+0.47}$ &  0.8&  1.3& 44.7& $ 44.57_{-0.03}^{+0.01}$  \\
egs9                      &  438.4& 459& $ 1.80_{-0.04}^{+0.06}$ & $ 22.72_{-0.02}^{+0.02}$ & $ <0.28$ &  1.3&  2.2& 44.1& $ 44.13_{-0.03}^{+0.00}$ \\ 
egs26\tablenotemark{e}    &  424.4& 429& $ 1.56_{-0.03}^{+0.05}$ & $ 21.06_{-0.16}^{+0.12}$ & $ <0.68$ &  1.4&  2.0& 43.4& $ 43.37_{-0.09}^{+0.00}$ \\ 
egs27                     &  424.9& 411& $ 2.30_{-0.02}^{+0.02}$ & $ <20.56$ & $  1.06_{-0.31}^{+0.21}$ &  0.9&  1.5& 44.4& $ 44.00_{-0.01}^{+0.01}$  \\
egs32                     &  458.6& 462& $ 1.66_{-0.03}^{+0.03}$ & $ 21.62_{-0.08}^{+0.09}$ & $ <0.07$ &  0.8&  1.4& 44.7& $ 44.71_{-0.01}^{+0.00}$ \\ 
ser37\tablenotemark{f}    &  223.3& 219& $  1.8                $ & $ <20.58$ & $  1.78_{-0.84}^{+0.55}$ & 17.7& 23.7& 42.7& $ 42.48_{-0.05}^{+0.06}$  \\
ser77                     &   64.9&  66& $  1.8                $ & $ 21.93_{-0.07}^{+0.09}$  & $  2.00_{-0.95}^{+1.05}$ &  3.6&  5.2& 43.6& $ 43.29_{-0.03}^{+0.03}$  \\
ser97                     &   93.7&  95& $ 2.25_{-0.18}^{+0.32}$ & $ 23.06_{-0.22}^{+0.24}$ & $ <1.40$ &  1.7&  3.1& 43.4& $ 43.39_{-0.11}^{+0.00}$ \\ 
ser107                    &   27.6&  38& $ 1.48_{-0.23}^{+0.36}$ & $ 23.21_{-0.18}^{+0.27}$ & $ <4.45$ &  1.4&  1.9& 43.4& $ 43.34_{-0.18}^{+0.00}$  \\
ser148\tablenotemark{f}   & 1338.4&1260& $ 2.51_{-0.01}^{+0.02}$ & $ <19.89$ & $  2.63_{-0.36}^{+0.41}$ & 47.5& 85.7& 44.1& $ 43.80_{-0.03}^{+0.02}$  \\
ser153                    &   98.3&  93& $ 1.84_{-0.19}^{+0.26}$ & $ 23.42_{-0.32}^{+0.27}$ & $ <1.14$ &  1.1&  1.8& 45.1& $ 45.11_{-0.20}^{+0.00}$  \\
ser184                    &  461.2& 538& $ 1.73_{-0.02}^{+0.03}$ & $ 20.88_{-0.22}^{+0.20}$ & $ <0.10$ &  0.9&  1.5& 44.3& $ 44.35_{-0.01}^{+0.00}$ \\ 
ser213                    &  183.5& 183& $ 1.96_{-0.15}^{+0.14}$ & $ 23.33_{-0.12}^{+0.13}$ & $  1.37_{-0.53}^{+0.76}$ &  5.5&  7.5& 44.3& $ 44.05_{-0.07}^{+0.06}$  \\
ser215                    &   36.8&  35& $ 1.97_{-0.08}^{+0.06}$ & $ <21.71$ & $<16.72$ &  0.9&  1.5& 44.5& $ 44.29_{-0.04}^{+0.06}$  \\
ser235                    &   64.8&  56& $ 1.79_{-0.10}^{+0.37}$ & $ <24.25$ & $ <0.70$ &  1.2&  2.0& 45.6& $ 45.65_{-0.28}^{+0.00}$ \\ 
ser243                    &  152.6& 145& $ 1.35_{-0.10}^{+0.11}$ & $ 23.02_{-0.05}^{+0.05}$ & $ <0.74$ &  3.3&  4.6& 44.4& $ 44.43_{-0.09}^{+0.00}$  \\
ser254                    &  261.7& 260& $ 2.28_{-0.11}^{+0.08}$ & $ 22.98_{-0.06}^{+0.05}$ & $  1.90_{-0.62}^{+0.74}$ &  3.7&  5.9& 43.6& $ 43.33_{-0.04}^{+0.04}$  \\
ser261                    &   15.5&  15& $  1.8                $ & $ 24.61_{-0.32}^{+0.51}$ & $  0.37_{-0.29}^{+0.95}$ &  0.9&  1.0& 44.6& $ 44.41_{-0.19}^{+0.14}$  \\
ser267                    &   14.8&  14& $ 2.13_{-0.23}^{+0.26}$ & $ <21.74$ & $  4.09_{-3.26}^{+23.63}$ &  1.6&  2.8& 43.0& $ 42.52_{-0.13}^{+0.06}$ \\ 
ser273                    &  221.0& 276& $ 1.85_{-0.03}^{+0.04}$ & $ <20.87$ & $ <0.19$ & 11.8& 21.1& 44.7& $ 44.67_{-0.04}^{+0.00}$  \\
ser285                    &    7.7&   9& $  1.8                $ & $ <23.99$ & $<17.31$ &  0.9&  1.4& 44.4& $ 44.09_{-0.89}^{+0.27}$  \\
ser318                    &   63.9&  56& $ 2.41_{-0.08}^{+0.15}$ & $ <21.79$ & $  0.56_{-0.30}^{+0.76}$ &  1.2&  2.2& 45.1& $ 44.74_{-0.05}^{+0.05}$  \\
ser335                    &  135.6& 189& $ 1.69_{-0.03}^{+0.08}$ & $ <20.82$ & $ <1.10$ &  1.4&  2.3& 43.9& $ 43.82_{-0.05}^{+0.01}$  \\
ser359                    &   23.8&  44& $ 1.46_{-0.20}^{+0.29}$ & $ 22.56_{-0.13}^{+0.18}$ & $<13.89$ &  2.2&  3.2& 43.9& $ 43.90_{-0.18}^{+0.00}$ \\ 
ser363                    &  129.4& 205& $ 1.64_{-0.07}^{+0.06}$ & $ 21.67_{-0.26}^{+0.14}$ & $ <1.56$ &  1.2&  1.9& 45.1& $ 45.10_{-0.07}^{+0.00}$  \\
ser382                    &   40.7&  42& $ 1.88_{-0.20}^{+0.24}$ & $ 22.19_{-0.12}^{+0.13}$ & $  0.75_{-0.72}^{+8.43}$ &  0.9&  1.5& 43.3& $ 43.18_{-0.16}^{+0.02}$  \\
ser401                    &   31.6&  31& $ 1.87_{-0.18}^{+0.20}$ & $ 22.01_{-0.20}^{+0.16}$ & $<12.95$ &  0.9&  1.5& 44.5& $ 44.33_{-0.22}^{+0.10}$  \\
ser409                    &   33.8&  30& $ 1.68_{-0.06}^{+0.11}$ &  ---                 & $  0.71_{-0.62}^{+2.99}$ &  0.9&  1.3& 44.9& $ 44.58_{-0.51}^{+0.17}$  \\
ser451                    &   52.2&  66& $ 1.81_{-0.10}^{+0.61}$ & $ 23.42_{-0.10}^{+0.36}$ & $  <1.11$ &  1.4&  2.1& 44.4& $ 44.41_{-0.15}^{+0.00}$  \\
\enddata
  \tablenotetext{a}{\nhsym\ in units of \nh.}
  \tablenotetext{b}{Units of $10^{-13}$~\fluxcgs.}
  \tablenotetext{c}{Unabsorbed luminosity in the 10-40~keV energy range in units of \lumcgs. See Section~\ref{lum1040} for details.}
  \tablenotetext{d}{Intrinsic luminosity in the 10-40~keV energy range in units of \lumcgs. Errors highlights the uncertainty associated to the reflection component modeling}. See Section~\ref{lum1040} for details.
  \tablenotetext{e}{For this source we further added a partial covering absorber by partially cold material (zpcfabs in XSPEC). See Appendix for further details}
  \tablenotetext{f}{For this source we further added a partial covering absorber by partially ionized material (zxipcf in XSPEC). See Appendix for further details}

  \end{deluxetable*}
  \end{center}

\subsection{The reflection component}\label{reflection}
We next estimate the significance of the reflection component in our sources. 
We first evaluated if for the obscured sources (${\rm log} [N_{\rm H}/{\rm cm^{-2}}]\ge 22)$ the absorbed spectral shape could be better modelled in the context of a CT scenario in which the primary continuum is completely suppressed and where the only dominant component other than the soft residual scattered one is the pure cold reflection component.
Hence we evaluated a reflection-dominated spectrum obtained by removing the absorbed primary power-law component from the baseline model. Since we are not using $\chi^2$ statitics, we are not able to use an F-test to evaluate the significance of the baseline model over the  simpler reflection-dominated one.  We therefore based our evaluation on the presence of: 1) a reasonable input power-law photon index for the {\sc pexrav} component of the best-fit parametrization of the reflection dominated model; 2) a large fraction of scattered flux at low energies for the baseline model\footnote{I.e., if we are modeling an intrinsic reflection-dominated source with the baseline model, we obtain an overestimate of this quantity. To check for this we tied $\Gamma$ of the scattered component to the primary one.}; 3) the presence of an Fe~K$\alpha$ emission line with a large equivalent width ($\rm EW\gs1$~keV); and 4) large residuals for the best-fit parametrization. Based on these criteria, we did not find clear cases of sources deviating from the baseline model or significantly better parametrized by a reflection-dominated model. 
Similarly we did not find scattered fractions in excess of a few percent, the value that is typically found in heavily obscured sources \citep[e.g.][]{L2015}. Moreover, only for {\it cosmos181} we obtained $\rm Fe~K\alpha$ $\rm EW\sim 1~keV$. Other sources show more moderate $\rm EWs$. We therefore are unable to discriminate between the two models.

\subsubsection{Reflection as a function of obscuration, slope and luminosity of the primary emission}\label{Rvs}
\begin{figure}[t!]
   \begin{center}
\includegraphics[height=0.45\textwidth]{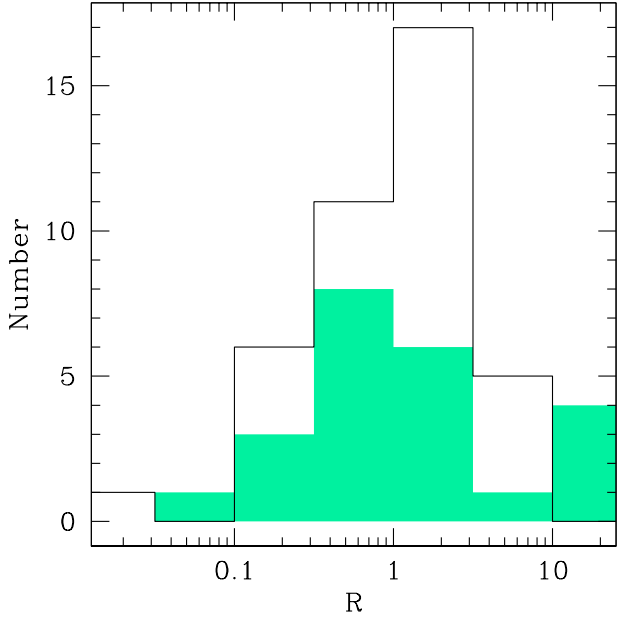}
   \end{center}
\caption{Distribution of measured reflection parameter (black histogram) as obtained from the baseline spectral modeling. The upper limits distribution is reported by the shaded green histogram. The binning is 0.5~dex.} 
   \label{rhisto}
\end{figure}

We measured $R$ for all the sources (see Table~\ref{specparambase} column 6) and obtained upper limits for 23 sources. We considered as upper limits all the best-fit values with $R<0.01$. 
In Fig.~\ref{rhisto} we report the distribution of $R$ in bins of 0.5~dex\footnote{Notice that the derived $R$ values are obtained by fixing the inclination angle ($\theta_{\rm incl}$) for the reflector to $\theta_{\rm incl}\sim63^\circ$ (default in Xspec). Assuming lower/larger inclination angles will decrease/increase $R$. For instance fixing $\theta_{\rm incl}=30^\circ$ ($\theta_{\rm incl}=85^\circ$) would lower (increase) our reported $R$ by 50\% (a factor of 2-3).}.

\begin{figure*}[t!]
   \begin{center}
\includegraphics[height=0.45\textwidth]{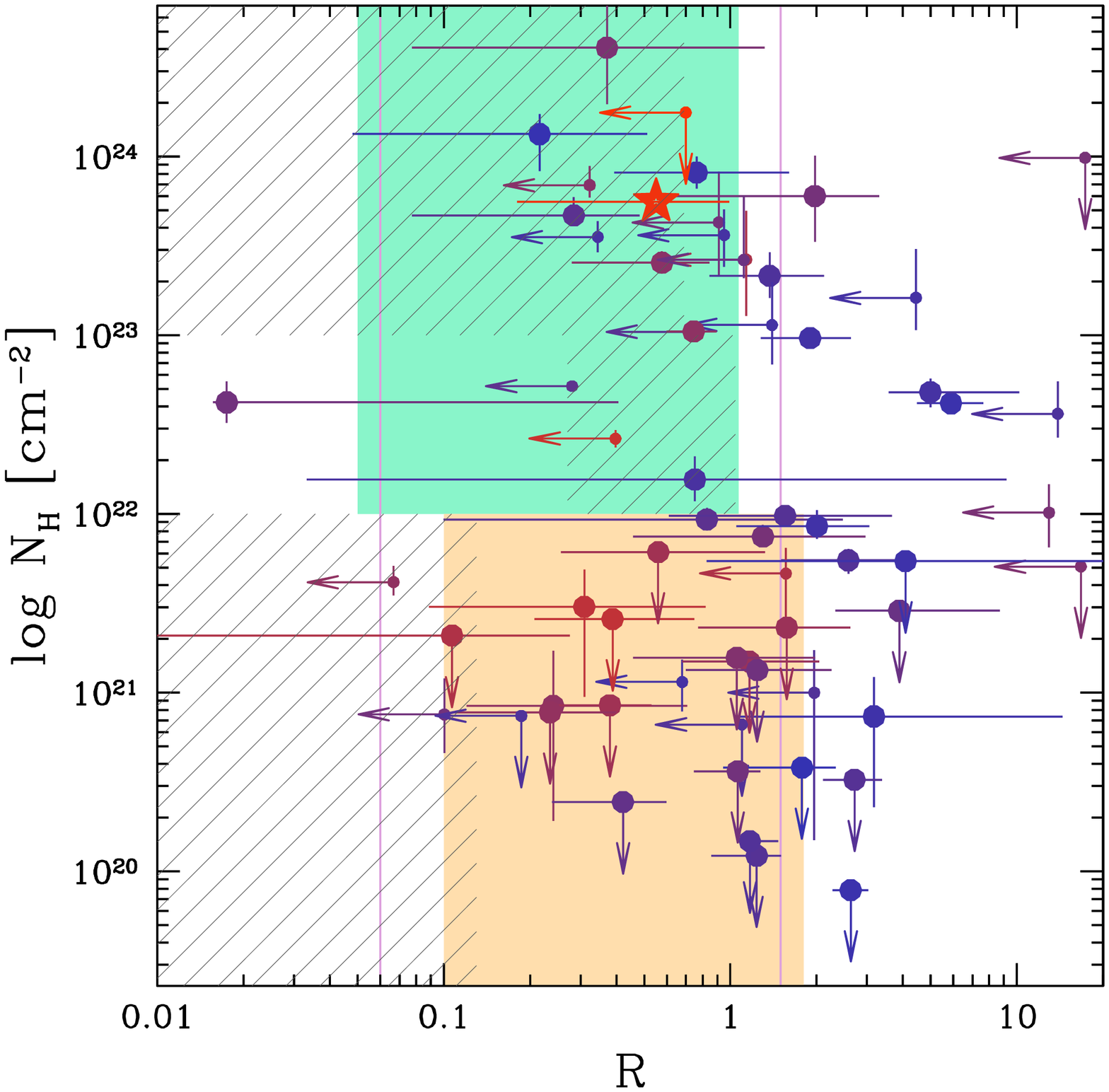}
\hspace{0.5cm}\includegraphics[height=0.45\textwidth]{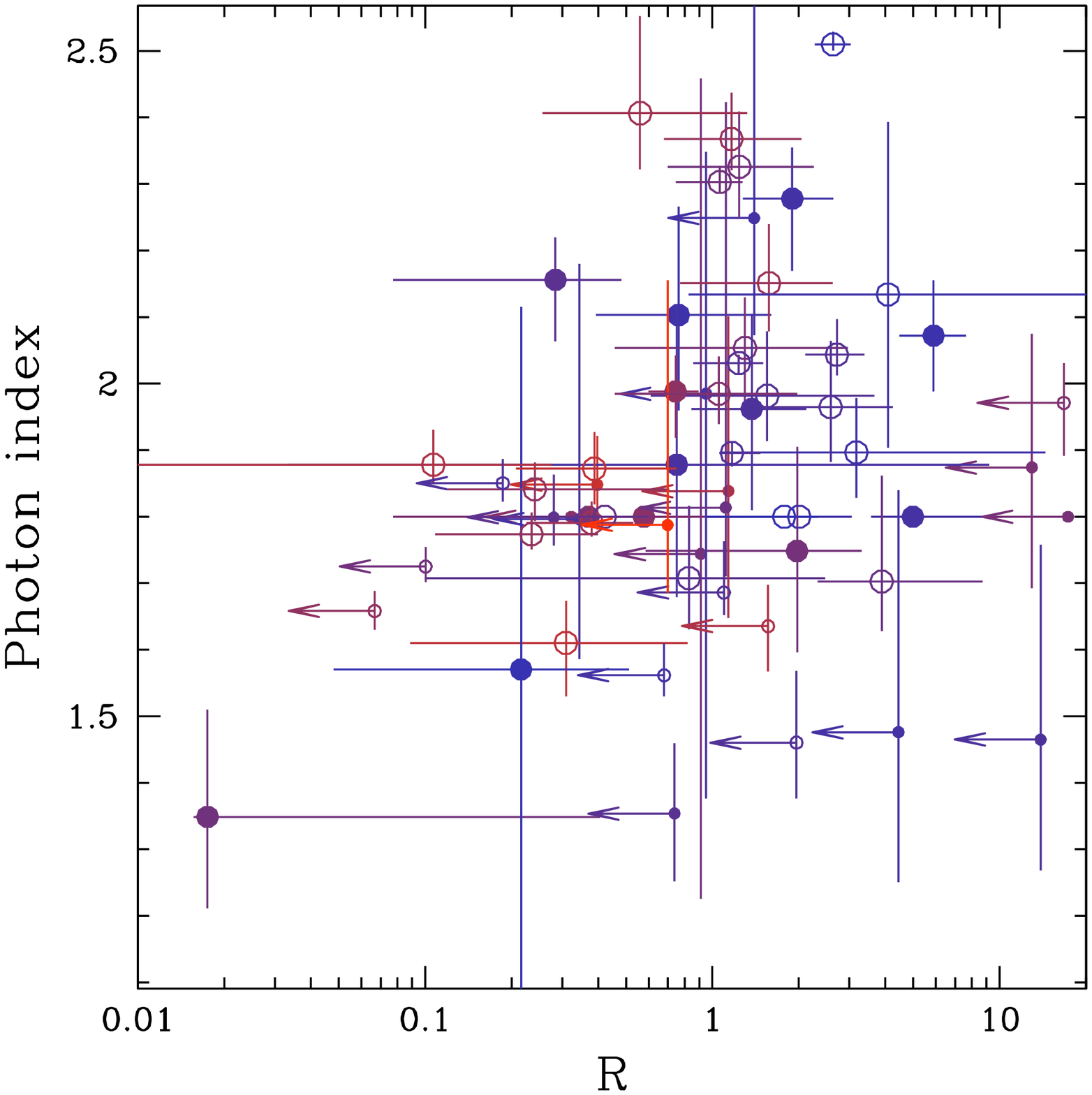}
\includegraphics[height=0.45\textwidth]{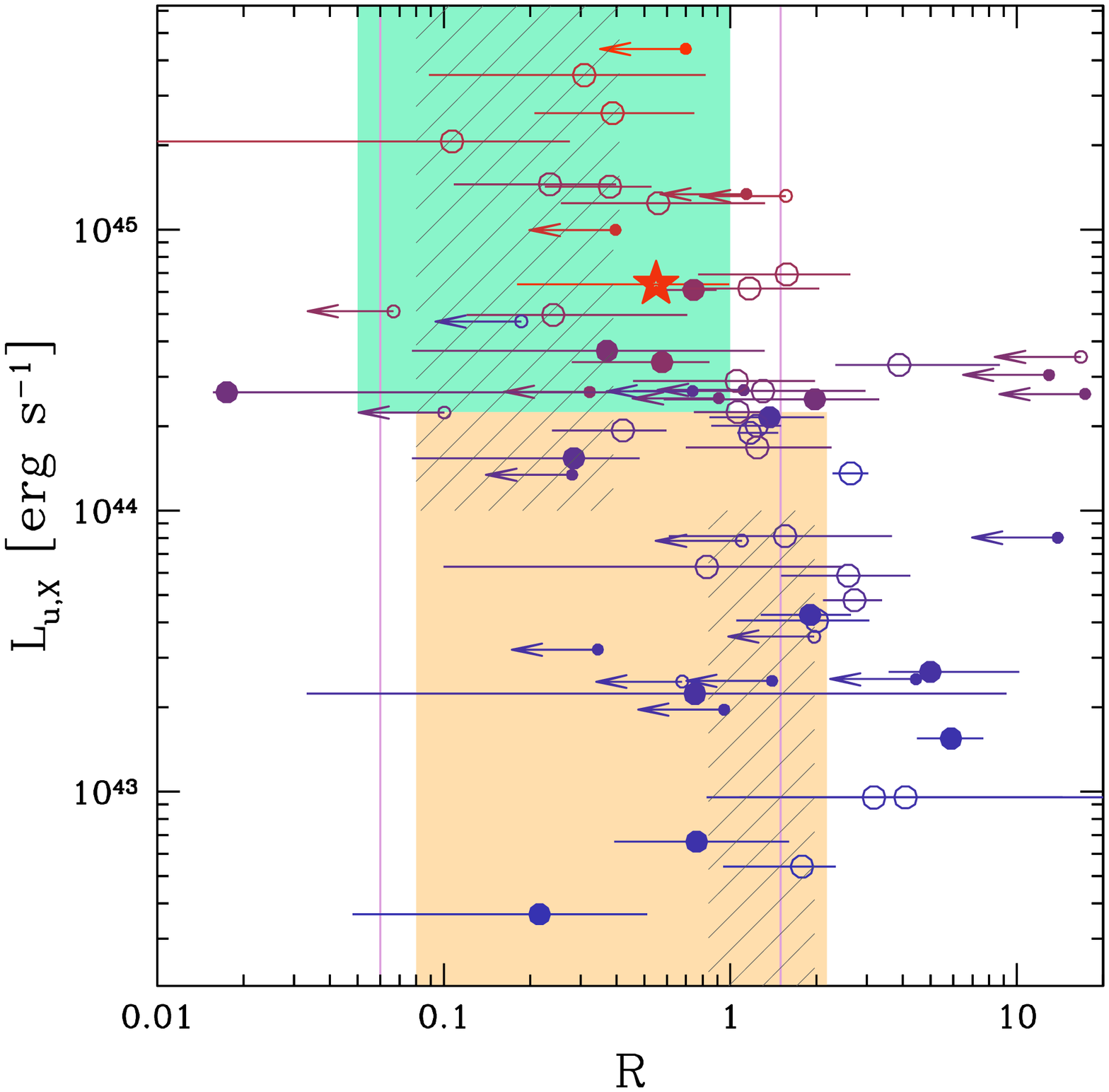}
\hspace{0.5cm}\includegraphics[height=0.45\textwidth]{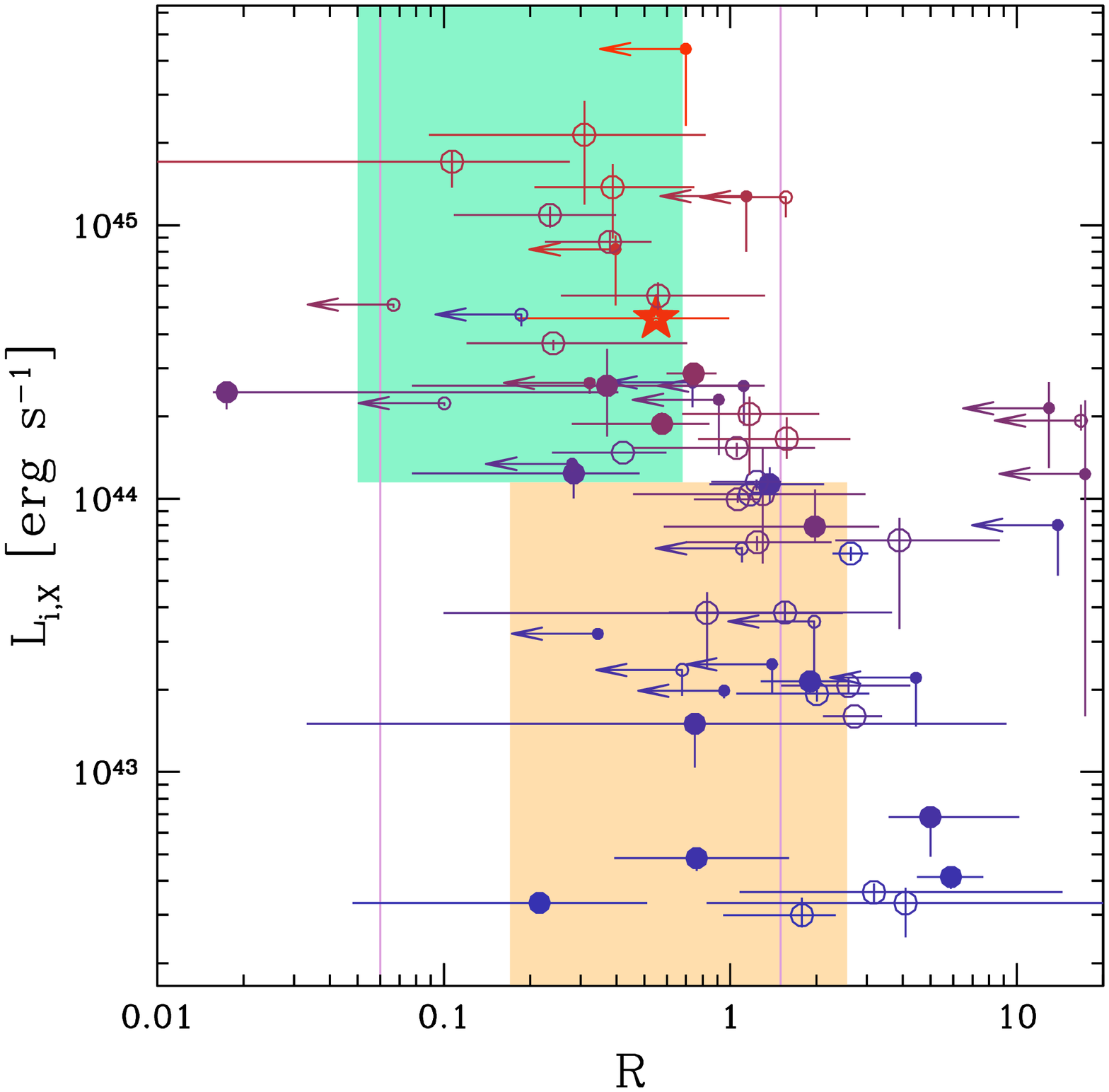}
   \end{center}
\caption{Reflection parameter against \nhsym\ (top-left panel), photon index (top-right panel), unabsorbed luminosity $L_{u,X}$  (bottom-left panel) and intrinsic coronal luminosity $L_{i,X}$ (bottom-right panel). Colors reflect the redshift of each source, with redder colors representing more distant objects. In the $R$ vs. $L_{\rm X}$ and $R$ vs. $\Gamma$ plots, empty (filled) circles represent unobscured (obscured) AGN. Vertical lines mark the interquartile interval for $R$ in the entire sample. Shaded green (yellow) regions represent the interquartile ranges for obscured (unobscured) and more (less) luminous sources (the latter being separated at a median luminosity). In the $R$ vs. luminosity  and  $R$ vs. \nhsym\ plots the red star represents the high-redshift quasar detected in the \ecdfs\ field and analyzed in \citet{DM2014}. The hatched regions in the $R$ vs. $L_{u,X}$ represent 90\% error range for bins in \nhsym\ and $L_X$ as measured from the stacking analysis performed by DM17 on a large sample of \nustar-detected sources. Values of interquartile ranges are reported also in Table~\ref{refltable}. See Section~\ref{lum1040} for the definition of $L_{u,X}$  and $L_{i,X}$.} 
   \label{R_vs_NH}
\end{figure*}

We investigated how  reflection correlates with  obscuration and luminosity for the whole sample. Fig.~\ref{R_vs_NH} presents the reflection parameter as a function of column density (top-left panel) and unabsorbed and intrinsic 10-40~keV luminosity (bottom panels). The color of each point corresponds to redshift with the redder colors representing the more distant sources.  Since ours is a flux-selected sample, more distant (i.e., redder) sources in the $R-L_X$ plane correspond to more luminous, less obscured sources (i.e., see $R-N_{\rm H}$ plane). 

There is an apparent tendency for obscured and luminous sources to have, on average, maximum $R$ values smaller than unobscured and less luminous sources.

We investigated and quantified these trends by: (1) computing the Spearman's rank correlation coefficient ($\rho$) for censored data using the ASURV package v.~1.2 \citep{asurv,FN1985,IFN1986} and (2) calculating the median $\langle R \rangle$ and its interquartile range (IQR) for the entire sample and the obscured/unobscured and luminous/less luminous sub-samples (the separation between the latter being dictated by the median luminosities of the sample, $\log [\langle L_{i,\rm X} \rangle/\rm{erg\,s^{-1}}]=44.06$ and $\log [\langle L_{u,\rm X} \rangle/\rm{erg\,s^{-1}}]=44.35$). 

\begin{center}
  \begin{deluxetable}{llclcc}
    \tablecaption{Reflection parameter: correlation coefficient values, median and interquartile ranges relative to column density and luminosities
      \label{refltable}
    }
\tablehead{ Parameters & $\rho$\tablenotemark{a}     & $p$\tablenotemark{a}           & Sample  &   $\langle R\rangle$\tablenotemark{b}  &  IQR\tablenotemark{b} }
\startdata
  $R$                          &  --                       &     --                           & $\rm All$                 &  0.43          &      0.06--1.50               \\
  \multirow{2}{*}{$\lognhonly$, $R$}  &  \multirow{2}{*}{-0.25}   & \multirow{2}{*}{0.05}           & unobscured\tablenotemark{c}                 &  0.67	    &      0.10--1.80 	            \\
                                   &                           &                                  & obscured\tablenotemark{c}               &  0.28          & 	   0.05--1.07               \\
  \multirow{2}{*}{${\rm log} L_{i,X}$, $R$}& \multirow{2}{*}{-0.59}    & \multirow{2}{*}{$<10^{-5}$} & low $L_{i,X}$\,\tablenotemark{d}  &  1.15          &     0.17--2.56  	            \\
                                   &                           &                                  & high $L_{i,X}$\,\tablenotemark{d}  &  0.25	    &    0.05--0.68             \\    
  \multirow{2}{*}{${\rm log} L_{u,X}$, $R$}& \multirow{2}{*}{-0.37}    & \multirow{2}{*}{0.0039}     & low $L_{u,X}$\,\tablenotemark{e}  &  0.73          &     0.08--2.17  	            \\
                                   &                           &                                  & high $L_{u,X}$\,\tablenotemark{e}  &  0.31          &     0.05--1.00  	            
\enddata
  \tablenotetext{a}{Spearman's rho correlation coefficient ($\rho$) and null-hypothesis probability ($p$) calculated for censored data from ASURV package (see Section~\ref{Rvs}).}
  \tablenotetext{b}{Median ($\langle R\rangle$) and Interquartile range (IQR) values for $R$ computed in each considered sample accounting for errors and upper limits as explained in Section~\ref{Rvs}.}
  \tablenotetext{c}{We used a $10^{22}$~\nh\ threshold  value.}
  \tablenotetext{d}{The median value $\log (\langle L_{u,X}\rangle/{\rm erg s^{-1}})=44.35$ is adopted as the threshold value.}
  \tablenotetext{e}{The median value $\log (\langle L_{i,X}\rangle/{\rm erg s^{-1}})=44.06$ is adopted as the threshold value.}
\end{deluxetable}
\end{center}

For the latter we accounted for measurement errors and upper limits in $R$, $\lognh$ and ${\rm log} L_{i,X}$ as follows: we performed 10000 realizations of the sample each time with Gaussian and uniform randomization for respectively each of the parameter best-fit values\footnote{We assumed a symmetric distribution centered on the  parameter value with 1$\sigma$ standard deviation as the mean of the lower and upper error-bars.} and the upper limits. In the case of $R$ and $\lognhonly$, the latter were randomized from their 90\% upper value down to a fixed minimum value of $R=0.01$ and $\lognh=20$.
We computed for each realization the median value and IQR, and adopted as representative for the sample the averaged values over all the realizations. 
The resulting values are reported in Table~\ref{refltable}. Note that accounting for the upper limits may lead to a shift of the lower interquartile bound toward smaller values. Therefore the lower interquartile range may not reflect the true relative $R$ distributions.  
The IQR values are reported as shaded areas in Fig.~\ref{R_vs_NH} for the sub-samples and vertical lines for the entire sample.

For the entire sample, the average median value is $\langle R\rangle=0.43$ with an interquartile range $0.06-1.50$.
We find a weak mildly significant anti-correlation between $R$ and $\lognhonly$ with $\rho=-0.25$ and a null hypothesis probability that the two quantities are not related to each other of $p=$0.05. The median $R$ values for unobscured and obscured samples are respectively of 0.67 and 0.28. Despite the apparent difference, their IQR have in common a quite large interval of values. The  difference in the lower values (with unabsorbed sources having larger values) may reflect the fact that the obscured sample has twice as many upper limits  as the unobscured sample.
We verified that the presence of such a large number of upper limits does not depend on the SNR of the \nustar\ spectra. The upper bounds of the interquartile range differ by a factor of $\sim1.7$. Both categories sample AGN with similar range in luminosities (interquartile range $\rm log (L_{u,X}/\rm{erg\,s^{-1}})=43.8-44.9$ and $\rm log (L_{i,X}/\rm{erg\,s^{-1}})=43.5-44.6$). Therefore the dependence of $R$ on luminosity (see below) should not affect our result. We checked for the possibility that the resulting trend is due to the covariance in our modeling at the highest column densities between the two quantities. We computed the confidence contours for \nhsym\
and $R$ for the heavily obscured sources (i.e., $\lognh\ge23$) and we find no significant covariance with few sources showing weak covariance which has been found to be either positive ({\it cosmos181} and {\it cosmos216}) or negative ({\it ser243} and {\it ser254}).
If the anti-correlation between $R$ and $\lognhonly$ is real, it could be explained by a configuration in which the obscurer absorbs also the reflected component. Hence a more pronounced reflection is necessary to reproduce the observed high energy spectral shape. This would bring the reflection parameters to higher values for the obscured AGN ($\lognhsq\ge22$) in agreement with those derived for the unabsorbed sources. Hence we estimated $R$ and $\lognhonly$ best-fit values by performing modeling on the obscured AGN with a modified baseline model in which the reflection component is absorbed by the same column density absorbing the power-law continuum. We obtained $\langle R\rangle=0.14$ and IQR$=0.02-0.93$ which is not dissimilar to the values obtained by the baseline model. We also estimated the correlation between the two quantities and obtained $\rho=-0.19$ with $p=0.147$. Hence although a weak anti-correlation still persists, it is not significant. We therefore cannot make significant claims about this hypothesis. 
We mention that accounting for Compton-scattering in our baseline model may affect the determination of $R$ for the most absorbed sources. We measured this for the most obscured sources in the \cosmos\ sub-sample (i.e., {\it cosmos330}, {\it cosmos181}, {\it cosmos216} and {\it cosmos129}) and found $R$ values which are lower but always in agreement with those from the baseline model within the $1\sigma$ uncertainties. The inclusion of a Compton-scattering term will, on average, lower the median $R$ value for the $\lognh\ge22$ sources and increase the disagreement compared to the unobscured sources (i.e., strenghten the anti-correlation). 

$R$ may be partially degenerate with $\Gamma$ in our modeling. We investigated this induced effect in the $R-\Gamma$ plane (see Fig.~\ref{R_vs_NH}, top right panel). There is a significant correlation between the two quantities with $\rho=0.54$ and a null-hypothesis probability $p=10^{-4}$. We find similar trends with obscuration at a lower significance level. Notice that our evaluation in this case is both affected by the small number of sources in the obscured and unobscured sample and for the obscured sources by the many sources with $\Gamma=1.8$ fixed which we had to exclude from the analysis.  Therefore we cannot draw firm conclusions on this point. A pronounced degree of degeneracy between the two parameters has also been  found through spectral stacking analysis in our companion paper (DM17) using a sample three times larger than ours.

As for the relation with luminosity, we find a significant anti-correlation for both $L_{u,X}$ and $L_{i,X}$ (see Fig.~\ref{R_vs_NH}, bottom panels). The correlation coefficient for the former quantity is $\rho=-0.37$ with $p=3.9\times10^{-3}$ while for the latter is $\rho=-0.59$ with $p<1\times10^{-5}$. The stronger correlation with $L_{i,X}$ is expected given that $L_{u,X}$ includes a contribution from the reflection component itself which partially mitigates the ``true'' relation. 
The stronger correlation of $L_{i,X}-R$ compared to $L_{u,X}-R$ reflects also in the median and IQR values as shown in Table~\ref{refltable}. 
The median and upper IQR bound values for luminous and less-luminous sources differ respectively by a factor of $\gs4$ and $\gs2$. 
These trends are not particularly sensitive to the luminosity value adopted to separate the two subsamples: less luminous sources always exhibit more pronounced reflection than luminous ones. For example, a change in  $L_{i,X}$ threshold values by $\pm$50\% translates into a $\gs3-5$ factor difference in median and upper bound $R$ values.
A $z=2$ quasar selected in the \nustar-ECDF-S field and analyzed in \citet{DM2014} is reported in the plots as the starred data point. This source shows a low degree of reflection, comparable to the luminous/obscured AGN in our sample.
Notice that we are using a flux-limited sample. Hence the more luminous sources are also on average the more distant ones. Indeed the sources brighter than the median intrinsic luminosity of the sample have median redshifts of $\langle z_{hi-z}\rangle \gs0.9$ while the less luminous ones have $\langle z_{hi-z}\rangle \gs0.25$. Hence it is also possible that the main driver of the correlation between luminosity and $R$ is the redshift. With this sample we are not able to break the degeneracy between luminosity and redshift in order to investigate this scenario.

In our companion paper (DM17) we analyze, through stacking techniques, the average spectral properties of the 182  AGN detected in the medium-deep \nustar\ surveys. This sample is three times larger than the one used on that work with a slight overlap (for one sixth of the sources) with our sample.  
The  average reflection strength is found to be $R\approx0.5$ ($\Gamma=1.8$ fixed) with hard-band detected sources showing a slightly higher value $R\approx0.7$ (for $\Gamma=1.8$ fixed, $R<0.4$ when leaving $\Gamma$ free to vary).  These values are in good agreement with our $\langle R \rangle$ and within the scatter suggested by the interquartile range.  
The derived $R$ values as a function of unabsorbed luminosity $L_{u,X}$ and column density are in good agreement with those inferred for our sample.  This can be seen in Fig.~\ref{R_vs_NH} (bottom left panel) where the  $90$\% error range in $R$, accounting for the degeneracy with $\Gamma$ (i.e., reporting ranges according to the average $\Gamma$ we measured in our sub-samples), is reported as hatched grey regions.

Our combined analysis of the \nustar\ sources gives an indication  of the average $R$ values (from DM17) and of their dispersion  among different sources (from this work) as a function of luminosity and column density for the intermediate redshift ($\langle z\rangle=0.5-1$) AGN population.

Our findings are also consistent with previous results showing low levels of reflection for high-redshift quasars \citep[][]{V1999,RT2000,Pa2005} though note that \citet{R2011}  found higher levels of reflection from Seyfert~2s compared to Seyfert~1s.  Fig.~\ref{R_vs_NH} (bottom panels)  shows unobscured and obscured sources as  empty and filled data points, respectively, and does not allow a firm conclusion on this point for Seyfert-like luminosity sources.
%
%
%
\smallskip
\smallskip
\begin{figure}[t!]
   \begin{center}
\includegraphics[width=0.45\textwidth]{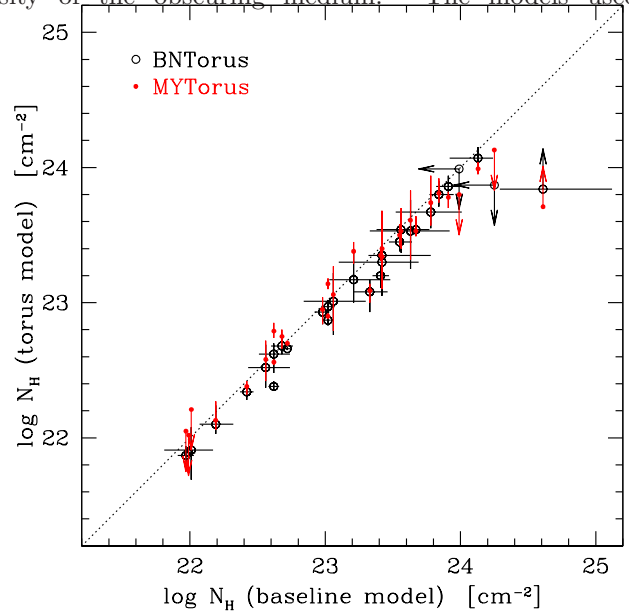}
   \end{center}
\caption{Comparison of column density values from the best-fit baseline model  and torus models for the obscured sources. Black hollow circles report values from the {\sc BNTorus} model. Red dots report values from the {\sc MYTorus} model.} 
   \label{nh_base_torus}
\end{figure}

\subsection{Physically motivated models for the obscured sources}\label{torus}
To constrain the spectral parameters we also adopted two physically motivated, Monte Carlo models which self-consistently account for the toroidal geometry of the obscuring/reflecting medium and properly treat continuum suppression due to the additional contribution of Compton-scattering at the highest column densities. The latter contribution, if neglected, can lead to an overestimation of the column density of the obscuring medium. The models used are {\sc MYTorus}  \citep[][]{MY2009} and the \citet{BN2011} model (hereafter {\sc BNTorus}). {\sc MYTorus} assumes a proper torus geometry with a half-opening angle $\theta_{\rm oa}=60$~deg (i.e., a covering factor of 0.5).  The torus geometry in {\sc BNTorus} is approximated as a sphere with variable polar conical openings.  We apply these models to the obscured sources (i.e., those found to have column density values or upper/lower limits consistent with $\lognhsq\gtrsim22$, including {\it ser409} for which \nhsym\ was unconstrained) as estimated by the baseline model (see Table~\ref{specparambase}) with the aim of comparing the \nhsym\ values.
We assume and edge-on orientation with inclination angle 85~deg. For the low SNR spectra with unconstrained photon indices we fix $\Gamma=1.8$.
\begin{center}
\tablewidth{0pt}
\tabletypesize{\scriptsize}
  \begin{deluxetable*}{l|rrrrccc|rrrrccc}

    \tablecaption{Best-fit parameters for the Torus models
\label{specparamtorus}
    }
      \tablehead{   & \multicolumn{7}{c|}{{\sc BNTorus}}  &  \multicolumn{7}{c}{{\sc MYTorus}} \\
        ID &     stat &   dof &    $\Gamma$          &   $\lognhonly\,$\tablenotemark{a}               &  $S_{\rm 8-24}$\,\tablenotemark{b}    &  $S_{\rm 3-24}$\,\tablenotemark{b}   &  $\log L_{u,X}$\,\tablenotemark{c}  &     stat &   dof &    $\Gamma$          &   $\lognhonly\,$\tablenotemark{a}               &  $S_{\rm 8-24}$\,\tablenotemark{b}    &  $S_{\rm 3-24}$\,\tablenotemark{b}   &  $\log L_{u,X}$\,\tablenotemark{c}   }
\startdata
cosmos97    &     271.1  &   264   &  $ 1.33_{-0.07}^{+0.06}$   &  $ 22.38_{- 0.03}^{+ 0.04}$   &   3.8  &   5.4  &  43.2  &     291.0  &   265   &  $ 1.80              $   &  $ 22.56_{- 0.02}^{+ 0.02}$   &   2.8  &   4.6  &  43.0 \\ 
cosmos107   &      19.6  &    17   &  $ 1.73_{-0.56}^{+0.56}$   &  $ 23.53_{- 0.28}^{+ 0.23}$   &   0.9  &   1.2  &  44.4  &      19.5  &    17   &  $ 1.77_{-1.77}^{+0.58}$   &  $ 23.61_{- 0.29}^{+ 0.22}$   &   0.9  &   1.3  &  44.5 \\ 
cosmos129   &      18.5  &    25   &  $ 1.80              $   &  $ 23.80_{- 0.09}^{+ 0.10}$   &   0.5  &   0.7  &  44.6  &      18.5  &    25   &  $ 1.80              $   &  $ 23.81_{- 0.07}^{+ 0.11}$   &   0.5  &   0.7  &  44.6 \\ 
cosmos145   &     151.2  &   171   &  $ 1.59_{-0.08}^{+0.12}$   &  $ 21.87_{- 0.07}^{+ 0.07}$   &   0.6  &   0.9  &  43.7  &     152.0  &   171   &  $ 1.67_{-0.05}^{+0.07}$   &  $<22.05                $   &   0.5  &   0.8  &  43.7 \\ 
cosmos154   &      75.4  &    72   &  $ 1.73_{-0.22}^{+0.26}$   &  $ 23.49_{- 0.06}^{+ 0.09}$   &   1.6  &   2.0  &  43.6  &     125.2  &    81   &  $ 1.68_{-0.21}^{+0.25}$   &  $ 23.50_{- 0.08}^{+ 0.08}$   &   1.4  &   1.9  &  43.5 \\ 
cosmos181   &      49.3  &    48   &  $ 1.80              $   &  $ 23.89_{- 0.05}^{+ 0.10}$   &   1.2  &   1.4  &  43.1  &      88.1  &    52   &  $ 2.27_{-0.08}^{+0.16}$   &  $ 23.78_{- 0.08}^{+ 0.04}$   &   1.0  &   1.3  &  42.9 \\ 
cosmos216   &      30.3  &    44   &  $ 1.80              $   &  $ 23.67_{- 0.13}^{+ 0.06}$   &   0.8  &   1.1  &  44.4  &      29.8  &    43   &  $ 2.07_{-0.45}^{+-2.07}$   &  $ 23.74_{- 0.17}^{+ 0.20}$   &   0.6  &   0.9  &  44.4 \\ 
cosmos217   &     321.7  &   268   &  $ 1.76_{-0.05}^{+0.07}$   &  $ 22.34_{- 0.06}^{+ 0.04}$   &   0.5  &   0.8  &  45.0  &     321.6  &   268   &  $ 1.74_{-0.06}^{+0.08}$   &  $ 22.38_{- 0.06}^{+ 0.04}$   &   0.5  &   0.8  &  45.0 \\ 
cosmos232   &      64.6  &    77   &  $ 1.32_{-0.18}^{+0.24}$   &  $ 23.20_{- 0.09}^{+ 0.07}$   &   0.8  &   1.0  &  44.6  &      69.2  &    78   &  $ 1.80              $   &  $ 23.35_{- 0.04}^{+ 0.05}$   &   0.5  &   0.8  &  44.5 \\ 
cosmos249   &      23.1  &    29   &  $ 1.80_{-0.46}^{+0.35}$   &  $ 23.54_{- 0.17}^{+ 0.13}$   &   1.1  &   1.4  &  43.4  &      23.7  &    29   &  $ 1.52_{-1.52}^{+0.58}$   &  $ 23.53_{- 0.13}^{+ 0.17}$   &   1.3  &   1.6  &  43.5 \\ 
cosmos253   &      61.2  &    53   &  $ 1.80              $   &  $ 22.68_{- 0.06}^{+ 0.09}$   &   0.8  &   1.3  &  43.2  &      61.9  &    53   &  $ 1.80              $   &  $ 22.75_{- 0.09}^{+ 0.05}$   &   0.9  &   1.4  &  43.2 \\ 
cosmos272   &      85.7  &    70   &  $ 1.39_{-0.14}^{+0.14}$   &  $ 22.62_{- 0.14}^{+ 0.08}$   &   1.0  &   1.5  &  44.4  &      88.2  &    71   &  $ 1.80              $   &  $ 22.79_{- 0.05}^{+ 0.06}$   &   0.8  &   1.3  &  44.3 \\ 
cosmos297   &     277.3  &   249   &  $ 1.80_{-0.08}^{+0.08}$   &  $ 21.89_{- 0.05}^{+ 0.04}$   &   1.0  &   1.7  &  43.8  &     295.8  &   249   &  $ 1.82_{-0.05}^{+0.05}$   &  $<22.02                $   &   0.6  &   1.0  &  43.6 \\ 
cosmos299   &     100.7  &    73   &  $ 1.80_{-0.05}^{+0.07}$   &  $ 23.51_{- 0.04}^{+ 0.07}$   &   1.7  &   2.4  &  44.3  &     107.9  &    73   &  $ 1.81_{-0.15}^{+0.56}$   &  $ 23.53_{- 0.04}^{+ 0.11}$   &   1.7  &   2.4  &  44.3 \\ 
cosmos330   &      83.5  &    84   &  $ 1.86_{-0.19}^{+0.16}$   &  $ 24.07_{- 0.05}^{+ 0.08}$   &   3.2  &   3.5  &  42.7  &      90.8  &    84   &  $ 1.46_{-1.46}^{+0.08}$   &  $ 23.99_{- 0.04}^{+ 0.05}$   &   3.3  &   3.6  &  42.7 \\ 
ecdfs20     &     706.1  &   718   &  $ 1.61_{-0.07}^{+0.07}$   &  $ 22.87_{- 0.04}^{+ 0.02}$   &   1.0  &   1.6  &  44.8  &     706.8  &   718   &  $ 1.57_{-0.06}^{+0.05}$   &  $ 22.90_{- 0.02}^{+ 0.02}$   &   1.1  &   1.6  &  44.8 \\ 
egs9        &     443.1  &   461   &  $ 1.80_{-0.03}^{+0.07}$   &  $ 22.66_{- 0.01}^{+ 0.01}$   &   1.3  &   2.2  &  44.1  &     444.5  &   461   &  $ 1.79_{-0.05}^{+0.05}$   &  $ 22.70_{- 0.02}^{+ 0.02}$   &   1.3  &   2.2  &  44.1 \\ 
ser97       &      94.2  &    96   &  $ 2.31_{-0.24}^{+0.22}$   &  $ 23.01_{- 0.25}^{+ 0.21}$   &   1.7  &   3.1  &  43.4  &      94.2  &    96   &  $ 2.29_{-0.24}^{+0.21}$   &  $ 23.06_{- 0.27}^{+ 0.21}$   &   1.7  &   3.1  &  43.4 \\ 
ser107      &      27.6  &    40   &  $ 1.49_{-0.28}^{+0.26}$   &  $ 23.17_{- 0.17}^{+ 0.23}$   &   1.4  &   1.9  &  43.4  &      27.7  &    41   &  $ 1.80              $   &  $ 23.38_{- 0.09}^{+ 0.07}$   &   1.3  &   1.7  &  43.4 \\ 
ser153        &      99.5  &    95   &  $ 1.87_{-0.19}^{+0.24}$   &  $ 23.30_{- 0.25}^{+ 0.24}$   &   1.1  &   1.8  &  45.2  &      99.9  &    95   &  $ 1.86_{-0.20}^{+0.24}$   &  $ 23.34_{- 0.24}^{+ 0.25}$   &   1.1  &   1.8  &  45.2 \\ 
ser213        &     187.4  &   184   &  $ 1.43_{-0.13}^{+0.13}$   &  $ 23.08_{- 0.15}^{+ 0.09}$   &   5.5  &   7.5  &  44.4  &     187.4  &   184   &  $ 1.80              $   &  $ 23.09_{- 0.09}^{+ 0.05}$   &   5.6  &   7.6  &  44.4 \\ 
ser235       &      64.9  &    57   &  $ 1.79_{-0.31}^{+0.35}$   &  $<23.87                $   &   1.2  &   2.0  &  45.6  &      64.8  &    57   &  $ 1.80_{-0.33}^{+0.31}$   &  $<24.13                $   &   1.1  &   2.0  &  45.6 \\
ser243       &     153.2  &   147   &  $ 1.38_{-0.09}^{+0.11}$   &  $ 22.97_{- 0.05}^{+ 0.05}$   &   3.3  &   4.6  &  44.4  &     158.8  &   148   &  $ 1.80              $   &  $ 23.14_{- 0.04}^{+ 0.04}$   &   3.4  &   4.7  &  44.5 \\ 
ser254       &     260.0  &   260   &  $ 2.17_{-0.53}^{+0.16}$   &  $ 22.93_{- 0.09}^{+ 0.05}$   &   3.8  &   6.0  &  43.7  &     260.1  &   260   &  $ 2.05_{-0.32}^{+0.33}$   &  $ 22.95_{- 0.11}^{+ 0.09}$   &   3.6  &   5.8  &  43.7 \\ 
ser261      &      15.7  &    16   &  $ 1.80              $   &  $>23.84                $   &   1.1  &   1.2  &  45.0  &      16.2  &    16   &  $ 1.80              $   &  $>23.71                $   &   0.9  &   1.0  &  44.9 \\ 
ser285       &       8.3  &     9   &  $ 1.57_{-0.33}^{+0.07}$   &  $<23.99                $   &   1.0  &   1.5  &  44.5  &       8.3  &     9   &  $ 1.59_{-1.59}^{+0.71}$   &  $<23.80                $   &   1.0  &   1.5  &  44.5 \\ 
ser359       &      23.7  &    45   &  $ 1.52_{-0.22}^{+0.27}$   &  $ 22.52_{- 0.15}^{+ 0.17}$   &   2.3  &   3.3  &  43.9  &      23.7  &    45   &  $ 1.51_{-1.51}^{+0.28}$   &  $ 22.58_{- 0.16}^{+ 0.14}$   &   2.3  &   3.4  &  43.9 \\ 
ser382       &       41.3  &     43   &  $ 1.75_{-0.12}^{+0.20}$   &  $ 22.10_{- 0.07}^{+ 0.14}$   &   0.8  &   1.4  &  43.3  &       41.5  &     43   &  $ 1.70_{-0.13}^{+0.19}$  &  $ 22.13_{- 0.07}^{+ 0.14}$   &   0.8  &   1.3  &  43.3 \\ 
ser401       &      31.8  &    33   &  $ 1.76_{-0.18}^{+0.19}$   &  $ 21.91_{- 0.22}^{+ 0.17}$   &   0.9  &   1.4  &  44.5  &      31.9  &    33   &  $ 1.80_{-0.13}^{+0.15}$   &  $<22.21                $   &   0.8  &   1.4  &  44.4 \\ 
ser409      &      35.7  &    32   &  $ 1.80              $   &  $<24.08                $   &   0.7  &   1.2  &  44.9  &       8.2  &     5   &  $ 1.80              $   &  $<23.72                $   &   0.9  &   1.3  &  44.9 \\ 
ser451        &      52.0  &    68   &  $ 1.82_{-0.10}^{+0.36}$   &  $ 23.35_{- 0.09}^{+ 0.29}$   &   1.4  &   2.1  &  44.5  &      52.0  &    68   &  $ 1.92_{-0.14}^{+0.41}$   &  $ 23.40_{- 0.10}^{+ 0.28}$   &   1.3  &   2.0  &  44.5  
  \enddata
  \tablenotetext{a}{Line of sight column density in units of \nh.}
  \tablenotetext{b}{Units of $10^{-13}$~\fluxcgs.}
  \tablenotetext{c}{Unabsorbed 10-40~keV luminosity in units of \lumcgs.}
  \end{deluxetable*}
  \end{center}

The best-fit parameters are reported in Table~\ref{specparamtorus}. For both models there is broad agreement in \nhsym\ and $\Gamma$ with the baseline model. Indeed we have 15 and 16 sources, $\sim$25\% of the sample, in the $\lognh=23-24$ bin for {\sc BNTorus} and {\sc MYTorus} respectively\footnote{Furthermore both modelings have an additional source with estimated \nhsym\ consistent within 1$\sigma$ with $\lognh\geq 23$.}.
For the two sources having best-fit \nhsym\ values in the CT regime from the baseline model, {\it cosmos330} and {\it ser261}, the latter has the lowest SNR \nustar-only spectrum in the sample, and it is not confirmed to be CT. Indeed the estimated lower limits on \nhsym\ for both torus models are in the heavily obscured range $\lognh= 23-24$. Therefore, the source could still be CT. The other two sources having upper limits in the CT regime from the baseline model, {\it ser285} and {\it ser235}, typically have upper limits in the heavily obscured range for both models,the exception being {\it ser235} for which {\sc MYTorus} gives an upper limit in the CT range.  
Figure~\ref{nh_base_torus} shows a comparison between column densities derived from both models. For the {\sc BNTorus} model   (empty circles) there is good agreement, although with a tendency to estimate systematically slightly lower \nhsym\ values\footnote{This could be due to the tendency of BNTorus to overestimate the low-energy reflection component for edge-on orientations as pointed out by \citet{LL2015}. This would require a lower column density to explain the observed absorption cut-off.}. Red dots report the best fit \nhsym\ values from {\sc MYTorus}, which also show good agreement. Given the mild disagreement regarding \nhsym\ for the two CT sources as estimated by the baseline model, we further modify the primary power-law in the baseline model with an approximated Compton-scattering term parametrized in XSPEC with the {\sc cabs} model. Both sources are still reported to be CT. This highlights the need to account for accurate Compton scattering treatment and geometry-dependent effects, as provided by the torus modelings, in order to get an accurate estimate of the column density for the most obscured AGN.

\section{Observed \nhsym\ distribution}\label{NHdistribution}
\begin{figure}[t!]
   \begin{center}
     \includegraphics[width=0.45\textwidth]{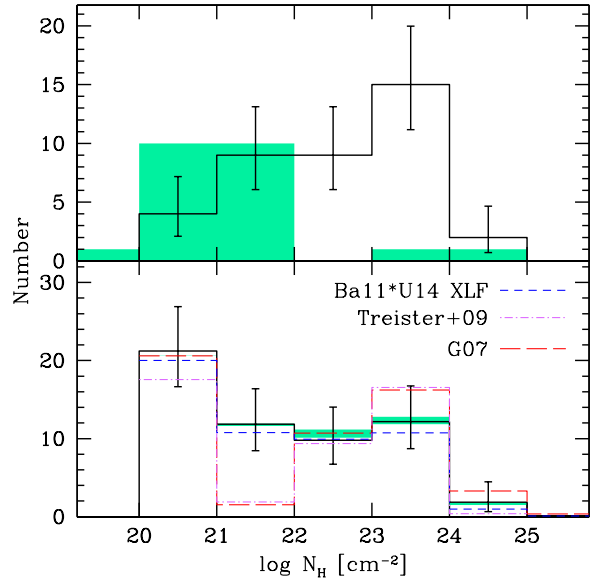}
   \end{center}
   \caption{Upper panel: column density distribution from the baseline spectral model (black histogram) with error bars calculated assuming low counting statistics. Shaded green histogram reports 90\% upper limits for data sets with limited constraining power (i.e., high-redshift, low \nhsym\ or sources with only \nustar\ spectra).
     Lower panel: Black line shows the distribution in which the \nhsym\ best-fit values and upper limits from the baseline model have been randomized as described in Section~\ref{NHdistribution}. In green is shown the range of \nhsym\ distribution by substituting the baseline-derived measurements for the sources with $\lognh\geq23$  with those derived by adopting the torus models.  Red (long dashed), blue (dashed) and magenta (dot dashed) histograms are model predictions from G07 and BA11 using the U14 XLF and \citet{TUV2009}.}  
   \label{nh_histo}
\end{figure}

The distribution of \nhsym\ derived from photoelectric absorption as evaluated by the baseline model (Section~\ref{broadband}) for 62 out of 63 sources is shown in the histograms reported in Fig.~\ref{nh_histo} (upper panel). The empty histogram reports sources for which \nhsym\ is constrained at least at the $1\sigma$ level with errorbars computed following Poisson counting statistics \citep[][]{G1986}. The filled green histogram reports sources with 90\% upper limits on \nhsym. For the latter, two sources have very high (i.e., poorly constrained) column density upper limits in the range $\lognh=23-25$ (one in the CT range). These are sources with \nustar-only X-ray data available and for which the missing coverage at lower X-ray energies limits our ability to measure lower column densities.
We find two sources ({\it cosmos330} and {\it ser261}) in the CT range. However the torus modeling finds a lower number of CT AGN.  Only {\it cosmos330} is formally considered as such by {\sc BNTorus}. {\sc MYTorus} finds it to be slightly below the the CT range. For  {\it ser261} both models place a lower limit in the heavily obscured regime ($\lognhsq=23-24$) for the low-quality spectrum of {\it ser261} (see Fig.~\ref{nh_base_torus}). Both models place an upper limit in the CT regime for only one source: {\it ser409} for {\sc BNTorus} (this source has \nhsym\  unconstrainted by the baseline modeling) and {\it ser235} for {\sc MYTorus}.

We build a histogram of the column-density distribution of the sample by folding in the error information in \nhsym\ best-fit values and the 90\% upper limits. In order to do this, in analogy with Section~\ref{reflection},  we performed 1000 random realizations of the sample. We assume symmetric Gaussian distributions in $\lognhonly$ with 1$\sigma$ standard deviations as the mean of the lower and upper errors. For sources with upper limits in $\lognhonly$ we assume a smooth uniform random distribution down to $\lognh=20$~\nh. We then averaged the resulting \nhsym\ distributions and obtained the randomized histogram reported in black in Fig.~\ref{nh_histo} (lower panel). Notice that this procedure, because of the upper limits, may lead to an overestimate of the $\lognh=20-21$ numbers due to the contributions from the sources with upper limits in the $\lognh=21-22$ bin (some may really belong to this bin but are averaged over all the $\lognhsq=20-22$ range). 
We also accounted for the \nhsym\ values obtained by the torus modeling. We substituted them for the sources with baseline-derived \nhsym\ measurements in excess of $10^{23}$~\nh. Fig.~\ref{nh_histo} shows the torus-derived \nhsym\ distribution (and its range) in green. There is very little difference with those derived with the baseline values. 

We compared the randomized \nhsym\ distributions with the prediction from theoretical models (dashed histograms in figure) by G07 (red long dashed), \citet{TUV2009} (magenta dot dashed) and BA11 (blue short dashed), updated with the X-ray Luminosity Function (XLF) of \citealt{U2014} (hereafter BA11+U14) folded with the sky coverage of our survey (Fig.~\ref{areas}) at fluxes above the threshold set for selecting the sample ($S_{8-24}\geq7\times10^{-14}$~\fluxcgs).
All models predict a  total number of sources which is slightly smaller compared those in the randomized histogram. However they are all roughly consistent within the Poissonian errors.  
Specifically we have above threshold a total of 57 sources, while G07, \citet{TUV2009} and BA11+U14 predict  52.7, 46 and 52.5, respectively. 
There is fair agreement within the uncertainties between the models and our randomized histogram for the obscured sources.  As for the unobscured sources ($\lognhsq\leq22$) the anomalous low value of the G07 model at $\lognh=21-22$, as reported by \citet{G2007} themselves,  seems to be due to the assumed XLF for type-1 AGNs \citep[from][]{H2005} which probably is contaminated at the level of 10-20\% by  mildly obscured ($\lognhsq=21-22$) sources and makes the transition to the $\lognh\ge22$ sources unrealistically steep. Correcting for a 10-20\% contamination alleviates the disagreement with our data at 1.5-2.1$\sigma$ level. Also \citet{TUV2009} predicts a very small number of sources at $\lognh=21-22$. This may probably be due to the fact that in this regime the host galaxy obscuration plays a non-negligible role. This further extra-nuclear absorption component is not accounted for in the model.  

\section{Intrinsic \nhsym\ distribution and fraction of CT AGN}\label{CTfrac}
\begin{figure}[t!]
   \begin{center}
\includegraphics[width=0.45\textwidth]{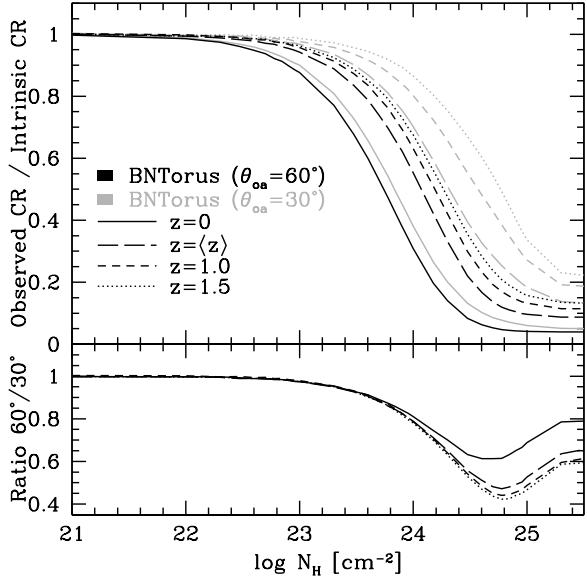}
   \end{center}
\caption{Upper panel: absorption bias in the 8-24~keV band assuming a {\sc BNTorus} model with $\Gamma=1.8$ for different redshifts as indicated. Black and light grey are for respectively $\theta_{oa}=60$~deg and 30~deg ({\sc BNTorus30} case). Lower panel: ratio between the 8-24~keV absorption bias between 60~deg and 30~deg.}  
   \label{absbias}
\end{figure}

Both models in Fig.~\ref{nh_histo} agree with our data and predict a very low number ($\sim$1-4) of CT AGNs in our hard selected sample. 
This is a consequence of the fact that the $8-24$~keV \nustar\ band is still biased against extremely obscured sources. This bias depends primarily on the redshift of the sources, on the width and high-energy bound of the range used for selecting the sources and on the value of the instrumental effective area at the energies where CT sources mostly emit. Fig.~\ref{absbias} highlights this for a {\sc BNTorus} model\footnote{The adoption of BNTorus (which is assumed representative for the toroidal models) instead of the baseline model is due to the fact that at high \nhsym\ it allows a more accurate estimate of the bias. At low \nhsym\ they both predict very negligible bias.} with $\Gamma=1.8$ by showing the ``absorption bias'' $B(\lognhonly,z)$, i.e. the ratio between observed and intrinsic 8-24~keV \nustar\ count-rates  as a function of the absorbing column density. The different black curves show this bias for the 8-24~keV band at redshifts 0, 0.58 (the median redshift of our sample), 1.0, 1.5 (respectively solid, long dashed, short dashed and dotted lines) for a torus with $\theta_{\rm oa}=60$~deg. The bias for a fixed \nhsym\ decreases with redshift.
In grey we show the bias for  $\theta_{\rm oa}=30$~deg (hereafter {\sc BNTorus30}). In the lower panel we report the ratio between the bias among the two opening angles. A smaller opening angle of the torus tends to give less bias. Given the fact that \nustar\ probes non-local sources up to $z\approx2-3$ we infer that \nustar\ has, on average, roughly the same absorption bias as {\it Swift}-BAT, at these flux levels.

For the unobscured sources the hard \nustar\ band is not biased at any redshift. The following considerations take as reference the absorption bias with $\theta_{\rm oa}=60$~deg. This will weight more the heavily absorbed sources when correcting for it. We later discuss possible changes in our estimates when using a bias given by the {\sc BNTorus30} case. 
At larger column densities the bias is more evident for increasing \nhsym. It becomes more pronounced for the CT sources (from 0.4-0.7 depending on redshift to less than 0.2). This means that while we detect all the unobscured sources at intrinsic fluxes down to $S_{8-24}\approx7\times10^{-14}$~\fluxcgs, for the CT sources we are sensitive to sources with intrinsic fluxes in the range $\sim1-3\times10^{-13}$~\fluxcgs, therefore missing a sizable fraction of the CT AGN population at fainter intrinsic (i.e., unabsorbed) fluxes. We can recover the missing AGN population by computing the intrinsic \nhsym\ distribution down to a certain intrinsic (i.e. unabsorbed) flux level common to all the sources regardless of their column density. Following \citet{Bu2011}, we integrate the source counts dN/dS derived in each 1~dex $\lognhonly$ bin (see Section~\ref{counts}) from a minimum ($S^{\rm obs}_{\rm min}$) to a maximum ($S^{\rm obs}_{\rm max}$) observed flux as follows:
$$  
\frac{dN}{d\lognhonly}=\int^{S_{\rm max}}_{S^{\rm obs}_{\rm min}=S^{\rm intr}_{\rm min}\times B(\log N_{\rm H},z)} \frac{dN}{dS}(\log N_{\rm H}) dS
$$
where $B(\log N_{\rm H},z)$  depends on \nhsym\ and on the mean redshift of the sources in each $\lognhonly$ bin and $S^{\rm intr}_{\rm min}$ is the minimum intrinsic (i.e. absorption-corrected) source flux at which the intrinsic distribution is estimated. 
In order to derive the source counts in each $\lognhonly$ accounting for uncertainties in \nhsym, we applied, through 10000 realizations of the sample, the same randomization procedure for error and upper/lower limits as explained for the observed \nhsym\ distribution Section~\ref{NHdistribution}. We model the derived  log$N$-log$S$ in each realization as a power-law with  slope $\alpha$ and normalization $K$  and adopted as representative for each bin their median values among all the realizations. These best-fit values are reported in Table~\ref{alpha_vs_Nh}.  The derived slopes are all consistent within the uncertainties with the 3/2 Euclidean value for a uniform non-evolving Universe. 
A small amount of realizations (at percent level) gave no sources in the CT bin hence in these cases no source count could be computed. Given their paucity we neglected these cases\footnote{We verified that our adopted median values are stable and not very sensitive to these outliers.}.
We derived the distributions for \nhsym\ estimated by (i) baseline-only modeling and (ii) baseline model for sources with $\lognh\le23$ and BNTorus model for more obscured sources. The latter case has been considered in order to include more accurate \nhsym\ estimates at the highest column densities (see Fig.~\ref{nh_base_torus}).
\begin{figure}[t!]
  \begin{center}
    \includegraphics[width=0.45\textwidth]{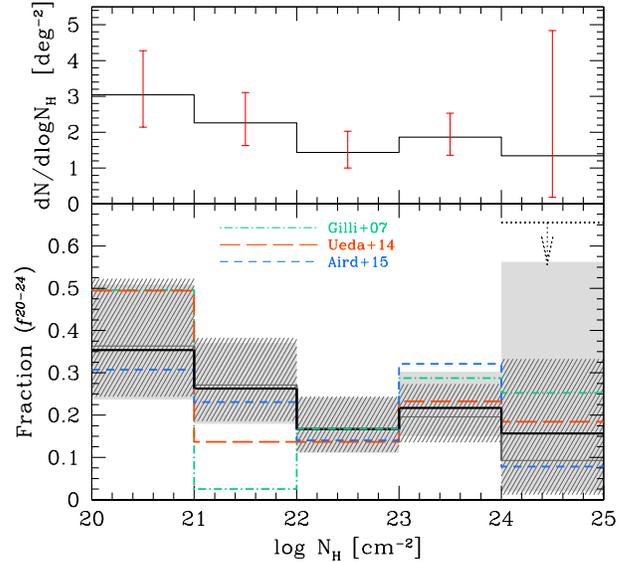}
   \end{center}
  \caption{Intrinsic \nhsym\ distributions (upper panel) and fraction  relative to the total number of objects in the intrinsic distribution (shaded grey, lower panel). The distributions presented here are for \nhsym\ derived by the baseline model except for the heavily obscured sources (i.e. $\lognhsq\ge 23$) for which the {\sc BNTorus} model estimates are adopted. Error-bars reflect the low counting statistics of the observed \nhsym\ distribution (Fig.~\ref{nh_histo}). The shaded grey and hatched dark grey regions represent the $1\sigma$ range of derived fractions assuming different absorption bias corrections with $\theta_{\rm oa}$ in the {\sc BNTorus} model of 60~deg and 30~deg, respectively (see Fig.~\ref{absbias}).  In the $\lognh=24-25$ bin we reported as dotted lines also the 90\% upper limits. Model predictions from G07, U14 and \citet{A2015b} are shown in dot-dashed green, long dashed orange and short dashed blue, respectively.
  }  
   \label{nh_histo_intrinsic}
\end{figure}
The value of $S^{\rm intr}_{\rm min}$ is chosen so that we are sensitive according to $B(\log N_{\rm H},z)$ to the same intrinsic flux to all the sources regardless of their \nhsym. This parameter is critical for deriving a CT fraction with reliable uncertainties. Indeed in the CT bin $B(\log N_{\rm H},z)$ may vary by a factor of 4-6 according to the exact value of \nhsym\ (see Fig.~\ref{absbias}). Given that we have only a couple of CT sources, one of which ({\it ser261}) has large errors and in the torus models is not even considered a bona fide CT, the choice of $S^{\rm intr}_{\rm min}$ is subject to large uncertainties. For determining it we therefore relied on the $\sim10$ sources which within their uncertainties have \nhsym\ compatible with $\lognh\approx24$. 
We verified, through 1000 realizations of the sample, that accounting for the flux uncertainties and adopting an absorption bias relative to the redshift of each source and its randomized \nhsym\ (we are excluding objects with upper limits whose absorption bias and therefore intrinsic flux cannot  be realiably estimated), a flux of $S^{\rm intr}_{\rm min}=10^{-13}$~\fluxcgs\ is adequate for our purposes.
In this way we estimate the intrinsic \nhsym\ distribution of the population of AGNs down to this intrinsic flux.
In the integration we are assuming, as representative for each bin, the value of the absorption bias relative to the central $\lognhonly$ value
at the mean redshift of each bin.  This sets in each bin a representative value of $S^{\rm obs}_{\rm min}$ to perform the integration of the source-counts.   This has non-negligible implications in the CT bin where the absorption bias, being strongly dependent on $\lognhonly$, makes the estimation of $S^{\rm obs}_{\rm min}$ highly uncertain. Given the paucity of possible CT AGN in our CT bin which possibly reflects the distribution of CT sources at ``intrinsic'' fluxes of $10^{-13}$~\fluxcgs, and given that they have column densities close to $\lognh=24$, we have decided to use for the CT bin an absorption bias relative to \nhsym$=1.5\times10^{24}$~\nh (i.e. the formal threshold for defining a source as CT).
The derived distribution is reported in the histograms of Fig.~\ref{nh_histo_intrinsic}, where the \nhsym\ estimates for the more obscured sources are from {\sc BNTorus}. Notice that the reported distributions for baseline-only \nhsym\ estimates are very similar providing only slightly higher estimates in the CT bin.
The error-bars reflect the low counting statistics on the number of sources whose unabsorbed flux is above $S^{\rm intr}_{\rm min}$ in each bin. In the upper panel  we report the intrinsic distribution using normalizations derived from the source counts fits to all the sources  in  1~dex $\lognhonly$ intervals (see Table~\ref{alpha_vs_Nh}) and assuming an Euclidean slope\footnote{The fraction of the CT sources (relative to all AGN population) obtained with the best-fit slopes is rather high and uncertain given its large uncertainties; see Table~\ref{alpha_vs_Nh}.}.
In the lower panel we report the fractional number of AGN in each bin relative to the total number of sources at $\lognh<24$ ($f^{20-24}$) from the intrinsic distribution. The shaded grey region is the $1\sigma$ range obtained through error propagation from the intrinsic \nhsym\ distribution uncertainties.  We report also in hatched dark grey regions the fractional distributions assuming an absorption bias derived by {\sc BNTorus30} (see grey lines in Fig.~\ref{absbias}) and fixing\footnote{The absorption bias in this case is smaller by a factor of about two compared to the adopted $\theta_{oa}=60$~deg, see Fig.~\ref{absbias} lower panel, however given the uncertainties in estimating the exact value of $S^{intr}_{min}$ due to the low number of sources, we prefer to assume the same value derived for the 60~deg case.}  $S^{\rm intr}_{\rm min}=10^{-13}$~\fluxcgs.
We obtain a fraction of CT sources $f^{20-24}_{\rm CT}=0.02-0.56$ (0.01-0.33 for {\sc BNTorus30}).
  The 90\% upper-limit is $f^{20-24}_{\rm CT}<0.66$ and is reported in figure as dotted line. Assuming \nhsym\ derived from baseline model only we obtain $f^{20-24}_{\rm CT}=0.08-0.66$.

\section{Fraction of absorbed sources}\label{fabs}
Clear trends in the fraction of absorbed AGNs ($f_{abs}$) compared to the whole population have been found with redshift and luminosity. Indeed several authors report 
a decrease in the absorbed AGN population with  source luminosity \citep[e.g., ][]{LE1982,S2003,LF2005,S2007,DC2008,Bu2011} and an increase with redshift at fixed X-ray luminosity \citep[e.g., ][]{LF2005,B2006,TU2006,V2014,U2014}.
Our spectral analysis effectively probes  rest-frame $\sim1.2-24$~keV for all the sources given the redshift distribution  from $z\gs0$ up to $z\approx1.5$. 
It can therefore account globally for the most relevant spectral complexity measurable in X-ray AGN spectra and allows an accurate determination of the absorbing column density.
For this reason we estimated the fraction of absorbed sources as a function of   X-ray luminosity and redshift. Given that at the highest absorptions (i.e. at CT column densities) our survey is biased, we chose to neglect these sources and estimate fraction of absorbed Compton Thin AGN.

\begin{figure}[t!]
   \begin{center}
\includegraphics[width=0.45\textwidth]{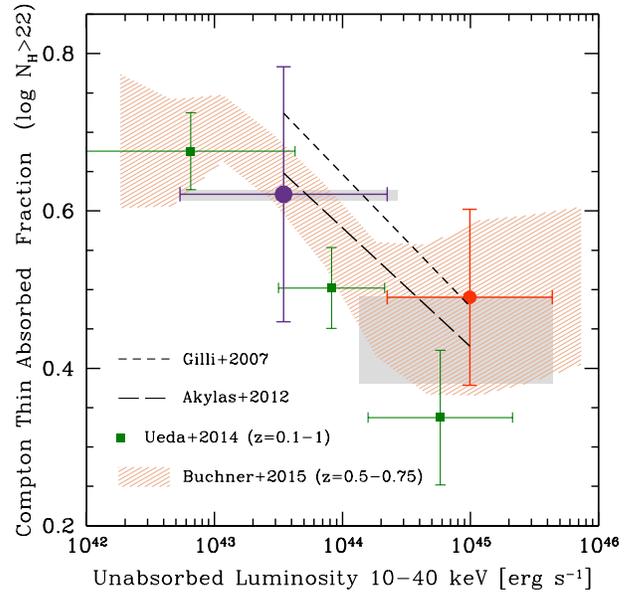}
   \end{center}
   \caption{Intrinsic absorbed fractions for Compton Thin sources with $\lognh>22$ as a function of unabsorbed 10-40~keV luminosity. The relative median redshift from the objects contributing to each bin is highlighted with a bluer-larger dot representing local lower luminosity sources and redder-smaller ones showing farther more luminous objects.
     Green data are fractions reported by \citet{U2014} for $z=0.1-1$ sources. Hatched orange region is the region determined by \citet{B2015} for $z=0.5-0.75$ AGN. 
         Short and long dashed lines represent the G07 and \citet{Ak2012} CXB synthesis model predictions, as indicated.
         Grey shaded regions relative to our two data points give an indication of the robustness of the absorbed fraction (see Section~\ref{fabslum} for details).    } 
   \label{obsabsfrac}
\end{figure}

Because of the selection of our sample we must correct our observed number of absorbed sources for the fact that AGN with a given intrinsic luminosity are progressively missed in the surveyed field for increasing column densities at larger distances. This translates to larger cosmic volumes sampled for unobscured sources compared to obscured ones. 
To account for this selection effect, for each source we computed the maximum surveyable comoving volume \citep[$V_{\rm max}$,][]{S1968,PC2000} accounting for its observed (i.e., absorbed) luminosity ($L_x^{\rm obs}$) and survey sky-coverage $\Omega(S)$ using the following formula:
$$
V_{\rm max}=\int^{z_{\rm max}}_{z_{\rm min}=0} \Omega(S(L_x^{\rm obs},z))\frac{dV}{dz} dz
$$
where $z_{\rm max}$ is chosen so that $S(L_x^{\rm obs},z_{\rm max}) = 7\times10^{-14}$~\fluxcgs\ in the 8-24~keV band.
The intrinsic fraction of obscured sources is therefore estimated as the ratio of the summed space densities of absorbed sources over the space densities of the total population:
$$
f_{N_{\rm H}}=\frac{\sum\limits^{N_{N_{\rm H}}}_{i=1} (V_{max}^i)^{-1}}{\sum\limits^{N_{\rm tot}}_{i=1}  (V_{\rm max}^i)^{-1}}
$$
where $N_{tot}$ and $N_{N_{\rm H}}$ are, respectively, the total number of sources and the number of sources characterized by a column density larger than a given \nhsym\ value. Notice that in our calculation we use parameters derived from the baseline modeling except for the heavily absorbed sources ($\lognhsq\ge23$) for which more accurate results from BNTorus have been incorporated.
The uncertainty on $F_{N_{\rm H}}$ is obtained by error propagation of the uncertainties on $(V_{\rm max})^{-1}$. The latter is usually estimated as $\sqrt{\sum_{i}(V_{\rm max}^i)^{-2}}$ \citep{M1985} by assuming Gaussian statistics in each bin. Since we are dealing with a relatively small number of sources per bin, this uncertainty estimate is not optimal in our case. We therefore  estimate uncertainties through bootstrap resampling\footnote{For each bin we performed 10000 random resamplings to derive the absorbed fractions and their standard deviation.} in each luminosity bin. Notice that in this case we are not correcting for the absorption bias as $\lognh=23-24$ sources show little bias (see Fig.~\ref{absbias}) and the small number of AGN in each bin is the dominant source of uncertainty. 

\subsection{Fraction as a function of luminosity}\label{fabslum}
Fig.~\ref{obsabsfrac} shows the fraction of absorbed sources with $\lognh\ge22$ ($f_{22}$) as a function of the unabsorbed 10-40~keV luminosity. We adopt as fiducial values those estimated from the bootstrap resampling procedure. We use the unabsorbed luminosity instead of the intrinsic coronal one in order to be consistent in comparing our results with those derived by previous works. Furthermore there is no comparable intrinsic luminosity quantity for the BNTorus-modelled heavily absorbed sources.  
In order to ensure good statistics and minimize the effects from single sources (i.e., outliers) we divide the sample in two bins, each containing a comparable number of objects. The size and color of each point gives an indication of the relative median redshift of the sources contributing to each bin, with the redder, smaller point sampling, on average, more distant sources. 

We see a hint of a decreasing trend of $f_{22}$ with luminosity. This dependence, is however not very significant being consistent within the uncertainties with no dependence with luminosity. Given the large range in redshift covered by our sample it is possible that the redshift evolution in the fractions act at the highest luminosities (where we have the more luminous sources) partially masking the luminosity dependence. The small number of sources prevent us from drawing firm conclusions on this point. In any case our estimated values are perfectly consistent with  population-synthesis model predictions not incorporating redshift evolution (G07 and \citealt{Ak2012}). Given that 80\% of the sources are at $z\ls1$  and that 90\% of the contribution to the high luminosity bin comes from sources at $z\approx0.3-1.1$ we can compare our results with recent determinations at similar redshifts. We find broad consistency with the estimated fractions for Compton Thin sources reported by \citet{U2014} and \citet{B2015} for $z=0.1-1$ and $z=0.5-0.75$, respectively.

To give an idea of the variance of our results on the adopted binning, we adjusted the bin width in order to include  up to 8 more sources (i.e. increasing the number of sources by $\sim$25\%) and reported the range of the corresponding variations as grey regions. We find very stable results on the fraction for low luminosity sources and a much broader range for high luminosity sources. Notice though that our nominal high luminosity fractions are at the upper end of this range. This is an indication that possibly the fractions at higher luminosities are somewhat smaller than estimated.

Our absorption fraction calculation has been obtained through the estimation of the space densities in the two bins. For the low luminosity bin including sources in the redshift range\footnote{We calculate this range as the interval in redshift for the sources contributing at 90\% of the total space densities.} $z\approx0.1-0.25$ we obtain total and absorbed space densities of $4.7\pm1.5\times10^{-5}$~Mpc$^{-1}$ and $2.9\pm1.3\times10^{-5}$~Mpc$^{-1}$. For the high luminosity bin including sources at $z\approx0.3-1$ we obtain  $1.8\pm0.2\times10^{-6}$~Mpc$^{-1}$ and $8.6\pm2.5\times10^{-7}$~Mpc$^{-1}$ for total and absorbed sources. It is difficult to compare these values to other results given the small number of sources we have and that our choice of luminosity intervals is driven by the need to maximize the statistics in each bin making also the redshift intervals equally poorly defined and luminosity dependent. In any case the total values are in fair agreement with the models from U14 and \citet{A2015}.

\subsection{Fraction as a function of redshift}\label{fabsz}
\begin{figure}[t!]
   \begin{center}
\includegraphics[width=0.45\textwidth]{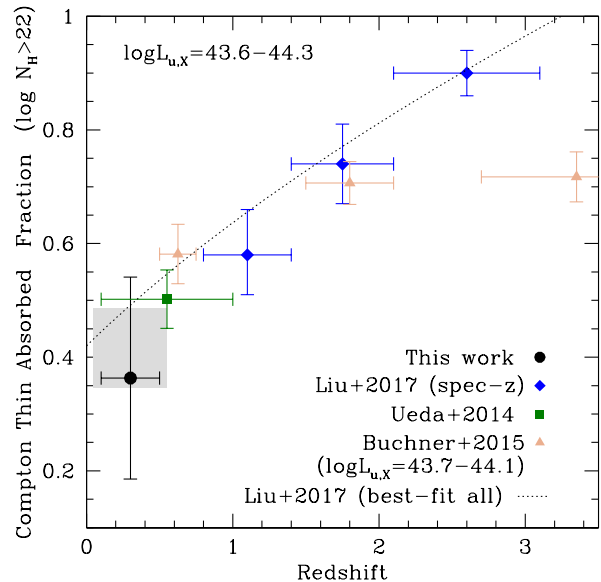}
   \end{center}
   \caption{Compton thin absorbed fractions as a function of redshift. For $\log(L_{u,X}/\lumcgsin)\approx43.6-44.3$ (corresponding to the 2-10~keV $\log(L_{\rm 2-10}/\lumcgsin)\approx43.5-44.2$, assuming for the conversion an unabsorbed power-law with $\Gamma=1.8$). Black is the absorbed fraction estimated from our sample. Blue, green and orange are absorbed fractions obtained by the spectroscopic sample in \citet{Liu2017}, U14 and \citet{B2015}, respectively (the latter reported at $\log(L_{u,X}/\lumcgsin)\approx43.7-44.1$). The grey shaded region is relative to our data point and gives an indication of the robustness of the estimated absorbed fraction (see Section~\ref{fabsz} for details).  Dotted line is the best-fit model reported by \citet{Liu2017} including in the absorbed fractions also sources with photometric redshift estimates.}  
   \label{absfracz}
\end{figure}

Recently \citet{Liu2017} performed a spectral analysis of the brightest AGN in the 7~Ms \cdfs\ and investigated $f_{22}$ as a function of redshift. They divided their sample in redshift bins in the range $z=0.8-3.5$ at fixed 2-10~keV luminosities $\log(L_{\rm 2-10}/\lumcgsin)=43.5-44.2$. They evaluated that these objects are not biased by absorption up to $\lognh=24$. In order to compare with their estimates we measured the absorption fraction in the same luminosity range for sources in the redshift interval z=0.1-0.5. According to their criterion this sub-sample is not biased for $\lognh<24$\footnote{They evaluated their completeness down to a certain column density by assuming a power-law with $\Gamma=1.8$ modified by photoelectric absorption with and empirically self-absorbed ($\lognh=23$) reflection component parametrized by {\sc pexrav} with $R=0.5$ and $E_c=300$~keV and a scattered fraction set to 1.7\% of the primary continuum level.}. We obtain an absorption fraction of $f_{22}=0.30\pm0.17$. 
  In Fig.~\ref{absfracz} we compare our value with those measured by \citet{Liu2017} for the sub-sample of sources with spectroscopic redshift determinations. We report also the best-fit relative to their whole sample. Our low value nicely follows the monotonic trend with redshift reported by \citet{Liu2017} at higher redshifts. Notice that they had in each bin only 12 sources. In any case their estimates broadly agree by those inferred by the entire sample (which includes sources with photometric redshifts) for which they have bins populated by 17-26 sources.  Given that our sub-sample consists of only 11 sources, we decided to explore the robustness of our measure by including neighbouring sources (enlarging the sub-sample to up to 16 sources) by modifying the luminosity and redshift boundaries by $\pm0.15$ dex and $\pm0.05$, respectively. The range of possible variation of $f_{22}$ is reported in Fig.~\ref{absfracz} by the shaded grey area. We also report relatively local (i.e. $z<1$) fractions by U14 and \citet{B2015}. Their fractions measured at, on average, higher redshift ranges, lie at slightly higher $f_{22}$ than our point.

\section{Source counts as a function of \nhsym}\label{counts}
  Given that population-synthesis models reproduce the CXB as a mixture of AGN with different column densities we can analyze the source counts as a function of \nhsym\ and compare them with model predictions. The latter starts to noticeably differentiate at higher column densities, with the largest difference in the CT regime.

  Using the $8-24$~keV sensitivity curves reported in Fig.~\ref{areas} we first produced the total source counts\footnote{We used equations 5 and 6 in \citet{C2007} for estimating cumulative number counts and relative uncertainties.} for our sample.
  We model them as a simple power-law $K\, S^{-\alpha}$ given that our sources, having the highest hard-band fluxes in the \nustar-Extragalactic Survey program (H16), still probe fluxes well above  the break at $\rm log (S_{8-24}/erg\,s^{-1}\,cm^{-2})\simeq-14$  predicted by background synthesis models (e.g., G07, BA11). 
  We employed a maximum likelihood estimator \citep[][]{C1970} to obtain, through a fit to the unbinned differential counts, the best-fit value of the slope $\alpha$ of the integral counts. The normalization of the power-law is fixed by the total number of estimated sources which should be matched by the best-fit model at the catalog lowest flux. 
  The best-fit power-law  has a slope value of $\alpha=1.36\pm0.28$ which is flatter but still consistent with the Euclidean 3/2 value for the integral distribution and a normalization of $10.1\pm1.3\,\rm deg^{-2}$ at $10^{-13}\, \cgs$ (where the error is Poissonian from the total number of sources employed in the source counts).

  \begin{figure}[t]
   \begin{center}
     \includegraphics[width=0.49\textwidth]{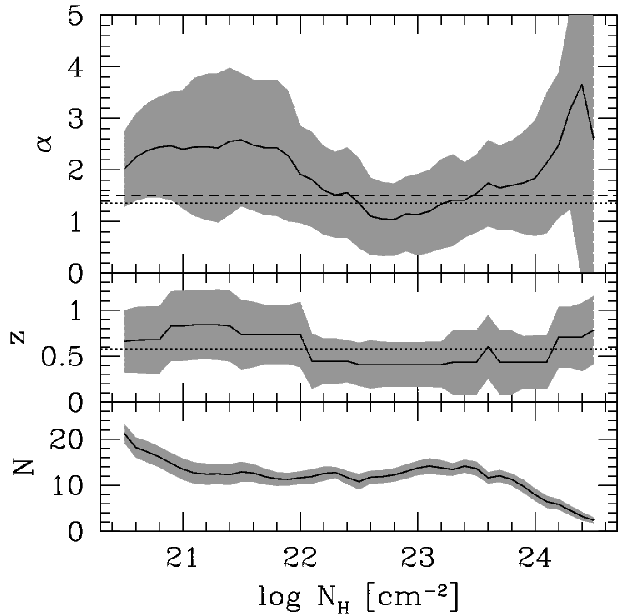}
   \end{center}
   \caption{Upper panel: best-fit power-law slopes to the logN-logS produced in 0.1 dex $\lognhonly$ steps within 1 dex $\lognhonly$ bins. The slope and the relative uncertainty (shaded grey region) are the median value of the distribution of the best-fit $\alpha$ from 1000 realizations of the sample obtained by randomizing $\lognhonly$ according to their errors and upper limits. Dotted and dashed lines are the best-fit value fo the logN-logS of the whole sample and the Euclidean value.  Middle panel: is the median redshift and relative interquartile range (shaded grey region) values obtained from the 1000 random realizations of the sample. Dotted line represents the median redshift value. Lower panel: mean value of the number of the sources considered with relative dispersion (shaded grey region) obtained from the random realizations.} 
   \label{lognlogs_vs_nh}
\end{figure}

  We  analyzed the variation of the log$N$-log$S$ slope as a function of column density. 
In order to do this we performed a scan in \nhsym\ with step 0.1 dex in $\lognhonly$. For each value of column density we estimate the value of the log$N$-log$S$ slope in an interval of 1 dex centered in it.
We performed the usual 1000 realizations of our sample randomizing \nhsym\ according to their errors and upper limit values. From each of these realizations we produce a log$N$-log$S$ and obtain a best-fit value of $\alpha$ in each $\lognhonly$ bin. 
We construct a distribution of $\alpha$ and its uncertainty and assign to the bin their median values. In Fig.~\ref{lognlogs_vs_nh} (upper panel) we report $\alpha$ as a function of \nhsym. The average number of objects included in each log$N$-log$S$ interval, which is always larger than 10 for $\lognh\ls 24$, is plotted in the lower panel (along with its $1\sigma$ dispersion).  Although the values calculated in adjacent steps are correlated, this plot illustrates the robustness of the best-fit $\alpha$ value and its dependence on outlier objects. On scales larger than $\sim1$~dex in $\lognhonly$ we have an indication of  uncorrelated variations of the slope as a function of \nhsym.  Throughout the range there is always consistency within the large errors with an Euclidean slope value.
Table~\ref{alpha_vs_Nh} reports values for $\alpha$ and normalization of the best-fit power-law in several $\lognhonly$ intervals  from $\lognh=20$ to $\lognh=25$. Notice that fitting in separate column density bins and randomizing $\lognhonly$ within the uncertainties results in the sum of the normalizations from all the bins, $10.6\pm1.4 \,\rm deg^{-2}$, 
which is slightly higher than the normalization obtained from the fit to the whole dataset but is consistent  within the  uncertainties.

\begin{table}
  \begin{center}
    \caption{Best-fit values of the slope and normalization of a power-law model to the source counts as a function of column density. Values are presented for baseline-only modelling and baseline + {\sc BNTorus} (the latter used for only $\lognh\ge23$ sources).
      \label{alpha_vs_Nh}
    }
  \begin{tabular}{ccc}
    \hline
    \hline
    $\Delta \log N_{\rm H}$    &    $\alpha$     &    $K$ [$\rm deg^{-2}$] \\
    ($\rm cm^{-2}$)           &                &  at $10^{-13} \cgs $   \\
    \hline
    \multicolumn{3}{c}{Baseline} \\
    \hline
    20-21                  & $2.02 \pm 0.73$ &    $3.93\pm0.85$ \\
    20-22                  & $2.22 \pm 0.61$ &    $5.88\pm1.00$ \\    
    21-22                  & $2.63 \pm 1.30$ &    $2.00\pm0.56$\\
    22-23                  & $1.33 \pm 0.86$ &    $1.94\pm0.59$\\
    23-24                  & $1.50 \pm 0.74$ &    $2.25\pm0.60$\\
    24-25                  & $2.62 \pm 3.96$ &    $0.49\pm0.36$\\
    \hline    
    \hline
    \multicolumn{3}{c}{Baseline+BNTorus} \\
    \hline
    20-21                  & $ 2.04\pm0.74$ &    $ 3.91\pm0.85$ \\
    20-22                  & $ 2.24\pm0.62$ &    $ 5.83\pm0.99$ \\    
    21-22                  & $ 2.68\pm1.30$ &    $ 1.99\pm0.55$\\
    22-23                  & $ 1.28\pm0.77$ &    $ 2.31\pm0.65$\\
    23-24                  & $ 1.63\pm0.83$ &    $ 2.18\pm0.61$\\
    24-25                  & $ 2.19\pm4.33$ &    $ 0.62\pm0.46$\\
    \hline    
  \end{tabular}
  \end{center}
  \end{table}
We notice that, with the exception of the CT sources,  unobscured AGN have, on average, higher best-fit slope values ($\alpha\approx2-2.4$) than obscured objects ($\alpha\approx1.3-1.5$). In the middle panel of Fig.~\ref{lognlogs_vs_nh} we report the distribution of the median redshift values along with their interquartile range. Both $\alpha$ and the median redshift appear to be  correlated, showing similar variations as a function of \nhsym. Indeed, they have a Spearman correlation coefficient of $\sim$0.9. As expected, the median redshift of unobscured sources  is larger than the sample median, meaning they comprise a large fraction of high-redshift sources.
We investigated this trend by progressively removing high-redshift sources. To preserve a sufficiently high number of sources ($>10$) for each \nhsym\ interval, we only investigated  unobscured ($\lognhsq=20-22$) and obscured ($\lognhsq=22-24$) bins. 
For the unobscured sources, we find that $\alpha$ progressively decreases to values around $\sim1.4$ when we only consider sources at $z\ls 0.4$. For the obscured sources we notice variation around the Euclidean value across the redshift range within $\sim \pm0.5$, by $z=0.4$. Despite models predicting that at the fluxes probed by \nustar\  log$N$-log$S$ progressively flattens by going down to lower redshifts this effect is mild and not as strong as we are finding for the unobscured sources. Indeed the G07 model predicts in the flux range $S_{8-24}=5\times10^{-14}-10^{-12}$~\fluxcgs\ the slope of the source counts  flattens from $z=1.1$ to $z=0.4$, with values going from 1.6 to 1.2. 

\begin{figure}[t]
   \begin{center}
     \includegraphics[width=0.45\textwidth]{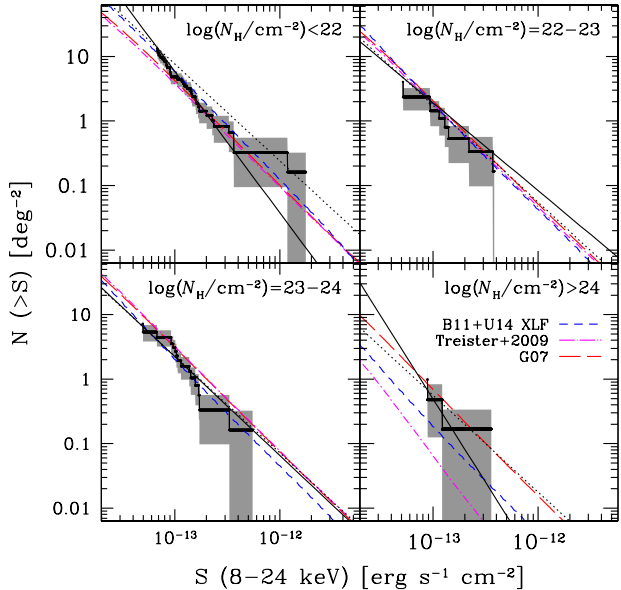}
   \end{center}
   \caption{Unbinned source counts for the baseline model case for different $\lognhonly$ bins. For display purposes we show the counts obtained by including the sources with column density upper limits in the bin relative to their upper limit value. Solid line show the best-fit power-law models reported in Table~\ref{alpha_vs_Nh} and obtained by accounting for the column density uncertainties as described in the Section~\ref{counts}. Dotted line is the Euclidean power-law model whose normalization at the lowest flux equals that estimated for the best-fit model. Red long dashed, blue short dashed and magenta dot dashed lines are predictions from G07, BA11 and \citet{TUV2009} models, respectively.} 
   \label{lognlogs_vs_models}
\end{figure}

Fig.~\ref{lognlogs_vs_models} reports log$N$-log$S$ for unobscured ($\lognhsq<22$), mildly obscured ($\lognhsq=22-23$), heavily obscured ($\lognhsq=23-24$) and CT sources ($\lognhsq>24$). The best-fit power-law models (accounting for \nhsym\ uncertainties) for each are reported (see also Table~\ref{alpha_vs_Nh}) as solid lines. The dotted line represents the Euclidean power-law normalized so that it correctly predicts the total number of objects. Red long dashed, blue short dashed and magenta dot dashed lines show the predictions from G07, BA11 with U14 XLF and \citet{TUV2009} models, respectively. The discriminating power is stronger for the CT regime. However in this range we have only three objects, one of which has an upper limit and a minor contribution (during our randomization process) from less obscured objects with estimated \nhsym\ close to the $\lognh=24$. The uncertainties are very large (see Table~\ref{alpha_vs_Nh}). The models predictions are all roughly consistent with the source counts within the uncertainties, hence we cannot discriminate between models and draw firm conclusions.

\section{Discussion}\label{discussion}

\subsection{Sifting through candidate CT sources}
           Fig.~\ref{nh_histo} displays, regardless of the spectral modeling adopted, a high fraction ($\sim25$\%) of heavily absorbed ($\lognhsq \ge23$). Among these sources we report one source from the \nustar-\cosmos\ field ({\it cosmos330}) which is a bona fide CT and  one source  ({\it ser261}) from the Serendipitous fields which has an estimate compatible with being CT (i.e. either CT in the baseline model or 90\% lower-limit in the range $\lognhsq\approx23.7-23.8$ from the toroidal modelings). Accounting for the uncertainties we have an equivalent number of 1.5-2 CT which within the poissonian uncertainties is fully consistent with the $\sim$0.4-3.3 CT  sources ($\lognhsq=24-25$) predicted to be observed in our survey by CXB population synthesis models (Fig.~\ref{nh_histo}, bottom panel).

             \subsubsection{Claimed CT candidates in the \cosmos\ field and their impact on the observed CT budget}\label{CTcosmos}
Nonetheless in our sample we have sources from the \cosmos\ field classified as heavily absorbed which are in common with works focusing on \chandra\ \citep{B2014} and \xmm\ \citep{L2015} spectral analysis and which where claimed to be CT: {\it cosmos181} and {\it cosmos216}. The spectra and the best fit models for these sources are reported in Fig.~\ref{spectrasample}.  This disagreement is not unusual, several works using low-energy data have so far attempted the spectral identification of distant CT sources in the medium/deep survey fields \citep[e.g.][]{T2006,GGA2007,G2009,G2013,C2011,F2011,BU2012,BGN2014,B2014,L2015} 
           and given the range of possible spectral shapes for CT sources and the limited counting statistics (from few tens to $\sim100$ counts) for these faint sources especially at the highest energies $\gs5-6$~keV where the effective area starts to drop, they have always struggled to consistently identify CT sources \citep[see, e.g., ][]{CM2013}. Moreover, their analysis, at least for sources with redshifts $\ll2$ (i.e., the great majority of the potential contributors to the CXB according to population-synthesis models), have been  limited by  sampling of the restricted lower energy  part of the reflection component and of the heavily obscured primary emission. This may potentially add possibly non negligible systematic uncertainties to the statistical ones. Interestingly Marchesi et al. (2018, submitted) analyzing 30 local candidate CT AGN selected from the {\it Swift/BAT} 100-month survey, found that the addition of \nustar\ data allowed a re-classification of $\sim40\%$ of the sources as Compton thin. Source variability may furthermore  play a role since these surveys have gathered data over time scales of years \citep[e.g., ][]{L2014}.

           It is therefore worth to further investigate these sources in order to understand if they can possibly substantially change the CT budget in our investigation. Following  \citet{L2015} and  \citet{B2014}, we applied their same toroidal modeling\footnote{ They both used an edge-on {\sc BNTorus} model. \citet{L2015} adopted $\Gamma=1.9$ with a scattered power-law component with same photon index. \citet{B2014} adopted  $\Gamma=1.7$ and torus semi-opening angle of $30^\circ$ with negligible scattered component.}  to our joint \xmm/\chandra-\nustar\ datasets. The only source which resulted to have a different column density classification is {\it cosmos181}, for which we obtained  $\lognh=24.01_{-0.06}^{+0.05}$ but only with the \citet{L2015}  modeling. For the other modeling and for the source {\it cosmos216} the agreement between the measured column density is well within the $1\sigma$ ucertainties.
Therefore only {\it cosmos181} would nominally and potentially change its classification to CT due to a small increase in the estimated column density by a little more than 0.1~dex. However its addition to the CT bin do not appreciably change the column density distribution (Fig.~\ref{nh_histo}, bottom panel) as the equivalent number of CT sources would raise from $\sim2$ to $\sim2.3$.

\subsubsection{CT candidates in the Serendipitous catalog}
In the \ser\ catalog we are finding three candidates CT sources: {\it ser261}, {\it ser235} and {\it 409}. None of these   have archival low-energy data from  \chandra\ or \xmm. 

The heavily absorbed nature of  {\it ser261}  was  established already by a simple power-law fit which returns a very flat value of $\Gamma=0.68^{+0.28}_{-1.08}$.
The baseline model finds {\it ser261} to be CT. The torus models return a lower \nhsym\ estimate, with lower limits in the heavily absorbed range ($\lognhsq>23.8$ for {\sc BNTorus} and $\lognhsq>23.7$  for {\sc MYTorus}), i.e. they cannot confirm or reject the CT classification. Notice that from an SDSS optical spectrum taken back in 2002 the source counterpart shows evidence of broad-lines hinting to a Type~1 nature. The source X-ray spectrum is the lowest quality in our sample, with just 24 \nustar\ net-counts. 
Clearly a much better \nustar\ spectrum, low-energy X-ray data and newer optical spectra are needed to better understand the true nature of this source. 

As for the border-line source {\it ser235},  we mention that both {\it baseline} and {\sc MYTorus} models report upper limits in the CT regime. A power-law-only model shows a more canonical slope value with $\Gamma=1.65^{+0.19}_{-0.10}$. Its border-line nature is likely due  to the lack of low-energy data and its high-redshift, z=2.1, which makes  the \nhsym\ measure with \nustar-only data more uncertain. 
For the source {\it 409}, the {\sc BNTorus} model  is the only one reporting an estimated column density upper limit in the CT regime (the baseline model cannot constrain its value at all).  A simple power-law model returns $\Gamma=1.21^{+0.53}_{-0.27}$, which although a little flat is still consistent at $\sim1\sigma$ with the canonical value for an unabsorbed source.

\subsection{Characterizing the major contributors of the residual missing CXB flux}
\subsubsection{On the heavily absorbed populations}
Thanks to the high sensitivity of \nustar\ at high energies we have investigated with our flux-limited sample the numerical predominance of the heavily absorbed AGN populations and the prominence in their spectrum the reflection components. Indeed these have been identified as the main actors in reproducing the residual unaccounted 20-30~keV CXB flux \citep{Ak2012}. 
  The limited sample size (63 sources) coupled with the large redshift range ($z\approx0-1.5$) do not allow to obtain  stringent constraints. However we have obtained indicative estimates of these quantities from a more robust source-by-source broad-band (0.5-24~keV) modeling which is less prone to systematics
  and large statistical uncertainties\footnote{Notice that in our 3-24~keV {\sc BNTorus} joint spectral analysis for the $\lognh=23.5-24$ sources, the $\lognhonly$ is derived with an accuracy a factor of 2 higher than that derived for sources selected at the same column densities and comparable fluxes in the soft X-ray studies by \citet{L2015} and \citet{B2014}.}
We have estimated an intrinsic fraction of AGN as a function of \nhsym. Despite the low number of CT sources we were able to extract a fraction relative to the whole Compton Thin AGN population (i.e., $\lognhsq=20-24$) which is formally in the range $f^{20-24}_{\rm CT}=0.02-0.56$ with an upper limit of $f^{20-24}_{\rm CT}<0.66$. There are many assumptions affecting this value and its uncertainty to different extents: (i) the small number statistics (function of the small sample size and of the relatively high unabsorbed flux at which the distribution of the AGN is unbiased), (ii) the assumption of the most representative absorption bias value in the CT bin (which strongly varies within this interval), (iii) the particular model used to infer the absorption bias (we explored the {\sc BNTorus} model with two different choices of the opening angle)  and (iv) an accurate spectral modeling for the CT sources in order to obtain accurate source counts needed to extrapolate at the limited flux the unbiased contribution of the source in the CT bin. Many of these aspects required assumptions to be made and we tried to make the most reasonable ones. This CT fraction is broadly representative for sources at redshift $z\sim0.2-1.0$ and luminosities $\log(L_{u,X}/\lumcgsin)\approx43.4-44.6$, the intervals being the 15.9\% and 84.2\% percentiles\footnote{The range delimited by these values corresponds to the inclusion of $\sim68$\% of the sources assuming Gaussian distribution.} ranges of the sub-sample of sources brighter than $S^{intr}_{smin}$  (see Sect.~\ref{CTfrac}) and with constrained \nhsym\ value. Given the large number of $\lognh=23-24$ sources and the less pronounced bias in this bin, we obtain much better and stable constraints for these sources. Observational constraints for the local population tend to favour a larger fraction for these sources among the absorbed Compton Thin AGN for sources with comparable luminosity \citep{R2015}. Comparable but stronger conclusions have been drawn by \citet{Liu2017} at higher redshifts ($z=1.6-2.4$) for roughly similarly luminous sources (they includes less luminous, by a factor of $\sim2$, quasars). We cannot draw comparably strong conclusions on this point, we find indications for a more numerous population of $\lognh=23-24$ compared  to the $\lognh=22-23$ one but within the uncertainties it is consistent with a constant value. Furthermore we are not able to test and disfavour models at a confidence level higher than 90\%.
As for the absorption fraction ($f_{22}$) for Compton Thin sources as a function of unabsorbed luminosity, our estimated fractions (computed in two bins) do not significantly imply a decreasing trend although they are in good agreement with those derived by much larger soft X-ray analysis on sources at similar redshifts \citep[i.e. around or within $z\approx0.1-1$; U14, ][]{B2015}.
A general increasing trend of $f_{22}$ with redshift has also been measured in deep X-ray spectroscopic investigations of the \cosmos\ and \cdfs\ fields \citep{B2015,Liu2017} and large complete samples (U14). It is reported in Fig.~\ref{absfracz} for sources at $\log(L_{u,X}/\lumcgsin)\approx43.5-44.62$. We have estimated a fraction at $z=0.1-0.5$ for sources of similar luminosity obtaining a generic agreement with extrapolations at lower redshift of the trend reported by \citet{Liu2017} and a good agreement with values estimated at $z<1$ by U14 and \citet{B2015}.
In order to better constrain and independently evaluate the luminosity and redshift dependence of $f_{abs}$  a much larger sample (at least twice the size of the present sample) and better quality data with good low energy X-ray coverage are necessary.

\subsubsection{The importance of the reflection component}
In our spectroscopic analysis we quantify also the reflection strength in each source. It is therefore interesting to compare our results to the typical assumptions made by in CXB population synthesis models.  Indeed they generally implement relatively similar assumptions. The reflection is always assumed to have a constant value $R\approx1$ within each population \citep[e.g. ][]{B2006,TUV2009,Ak2012,U2014} or possibly a function of the degree of obscuration \citep[e.g., G07;][]{U2014,EW2016}  with no dispersion around a mean value. On average we find median reflection values which are: (1) significantly lower ($R\approx0.3-0.7$) than those assumed ($R\approx1$) by CXB models and (2) exhibit a rather broad distribution with a median value relative to the whole sample of $\sim0.4$ (Table~\ref{refltable}). Furthermore we measure a significant anti-correlation with unabsorbed and intrinsic luminosities (the latter being more pronounced, see Fig.~\ref{R_vs_NH}, lower panels and Table~\ref{refltable}). This trend is further confirmed by the findings of our companion paper on stacked \nustar\ spectra (DM17). In this context we are finding sources with lower unabsorbed luminosities to have a median reflection ($\langle R\rangle=0.73$) a factor of two stronger than more luminous ones ($\langle R\rangle=0.31$). The broad $R$ distribution reaches 50\% percentile values a factor of two larger. When using intrinsic coronal luminosities the differences exhacerbates further by a factor of about two. A similar trend has been largely ignored by models the one exception being G07 model for which QSOs have been assumed to have no reflection. G07 assumes also higher $R$ for Type-1 sources ($R=1.3$) compared to Type-2s ($R=0.88$) in order to mimick a orientation-dependent disk reflection. \citet{U2014} instead assumes a flat $R=0.5$ from the disk and a torus-based contribution in the context of a luminosity and redshift dependend unified scenario in order to reproduce a total $R=1$ for Seyfert galaxies.

\citet{A2015} presenting the first direct measurements of the 10-40~keV XLF derived from the \nustar\ extragalactic survey program pointed out a degeneracy in the models parameters (the distribution of \nhsym\ as a function of luminosity and $z$ for the most obscured AGN and $R$) in order to reproduce the XLF. In particular they show that the high energy XLF can be alternatively modelled by either a distribution of \nhsym\ derived by \citet{A2015b} (see Fig.~\ref{nh_histo_intrinsic}) and a spectral model with a uniform distribution of reflection strength in the range $0<R<2$ (i.e. $\langle R \rangle=1$) or a distribution of  \nhsym\ derived by \citet{U2014} (Fig.~\ref{nh_histo_intrinsic}) and a fixed $R=2$ at all luminosities. Our analysis do not allow to conclusively break this degeneracy as the fraction of CT AGN is poorly constrained. Nonetheless the \nustar-derived $\langle R \rangle$ values as measured in this work and in DM17 firmly exclude the high and fixed values of $R$ invoked in order to bring the \citet{U2014} model into full consistency with the 10-40~keV XLF. 

We find tentative evidence of an anticorrelation of $R$ with \nhsym, whereby more absorbed sources exhibit lower reflections. This is in qualitative in agreement with a disk reflection scenario although our median $R$ values are smaller than those assumed by G07 (Table~\ref{refltable}). This result however may also be indicative of a reflection component absorbed by the same medium obscuring the primary continuum in the hypothesis of no \nhsym-dependence. With our small sample unfortunately we cannot test at high significance this hypothesis. Furthermore we mention that local studies find mildly obscured Seyferts to have stronger or equally strong reflections than unobscured counterparts \citep{R2011,EW2016}. We have insufficient statistic to infer such similar trend at comparable low luminosities (i.e. $\log[L_{u,X}/\lumcgsin]<44$).

\section{Conclusions}\label{conclusions}
We focused on a sample of 63 bright 8-24~keV-selected AGN with $S(8-24)\geq7\times10^{-14}$~\fluxcgs, from the multi-tiered \nustar\ Extragalactic Survey fields. The sample spans a redshift range of $z=0-2.1$, with a median value $\langle z \rangle=0.58$. For the great majority of the sources (58) we performed spectral modeling in the broad 0.5-24~keV band by using archival low-energy spectra from \chandra and \xmm.  For five sources, selected from the Serendipitous fields, low-energy data are not available. We used both empirically and physically motivated models, where the latter assumed Monte Carlo  implementations of toroidal geometries.
The results of the broad band  spectral analysis  can be summarised as follow:
\begin{itemize}
\item About 25\% of the sample is comprised of heavily absorbed sources with $\lognh\geq23$ (see Fig.~\ref{g_vs_nh}). 
\item Depending on the details of the adopted modeling, the number of bona fide CT sources is 1-2 (Fig.~\ref{nh_histo}).
\item For the empirically motivated model we computed unabsorbed 10-40~keV and intrinsic coronal 10-40~keV luminosities (i.e., removing the reflection contribution from the unabsorbed luminosities) and found that the former can lead to a luminosity-dependent overestimation of the latter which is highest, a factor $\sim3-4$, at the lowest luminosities (i.e., $<10^{44}$~\lumcgs; see Section~\ref{lum1040} and Fig.~\ref{lumcfr}). 
\item The median reflection strength of the sample is $\langle R\rangle=$0.43, with a large scatter (interquartile range  0.06-1.50).
We find that $R$ significantly anti-correlates with unabsorbed 10-40~keV luminosity (in agreement with DM17) and intrinsic 10-40~keV AGN luminosity (see Table~\ref{refltable} and Fig.~\ref{R_vs_NH}).
\item The observed \nhsym\ distribution for the absorbed ($\lognhsq\geq22$) sources is in agreement with CXB population-synthesis model predictions (see Fig.~\ref{nh_histo}). The agreement persists when accounting for different spectral modelings or possibly misclassified AGN claimed as CT by previous soft X-ray studies (Section~\ref{CTcosmos}).
\item From the absorption-corrected number fraction we obtain a CT fraction broadly representative for $z\approx0.2-1.0$ and $\log(L_{u,X}/\lumcgsin)\approx43.4-44.6$ AGN. The estimated fraction, relative to the AGN population at $\lognh<24$, is $f^{20-24}_{\rm CT}=0.02-0.56$ ($<$0.66 at 90\% c.l.). This value drops by a factor of $\sim$1.7 if we assume a toroidal obscurer with a halved opening angle (Fig.~\ref{nh_histo_intrinsic}). 
\item We estimated the intrinsic fraction of obscured Compton Thin ($\lognhsq\ge22$) sources as a function of unabsorbed luminosity at 10-40~keV (Fig.~\ref{obsabsfrac}). The derived fractions cannot constrain a trend with luminosity however they are in good agreement with results reported by other authors at similar epochs (i.e. $z\ls1.0$). We further calculate the absorption fraction for sources  at $z=0.1-0.5$ and with $\log(L_{u,X}/\lumcgsin)\approx43.6-44.3$. The resulting $f_{22}=0.36\pm0.18$ agrees with extrapolated decreasing trends (from high to low z) from surveys covering the same luminosity range and higher redshifts ($z\approx3$) and with reported values at similar redshift range.
\end{itemize}

Clearly increasing the number of sources to spectroscopically study at
these flux levels and with good quality low-energy data would help to 
better characterize the \nustar\ hard-band selected AGN population at
moderate redshift ($z\approx0.5-1$). This will help in further
elucidating the hinted correlation between reflection strength and
column density and start a robust investigation of possible redshift
dependence of the absorbed fractions.

The main benefit of enlarging the sample size will be the increase in
the number of robust CT sources identified at redshifts and 
luminosities poorly probed currently. This will allow: 1)
a more robust and constrained
estimate of the CT fraction; 2) the discrimination of CXB population
synthesis models through the comparison of more accurate source counts
in the $\lognh=24-25$ range of column densities; 3) to start probing 
the $\lognh=25-26$ bin which is precluded in our analysis due to a
combination of strong absorption bias and small number statistics.  The
Serendipitous Survey will progressively increase its sky coverage, 
providing crucial help in this regard, especially if backed-up by
ancillary spectroscopic low-energy X-ray and optical data.
Three (likely four) CT AGN have already been
found in the 13~deg$^2$ area probed by the first 40-month
Serendipitous Survey sample in \citet{L2017b}. This
number has to be regarded as lower limit given the 70\% redshift
completeness and the hardness ratio approach used to find extreme
heavily obscured candidates.

To make significant progresses in this field, an X-ray observatory with more than one order of magnitude larger collecting area, sub-arcmin PSF and higher energy coverage such as the High-Energy X-ray Probe\footnote{https://pcos.gsfc.nasa.gov/physpag/probe/HEXP\_2016.pdf}, is required. It will allow to: (1) resolve the great majority ($\sim80-90$\%) of the CXB at its spectral energy density peak, (2) break the degeneracy between high energy spectral features and the abundance of CT sources and (3) perform detailed studies on the CT source population.  Despite the modest technology development required which is based on the \nustar\ heritage, such instrument is not currently planned for the next decade.

\begin{acknowledgments}
We thank the anonymous referee for the uselful comments which helped improve our analysis and the presentation of our results. 
We thank Johannes Buchner and Yoshihiro Ueda for providing machine readable results on the absorbed fractions. LZ thanks R.~Gilli, F.~Nicastro, E.~Piconcelli, R.~Valiante, F.~Vito and D.~Burlon for useful  discussions. 
LZ, AC, FF, GL and AM acknowledge financial support under ASI/INAF contract
  I/037/12/0. AC acknowledges the Caltech Kingsley fellowship program. GL acknowledges financial support from the CIG grant “eEASY” No. 321913. FEB acknowledges support from CONICYT-Chile (Basal-CATA PFB-06/2007, FONDECYT Regular 1141218), the Ministry of Economy, Development, and Tourism's Millennium Science Initiative through grant IC120009, awarded to The Millennium Institute of Astrophysics, MAS. 
  This    work    was    supported    under    NASA    Contract NNG08FD60C,  and  made  use  of  data  from  the  \nustar\  mission,  a  project led  by  the  California  Institute  of  Technology,  managed  by  the  Jet  Propulsion Laboratory, and funded by the National Aeronautics and Space Administration.
  We thank the \nustar\ Operations, Software and Calibration teams for support with the execution and analysis of these observations. 
  This  research  has  made  use  of  the \nustar\ Data  Analysis  Software (NuSTARDAS)  jointly  developed  by  the  ASI  Science  Data  Center  (ASDC,  Italy)  and  the  California  Institute  of Technology (USA).  
\end{acknowledgments}

\appendix
\section{Notes on single sources}\label{singlesource}
\subsection{COSMOS}
The evaluation of possible contribution to our spectra from additional flux from sources located within the spectral extraction radii is performed using the source information provided by the \cosmos-Legacy catalog \citep{C2016} and the low-energy spectra extracted by \citet{M2016a}. We did not try joint broad-band models with the \nustar\ data (except for the CT source {\it cosmos330}) through complicated models but rather used a flux estimate at low energy with a simple absorbed power-law model to  evaluate possible contamination in common energy bands.
\begin{itemize}
\item {\it cosmos107}: The \nustar\ extraction radius includes two sources of similar flux in the \cosmos-Legacy field: lid1689, lid1688. The \xmm\ spectrum includes them as well. The source lid1688 has twice the number of net-counts as lid1689 and is the obscured one (i.e., low energy counterpart).  It has an Fe~K$\alpha$ line at the optical spectroscopic redshift. The hard-band \nustar\ flux for {\it cosmos107} from the baseline model gives a flux of $9\times10^{-14}$~\fluxcgs. 
  Assuming the absorbed source accounts, as suggested from the number of \chandra\ detected net-counts, for 2/3 of the 8-24~keV \nustar\ flux, this source would have a \nustar\ flux of  $\sim6\times10^{-14}$~\fluxcgs\ (i.e., below the threshold defining our sample),  potentially drop it from the sample. 
\item {\it cosmos129}: The \nustar\ extraction radius includes two \chandra\ sources: cid284 (the low energy counterpart) and cid818. The latter is at the edge of the \nustar\ extraction radius and its spectrum has 30 net-counts (0.5-8~keV) and is consistent with being unabsorbed. The \xmm\ extraction radius does not include this source. Hence only the \nustar\ spectrum is potentially affected by cid818. However, at 5-7~keV cid818 flux is one to two orders of magnitudes fainter than cid284, so we can safely assume that the  \nustar\ measurement is not significantly affected by cid818. 
\item {\it cosmos154}: The scattered components in \chandra\ and \xmm\ seem to have different shapes ($\Gamma$) and normalizations, suggestive of source variability. The source is close (several arcsec) to another AGN (cid366). \chandra\ does not include the latter source while \xmm\ partially does (it is heavily blended). Moreover \chandra\ seems to exhibit diffuse emission around the source which is probably partially included in \xmm. We have decided to exclude the \xmm\ data and use only \chandra\ (which has the highest statistics). \nustar will include both CID366 emission and thermal diffuse emission. However they are relevant only at energies lower than 3-4 keV. The cid366 spectrum level at 3 keV is comparable with cosmos154 while at 4 keV there is a factor $\sim10$ difference. We decided to use \nustar\ only at energies above 4 keV. 
\item {\it cosmos178}: The \nustar\ extraction radius includes five sources. Three of them dominate in terms of counts: cid168 (the low-energy counterpart), cid190, cid192. The 0.5-8~keV (3-8~keV) flux of the first one, i.e., the low-energy counterpart, is $5.1\times10^{-14}$~\fluxcgs\ ($2.7\times10^{-14}$~\fluxcgs). The other two have lower fluxes,  $1.7\times10^{-14}$~\fluxcgs\ ($9.5\times10^{-15}$~\fluxcgs) and $2.2\times10^{-14}$~\fluxcgs\ ($6.8\times10^{-15}$~\fluxcgs). The \xmm\ spectrum includes only the cid168 source. The fluxes are comparable within a factor of  3. Therefore the \nustar\ spectrum includes flux from all the three sources. The 8-24~keV flux of this source is  $\sim7\times10^{-14}$~\fluxcgs, implying that the cid168 flux is very likely fainter. This source would potentially be dropped from the sample. 
\item {\it cosmos181}: The  \nustar\ extraction radius includes four sources. Out of these, two are very faint and the other two dominate the total number of counts. These are: cid482 (the low-energy counterpart) and cid484. Below 4-5~keV both sources have comparable \chandra\ fluxes although cid482 is a factor of few higher in flux; the contribution at 4-5~keV of cid484 is not negligible. Therefore we decided to limit the \nustar\ range for spectral fitting to 4.5-24~keV. The \xmm\ spectrum does not include cid484. 
\item {\it cosmos206}: The  \nustar\ extraction radius includes two sources, cid329 (the low-energy counterpart) and cid328. The latter is more than one order of magnitude fainter at all energies. Therefore it should not significantly affect our modelings. 
\item {\it cosmos207}: The \nustar\ extraction radius includes two sources separated by 33~arcsec: lid1645 (the low-energy counterpart) and lid1644. The counterpart has a factor of $\sim 5$ more counts than lid1644 in the 3-8~keV \chandra\ spectral range. Therefore the \nustar\ spectra should not be substantially contaminated by the latter source. 
\item {\it cosmos229}: The small \nustar\ extraction radius (25~arcsec) includes two sources separated by 26~arcsec: cid420 (the low-energy counterpart, offset 14~arcsec from the \nustar\ position of cosmos229) and cid 1120. In the \chandra\ data,  cid420 has a 3-8~keV flux which is $1.2-4$ ($1\sigma$ range) times higher than cid1120. Therefore it is likely that at least the soft \nustar\ band is contaminated to some extent by the fainter source. Because of this the source would potentially be dropped from the sample since its hard-band \nustar\ flux would potentially fall below threshold. 
\item {\it cosmos297}: The \chandra\ spectrum slightly differs at very low energies ($\sim$0.5-0.7~keV) from the \xmm\ spectrum, though has very little impact on our modeling;
\item {\it cosmos330}: The \nustar\ extraction radius includes two sources separated by 26~arcsec: lid1791 (the identified low-energy counterpart) and lid1792. In the 3-8~keV band the counterpart is a factor $\sim2-2.5$ (a factor of $\sim2$ in the \chandra\ collected net counts) brighter than the latter contaminant source. We fit the {\it cosmos330} spectra jointly with the \chandra\ spectrum of lid1792 to account for its contamination and recover the intrinsic spectral parameters for lid1791. We find that the spectral parameters of {\it cosmos330} do not appreciably vary and that the source is still classified as CT. 
\end{itemize}
\subsection{ECDFS}
\begin{itemize}
\item {\it ecdfs5}: This source does not have unique counterparts in M15. There is one at low redshift ($z=0.141$; \chandra\ ID~103) and one at high-redshift ($z=1.957$; \chandra\ ID~100). Their separation is $\sim22$~arcsec which is smaller than the \nustar\ spectral extraction radius. We therefore tried a joint modeling of these two sources with the \nustar\ data. Both sources are unabsorbed with $\Gamma\approx2.1$. The \nustar\ spectra in the common 3-8~keV band have normalizations which are  higher than the \chandra\ spectra by factors of $\sim2$ and $\sim5.4$ for  ID~103 and ID~100, respectively. In the 8-24~keV band, ID~103 has a flux $\sim3.5$ times larger than ID~100. We therefore assume that ID~103 is the correct low-energy counterpart and used it in our analysis.  
\end{itemize}
\subsection{EGS}
\begin{figure}[t!]
   \begin{center}
\includegraphics[height=0.285\textwidth,angle=0]{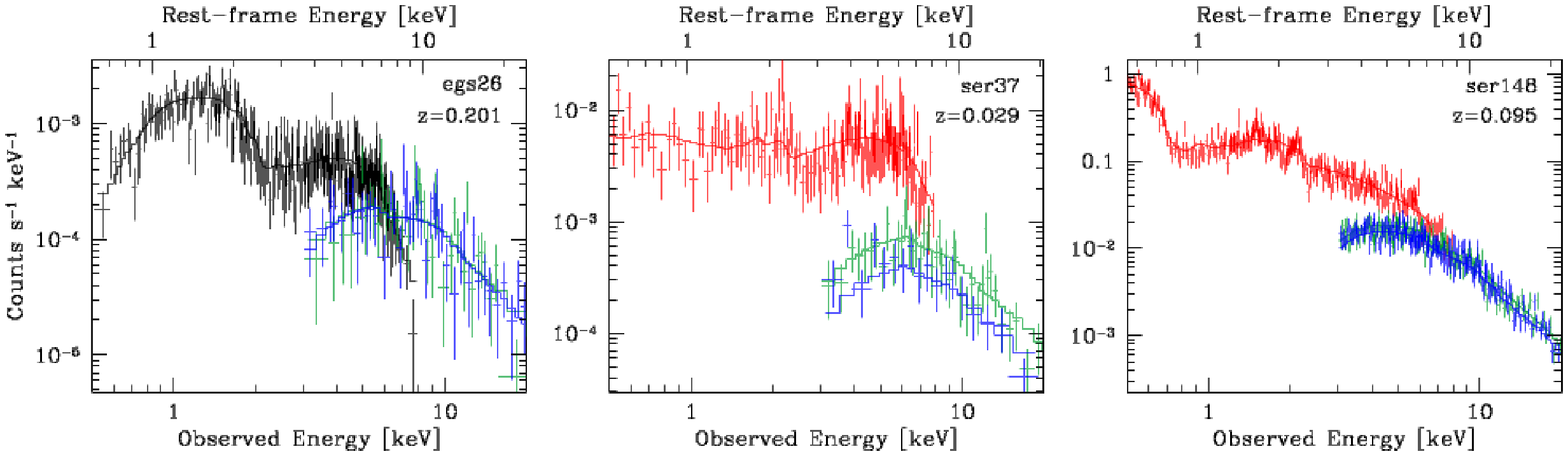}
   \end{center}
\caption{Broad-band spectra and relative best-fit model for sources {\it egs26}, {\it ser37} and {\it ser148}.  The adopted best-fit model is the baseline one modified with the addition of low-energy ionized/neutral partial covering absorber. Black, red, green and blue spectra refer to \chandra, \xmm, \nustar-FPMA and \nustar-FPMB, respectively.} 
   \label{spectra_addabs}
\end{figure}

\begin{itemize}
\item {\it egs26}: The spectrum of this source is flat and unabsorbed. A fit with an absorbed power-law returns a best-fit $\Gamma=0.9$ with negligible column density for which we place an upper limit at $\lognh\leq20.3$ with $Wstat/dof=472.31/433$. Our baseline parametrization returns an apparently better fit with  $Wstat/dof=442.9/431$ with $\Gamma=2.37$, $\lognh\leq21.3$ and $R\approx67$. The reflection parameter value is extremely high and unphysical. We therefore tried to add a dual-absorber modeling \citep[e.g., ][]{C1999,D2004} to the primary power-law, i.e., a further absorption component given by an inhomogenous cold medium at larger scales by employing the model {\sc zpcfabs}. With this parametrization, we obtained an even better fit ($Wstat/dof=424.4/429$) with more reasonable parameters (as reported in Table~\ref{specparambase}): $\Gamma=1.56$, $\lognh\leq21.1$ and $R<0.18$. For the second absorber we find $\lognh=23.40_{-0.02}^{+0.05}$ and a covering fraction $f_c=0.73\pm0.03$. Spectra and best-fit model are reported in Fig.~\ref{spectra_addabs}. 
\end{itemize}

\subsection{Serendip}
\begin{itemize}
\item {\it ser243}: This source has $\Gamma$ flatter than the canonical value at high significance (see Table~\ref{specparambase}). It also seems to require a  scattered component (in \chandra\ data at energies below 0.5 keV) with $\Gamma_{sc}\simeq2.4$. Freezing the primary photon index to the canonical  $\Gamma=1.8$ makes the scattered component steeper, $\Gamma_{sc}\simeq3$. In this case the reflection strength raises from $R<0.74$ to $R=1.1$.
\item {\it ser318}: The \xmm\ and \nustar\ data are simultaneous, though the spectra do not  to agree well at 3-4 keV:  \xmm\ has a factor of 7-8 fewer  counts in this range. \nustar\ shows a hint of Fe~K$\alpha$ at $\sim3$~keV.
\item {\it ser37}: The spectrum of this source is very flat and unabsorbed. A fit with an absorbed power-law returns $\Gamma=0.2$ and negligible column density. However the modeling is not acceptable, with strong residuals across the broad-band and fit statistics of $Wstat/dof=345.04/218$. The baseline modeling yields a much better parametrization with $Wstat/dof=225.77/217$. The best-fit parameters are $\Gamma\approx2.5$, $\lognh\leq19.9$ and $R=10.3$. Both $\Gamma$ and $R$ are too high. Therefore, as done with {\it egs26}, we tried a dual absorber model modifying the primary power-law with cold and partially ionized partial covering absorbers ({\sc zpcfabs} and {\sc zxipcf}, respectively). The model which gives the better parametrization in terms of fit statistics and reasonable parameters (see Table~\ref{specparambase}) is obtained using the warm ionized absorption model and imposing $\Gamma=1.8$. For the absorber we obtained the following parameters: $\lognh=22.80_{-0.03}^{+0.02}$, $f_c=0.969_{-0.006}^{+0.003}$ and a ionization parameter $\log(\xi/\rm erg\,cm\,s^{-1})=-0.55_{-0.19}^{+0.10}$. Spectra and best-fit model are reported in Fig.~\ref{spectra_addabs}. 
\item {\it ser267}: In our spectral analysis we treated this AGN as a canonical unabsorbed source. 
  The joint \chandra and \nustar\ FPMA+FPMB low quality spectra (i.e., 45, 27 and 38 total net-counts, respectively) are jointly modelled with an unabsorbed baseline model with primary continuum slope consistent with the canonical $\Gamma=$1.8-2 value. On the other hand, in the \nustar\ spectra (i.e., in both focal plane modules) we find evidence of a strong residual at an energy of $\sim5.7$~keV (10 net-counts in both focal plane modules). The significance of this feature if modelled with a Gaussian line is at the $\sim2\sigma$ level (based on $\Delta\chi^2$ confidence contours on line energy and normalization). The line can be modelled by a 6.4~keV Fe~K$\alpha$ at the redshift $z=0.131$ of the source. In our best-fit baseline parametrization the line has an observed $EW\approx1.4$~keV. The low quality \chandra\ spectrum is consistent with this best-fit line solution. This may be an  indication that the source  hosts an obscured AGN. We mention however that the optical spectrum from SDSS shows broad lines pointing to a Type~1 classification for this source. Clearly better X-ray data across the broad 0.5-24~keV band are needed to shed light on the nature of this source and properly assess the significance of the line as Fe~K$\alpha$. 
\item {\it ser148}: A simple power-law model shows a flat spectrum with $\Gamma=1.4$ with strong residuals at low energies ($Wstat/dof=3143.79/1261$). A simple cold absorption is not required. The baseline model does not improve the modeling. Reflection is not required as the large residuals are at soft energies ($<1-2$~keV). We therefore further tried additional warm absorption ({\sc zxipcf}) on the primary component and obtained a good representation of the spectrum (see Table~\ref{specparambase}) with the following warm absorber parameters: $\lognh=22.76\pm0.01$, $f_c=1.0$ and a ionization parameter $\log(\xi/\rm erg\,cm\,s^{-1})=1.39\pm0.05$. Spectra and best-fit model are reported in Fig.~\ref{spectra_addabs}. 
\item {\it ser261}:  No observations at low energy  are available for this sources from either  \chandra\ or \xmm. There are short observations ($<10 \rm ks$)  from {\it Swift}-XRT in which a source is barely detected  $10-20$~arcsec from the \nustar\ position. If this is the right low-energy counterpart it is difficult to model the joint {\it Swift}-XRT-\nustar\ spectrum. It results in a heavily absorbed source (from \nustar\ data) with a large scattered component (from {\it Swift}). We mention that the SDSS spectrum of the optical counterpart shows evidence of broad emission lines. 
\end{itemize}

\bibliography{nuspec}

\end{document}